\documentclass[12pt, preprint]{aastex}

\shorttitle{Photometry Local Group Dwarfs}
\shortauthors{Massey et al.}
\slugcomment{V3.5 SUBMITTED Dec 19}

\begin{document}
\title{A Survey of Local Group Galaxies Currently Forming Stars: \\II. 
{\it UBVRI} Photometry of Stars in 
Seven Dwarfs 
and a Comparison of the Entire Sample}

\author{Philip Massey,\altaffilmark{1, 2} 
K. A. G. Olsen,\altaffilmark{3}
Paul  W. Hodge,\altaffilmark{4}
George H. Jacoby,\altaffilmark{5}\\
Reagin T. McNeill,\altaffilmark{1,6}
R. C. Smith,\altaffilmark{3}
Shay B. Strong \altaffilmark{1,7}}

\altaffiltext{1}{Lowell Observatory, 1400 W. Mars Hill Rd., 
Flagstaff, AZ 86001; Phil.Massey@lowell.edu.}

\altaffiltext{2} {Visiting Astronomer, 
Kitt Peak National Observatory, National Optical 
Astronomy Observatory (NOAO), which is operated by the Association of
Universities for Research in Astronomy (AURA), Inc., under cooperative agreement
with the National Science Foundation (NSF).}

\altaffiltext{3} {Cerro Tololo Inter-American Observatory, NOAO, which is operated by AURA, Inc., under cooperative agreement
with the NSF; kolsen@noao.edu, csmith@noao.edu.}

\altaffiltext{4}{Department of Astronomy, University of Washington, 
Seattle WA 98195; hodge@astro.washington.edu.}

\altaffiltext{5}{WIYN Observatory, P. O. 26732, Tucson, AZ 85726-6732; jacoby@wiyn.org.}

\altaffiltext{6}{Participant in the Research Experiences for Undergraduates (REU)
program at Lowell Obs., summer 2006.  
Present address: Five College Astronomy Dept., 
Smith College, McConnell Hall, Northampton, MA 01063; rmcneill@email.smith.edu.}

\altaffiltext{7}{Participant, REU program at
CTIO, 2001. Present address: Department of Astronomy, RLM 16.318,
University of Texas,  Austin, TX 78712-1083; sholmes@astro.as.utexas.edu}

\begin{abstract}

We have obtained {\it UBVRI} images with the Kitt Peak and Cerro Tololo
4-m telescopes and Mosaic cameras of seven dwarfs in (or near) the Local Group,
all of which have known evidence of recent star formation: IC10, NGC 6822, WLM,
Sextans B, Sextans A, Pegasus, and Phoenix.  We construct color-magnitude diagrams
(CMDs) of these systems, as well as neighboring regions that can be used to evaluate
the degree of foreground contamination by stars in the Milky Way.  Inter-comparison
of these CMDs with those of M31, M33, the LMC, and the SMC permits us to determine
improved reddening values for a typical OB star found within these galaxies.  All of
the CMDs
reveal a strong or modest number of blue supergiants.  All but Pegasus and Phoenix
also show the clear presence of red supergiants in the CMD, although IC10 appears to be deficient in these objects given its large WR population.  
The bright stars of intermediate
color in the CMD are badly contaminated by foreground stars (30-100\%), and considerable
spectroscopy
is needed before statistics on the yellow supergiants in these systems will be known.
This study is intended to serve  both as the impetus and ``finding charts" for further space-based imaging,
and for many spectroscopic programs at large aperture.

\end{abstract}


\section{Introduction}
\label{Sec-intro}

Studies of the resolved stellar content of nearby galaxies provide the only
direct way of determining the effect that metallicity and other environmental
factors play in the formation and evolution of massive stars.  
For instance, determinations of the
initial mass function (IMF) in SMC, LMC, and Galactic clusters have
established that no statistically significant differences are seen in the
slope of the IMF (Massey 1998a; Elmegreen 1999; Kroupa 2001; Massey 2003).
Similarly, evolutionary models of massive stars do a good job of matching
the distribution of main-sequence stars in the H-R diagrams of  the Magellanic Clouds (Massey et al.\ 1995b), 
and the location of red supergiants in the SMC, LMC, and
Milky Way (Levesque et al.\ 2005, 2006).  
However, the range of metallicity
in such studies is only a factor of four (SMC to Milky Way).
With the advent of multi-object spectrographs on 8-m class telescopes it is now
possible to push such studies to the more distant galaxies of the Local Group and
beyond, where the range of metallicities is a factor of 20-30, 
and which include
examples of both relatively quiescent galaxies 
and energetic starburst systems.  (See Table~1 of Massey 2003, plus 
Table~\ref{tab:galaxies} of the present paper.)

This is the second of three papers  presenting the results of our survey
of the stellar content of Local Group galaxies currently forming stars.
We used the 4-m telescopes at Kitt Peak and Cerro
Tololo with the Mosaic CCD cameras to obtain images in {\it UBVRI}, as
well as 50\AA-wide interference filters centered on H$\alpha$, [OIII], and
[SII].  (Our survey excluded the SMC and LMC, which have been surveyed by
other programs; see Massey 2002 and references therein, and Smith
et al.\  2005.)
Our goal was to reach 2\% photometry in the broad-band filters
for massive stars, and to use the interference filter imaging to identify H$\alpha$ emission-lined stars.  
The first paper (Massey et al.\ 2006, hereafter Paper~I) dealt with {\it UBVRI} photometry of
stars in M31 and M33, the two spiral galaxies in our sample, and the third paper
(McNeill et al.\ 2007)
will discuss the identification and spectroscopy of the H$\alpha$ emission-lined stars in
all nine of the galaxies in our sample.

We note that our primary goal is to lay out the photometry and CMDs as a first
step for studying the massive stars of these systems.  Many of these galaxies have
had detailed studies (usually of small regions) with {\it HST} that have been used
to constrain the star formation histories (SFHs) of these objects.  While we provide
cursory mention of some of these, the interested reader is referred to van den Bergh (2000) and others for a more complete discussion.

In this paper, we 
continue with an analysis of the broad-band photometry of seven dwarfs:
IC10, NGC~6822, WLM, Sextans~B, Sextans~A, Pegasus, and Phoenix.
We list their properties in Table~\ref{tab:galaxies}.
Of these, Sextans~B and Sextans~A are located in the outer fringes of
the Local Group, and are probably not bound to the system (van den Bergh 1994),
while the others are all true members and show varying degrees of current star
formation (van den Bergh 2000).  Regretfully, poor weather and limited observing
time precluded the observation of IC1613, but we hope to rectify this omission
in the future.   Throughout this paper, we will
discuss these galaxies in order of decreasing luminosity.  In \S~\ref{Sec-data} we
describe our observations and reduction procedures, and present the photometry
catalogs.  In \S~\ref{Sec-results} we will compare the color-magnitude diagrams
(CMDs) in turn, and derive improved values of the reddenings of the complete
sample (including M31 and M33).  In \S~\ref{Sec-summary} we will summarize
our results.

\section{Observations and Data Reductions}
\label{Sec-data}

The observation and reduction
procedures for the Mosaic imaging data were described in considerable
detail in Paper~I, and we do not repeat these here, other than to note that the data
were taken using dithered exposures and that the photometry was performed 
chip-by-chip on each individual exposure rather than on the combined image.
Three southern galaxies
(WLM, NGC~6822, and Phoenix) were observed with the CTIO Blanco 4-m telescope; 
the other four (IC10, Sextans B, Sextans A, and Pegasus)
were observed at Kitt Peak with the Mayall 4-m.   The two Mosaic cameras are
nearly identical instruments, and consist of a two by four array
of 2048x4096 SITe CCDs, yielding images contain 8192x8192 pixels.
The scale of the final rectified images
is 0.27" pixel$^{-1}$. The field of view (FOV) of the Mosaic camera is 35' x 35', 
but for three of the
smaller galaxies (IC10, Sextans~B, and Sextans~A) observed at Kitt Peak
we centered the galaxy on just one of the eight chips in order
to avoid the effects of the ghost image caused by the Mayall's
prime focus corrector.  A fourth galaxy observed at Kitt Peak, Pegasus,
was centered on the array, but the calibration data extends only over the central 4 chips,
as described below.)  The journal of observations is given in Table~\ref{tab:journal},
where we have included the average delivered image quality (DIQ) on our frames.

\begin{figure}
\plotone{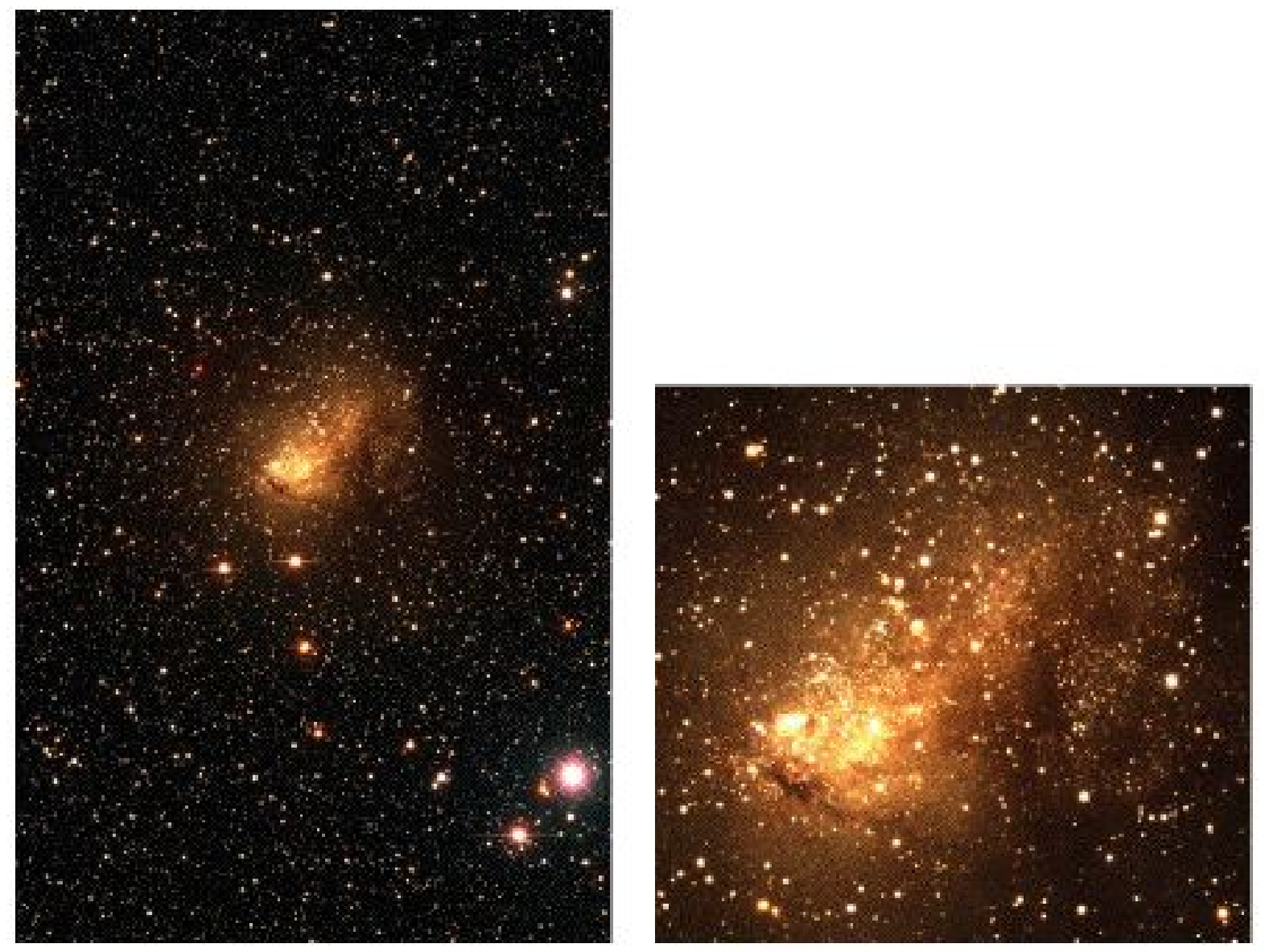}
\vskip -250pt
\caption{\label{fig:ic10} IC10 Mosaic field.  The region on the left shows the entire
20'x30' calibrated region. The region on the right shows an enlargement of a
6'x6' section near the center.}
\end{figure}

\begin{figure}
\plotone{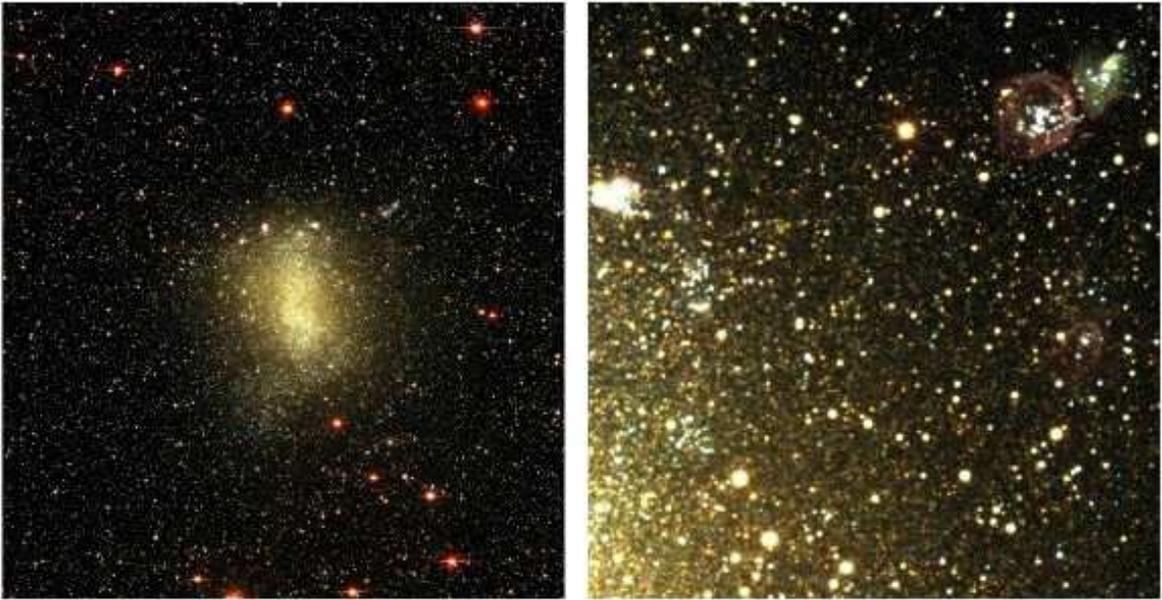}
\vskip -150pt
\caption{\label{fig:n6822} NGC 6822 Mosaic field.  The region shown on the left is the
entire 35'x35' FOV, all of which was calibrated.  The region on the right shows an enlargement of a 6'x6' section in the NW corner of the galaxy.}
\end{figure}

\begin{figure}
\plotone{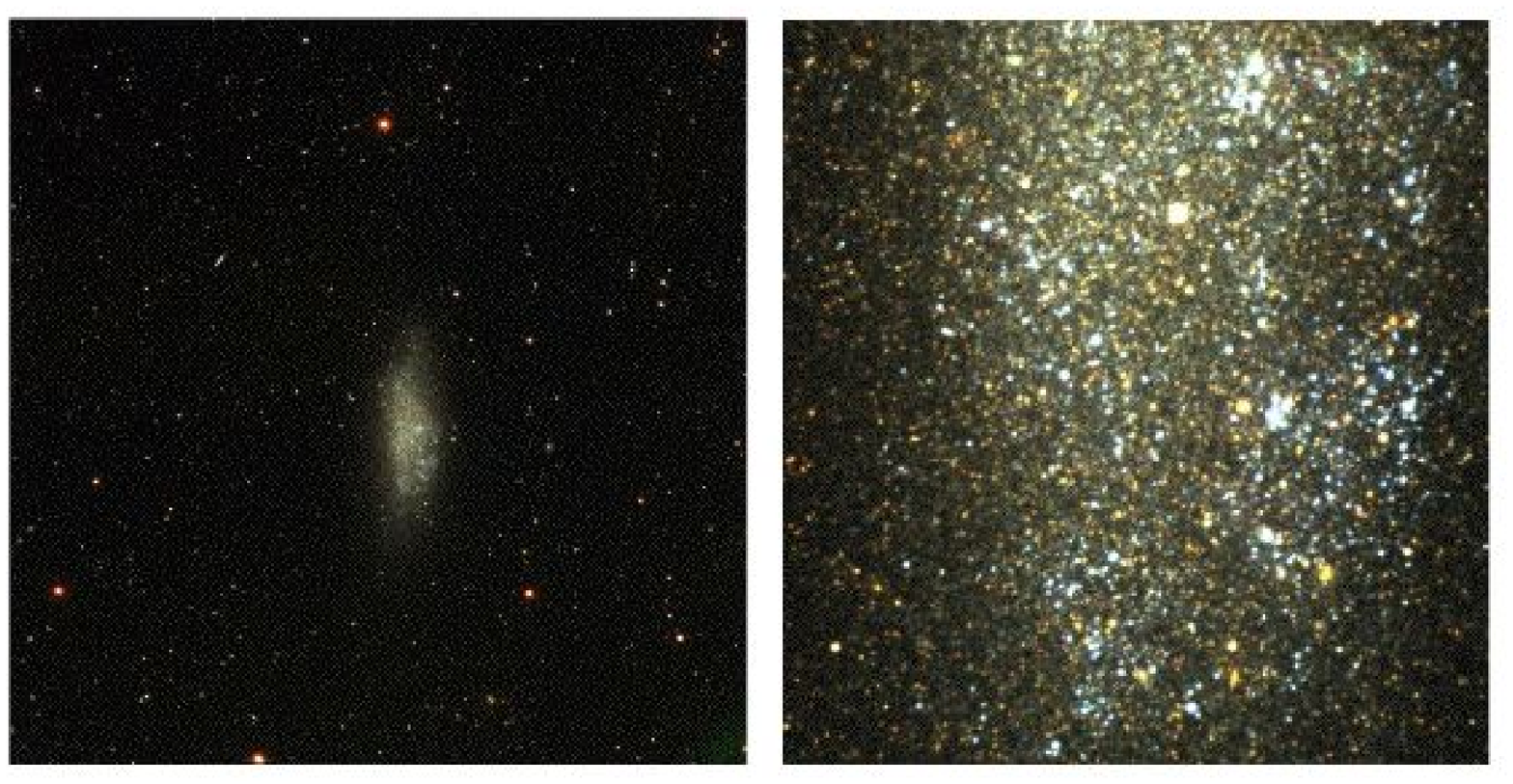}
\vskip -150pt
\caption{\label{fig:WLM} WLM Mosaic field. The region shown on the left is the
entire 35'x35' FOV, all of which was calibrated.  The region on the right shows an enlargement of a roughly 3.7'x3.7' section south of the galaxy's center. }
\end{figure}

\begin{figure}
\plotone{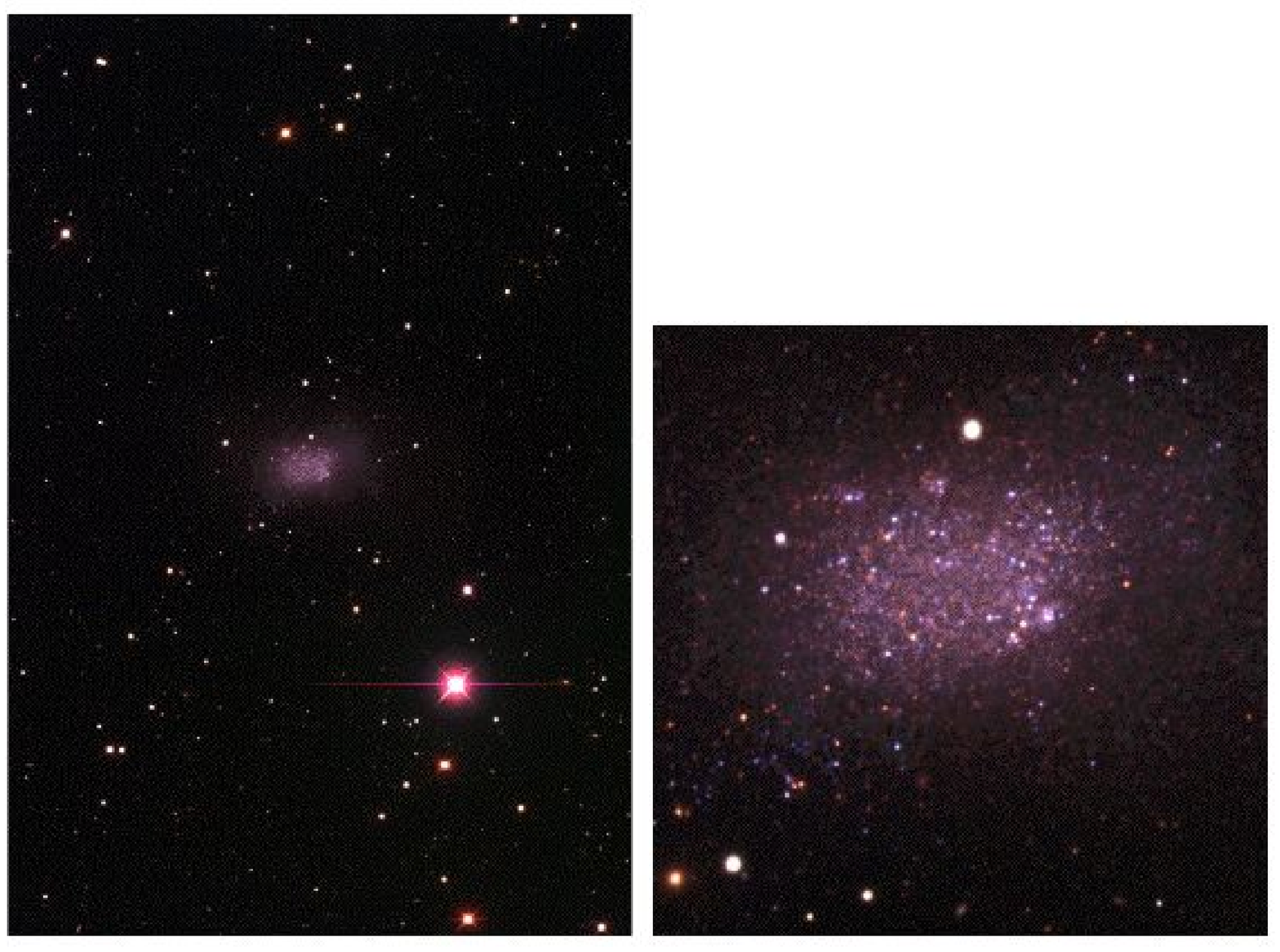}
\vskip -150pt
\caption{\label{fig:SexB} Sextans B Mosaic field.  The region shown on the left is
the 20'x30' calibrated region.  The region on the right shows an enlargement of a roughly 3.8'x3.8' section near the middle of the galaxy.}
\end{figure}

\begin{figure}
\plotone{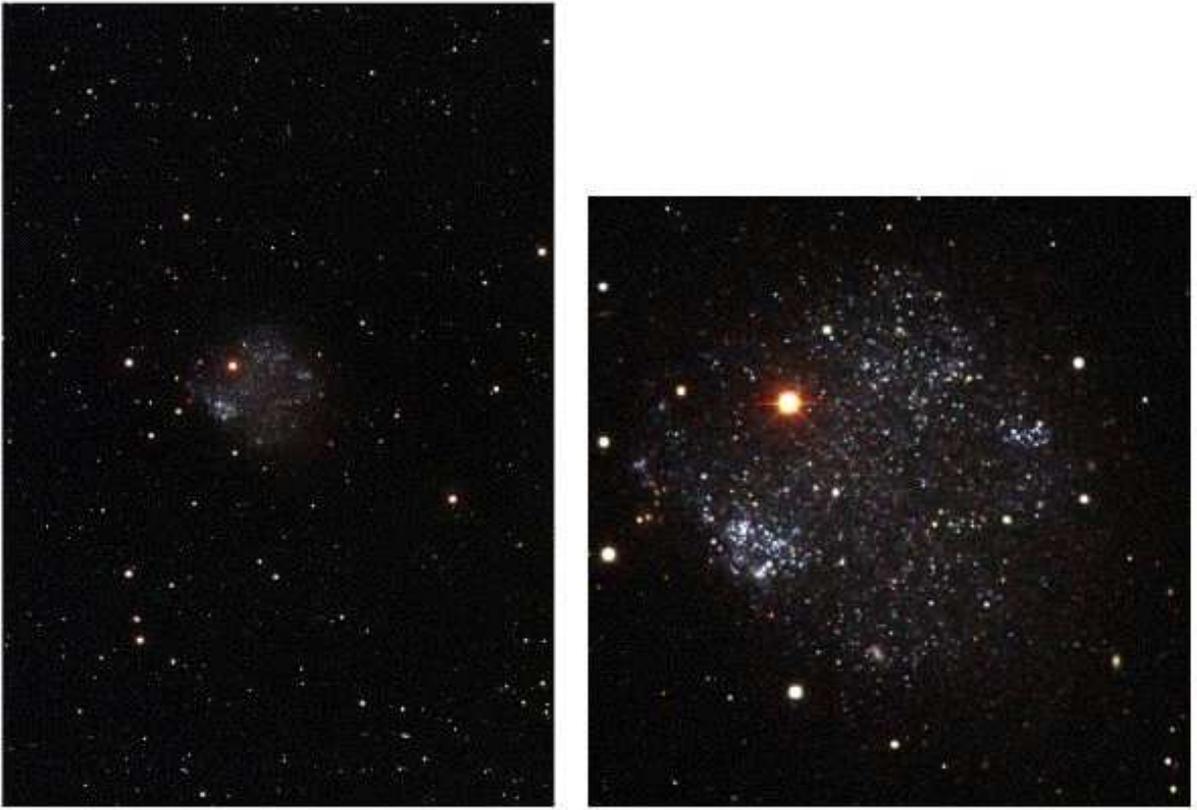}
\vskip -150pt
\caption{\label{fig:SexA} Sextans A Mosaic field. The region shown on the left is
the 20'x30' calibrated region.  The region on the right shows an enlargement of a roughly 6'x6' section centered on the galaxy.}
\end{figure}

\begin{figure}
\plotone{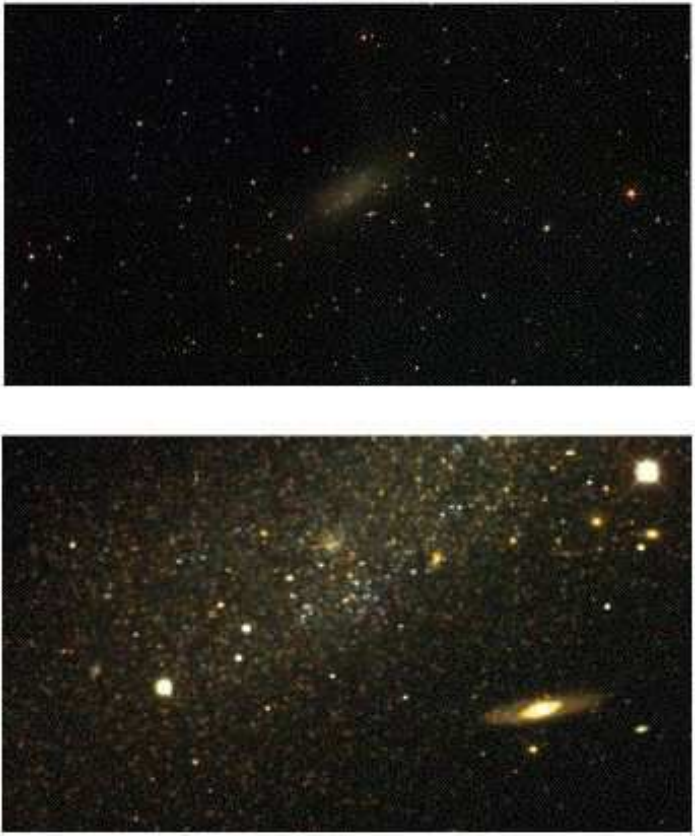}
\vskip -150pt
\caption{\label{fig:Peg} Pegasus Mosaic field.  The region shown on the top is
the 20'x30' calibrated region. The region below it an enlargement of a roughly 4.2'x2.4' section south of the center of the galaxy.}
\end{figure}

\begin{figure}
\plotone{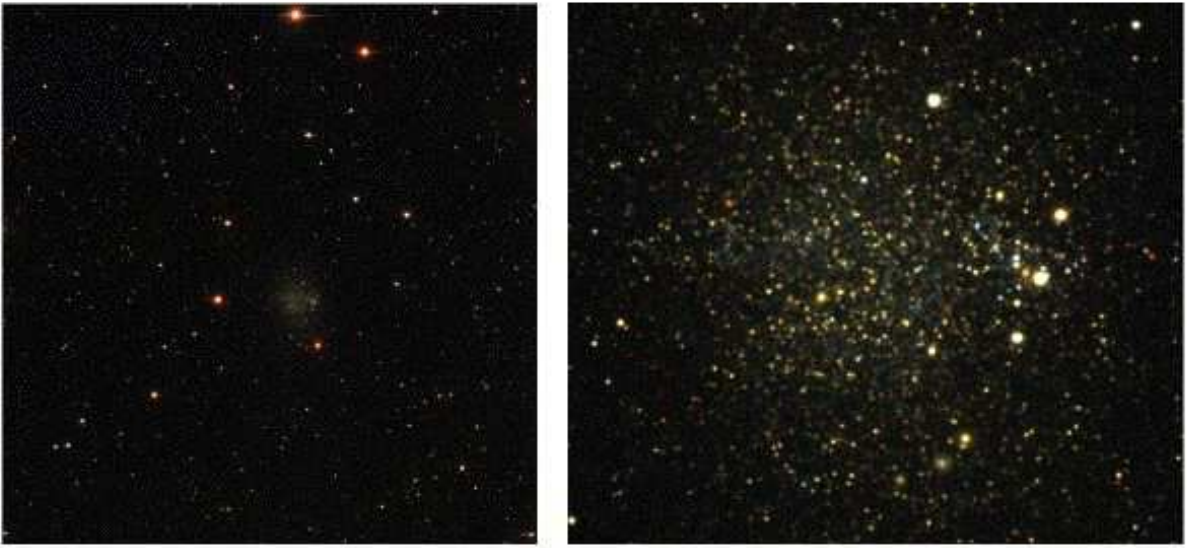}
\vskip -150pt
\caption{\label{fig:PHX} Phoenix Mosaic field. The region shown on the left is the
entire 35'x35' FOV, all of which was calibrated.  The region on the right shows an enlargement of a roughly 4.5'x4.5' section centered on the galaxy.}
\end{figure}

One of the strengths of our survey has been the large field of view
(35'x35') in a single field.  The spirals described in Paper I
required multiple fields (M31, ten fields; M33, three fields) for
good areal coverage. For the galaxies studied here, a single field
was sufficient, and even so, the large FOV might seem wasted, given
the small angular extent of some of these galaxies.  However, the
large areal coverage proved useful for determining the contribution
made by foreground Galactic stars to the color-magnitude diagrams
(CMDs).  In Figs.~\ref{fig:ic10} through \ref{fig:PHX} we show the
fields of view corresponding to the calibrated photometry for each
of the seven dwarfs.  For IC10 and NGC 6822 this provides both
complete coverage and a good sampling of the Milky Way foreground;
in the case of the smaller systems (such as Sextans B) the galaxies
are almost lost in the field.  Of course, another strength is
that the sampling is quite good (0.27" pixel$^{-1}$), and
to demonstrate this we have included  enlargements of interesting
areas\footnote{The imaging data are available through the NOAO
Science Archive at http://archive.noao.edu/nsa/, as well as the
Lowell ftp site ftp://ftp.lowell.edu/pub/massey/lgsurvey/datarelease/.
 In addition, we have made full resolution versions of Figs~\ref{fig:ic10}
 through \ref{fig:PHX} available from our web site,
  http://www.lowell.edu/users/massey/lgsurvey.}.

However, the primary strength of our survey is, we believe, in the high accuracy of our
photometric calibration.
Since each CCD has slightly different color responses, 
we decided to treat each of the eight CCDs in each camera as a separate instrument.  
The 150s readout time made it impractical to observe the large number of
standard stars required to determine satisfactory transformation equations for
each chip: simply observing one set of standards
 on each chip in all five filters would require 
$\sim$1.5 hrs just in reading out the array.  Instead, we chose the philosophy of providing an 
external calibration for each field using time on smaller telescopes.  That allowed us to devote most of each calibration night to observing standard stars, which
were carefully chosen from the best-calibrated stars in Landolt (1992).
This allowed extinction and color terms to be determined to high accuracy, and
allowed us to determine highly accurate
secondary calibration for each galaxy field.  
We found in Paper~I that the method worked well, as judged by the milli-magnitude agreement for the results found for stars in overlapping fields in the M31 and M33 data.  Furthermore, that study demonstrated that we were right to be concerned about the differences in the color-terms from chip to chip, as these would have introduced unacceptably large errors in the derived colors of stars had we simply used a single color term for each filter.  Of course, the calibration images need to include 
some of the area for all the chips of interest.  This decision also allowed
us to make use of mostly clear but not-quite photometric 
nights for the 4-m Mosaic imaging, vastly reducing the amount of 4-m time needed
for the project.  That said, a little time was spent on the second CTIO Mosaic
night observing a few Landolt fields; these observations served primarily as a ``reality
check" on the derived color terms, as shown below.

For the galaxies with Mosaic data from the CTIO 4-m (WLM, NGC 6822, and Phoenix),
we used the
CTIO 0.9-m telescope and Tektronix 2048x2048 CCD to obtain {\it UBVRI} images
of four 13.5'x13.5' fields for each galaxy.  The fields were each offset from the
galaxy coordinates by $\pm17.5'$ north-south and east-west in order to provide some overlap with all 8 chips in the Mosaic fields.  These
calibration  data were all taken on the
night of 2001 September 17 (UT).   For the galaxies with
Mosaic data from the Kitt Peak 4-m, 
we used the Lowell Observatory 1.1-m Hall telescope
and SITe 2048x2048 CCD  (FOV 19.4'x19.4')
to obtain {\it UBVRI} images of the galaxy field.
For M31 and M33 (Paper I) we had used two pointings,
offset by 500" north and 500" south of each galaxy field, so as to
provide overlap with the
regions covered  by all 8 Mosaic chips, but since most of the dwarfs are small,
we simply obtained a single 20'x20' field centered on the galaxy for Pegasus,
Sextans B, and Sextans A.  
We also included NGC 6822  (using the two pointings) and WLM
(using a single pointing)
in order to provide overlap in the calibration between the northern galaxies
and the southern.  The Lowell data were obtained on 2001 September 26 (IC10),
2001 October 14 (WLM and Pegasus), 2002 September 10 (NGC 6822), and 
2002 November 2 (Sextans A and Sextans B).

For the northern hemisphere Mosaic data we adopted the color terms from Paper I.
These were based upon averages of 13 independent calibrations (10 fields in
M31 and 3 in M33).  For the southern hemisphere
Mosaic data, we used both the CTIO 0.9-m
and Lowell 1.1-m calibrations of NGC 6822 to determine the color terms for all
8 chips.   We give these values in Table~\ref{tab:colorterms}.  Although these are
not as well determined as the Kitt Peak coefficients (as they are based upon fewer
data), the agreement was very good between the Lowell and CTIO calibration
data for NGC 6822 and WLM, with the exception of the $R-I$ color term.  For
those, we used observations with standards observed with the Mosaic camera
itself to determine that the CTIO 0.9-m calibration was better, and we adopted
those values.  Once the CTIO color-terms were fixed to the values in 
Table~\ref{tab:colorterms}, we then determined the photometric zero-point
in a chip-by-chip manner for each dithered Mosaic exposure for WLM, NGC 6822,
and the Phoenix dwarf using the combined Lowell and CTIO catalogs for WLM
and NGC 6822, and the CTIO catalog for the Phoenix dwarf.

We can test how well this procedure worked by comparing the agreements between
the final catalogs of WLM and NGC 6822 stars and the individual Lowell 1.1-m and
CTIO 0.9-m calibrating catalogs for these galaxies.  We show the median differences
in Table~\ref{tab:agree}, where we have restricted the sample to stars
with small photometric errors ($<$0.01 mag) and which were uncrowded.  We consider
the agreement quite good, although it is not of the same accuracy as that of the
M31 and M33 data in Paper~I, simply owing to the far sparser fields for the
dwarfs.

We present the final photometry catalogs in Tables~\ref{tab:ic10} through
\ref{tab:phx} for the seven dwarfs. As with Paper~I, we have chosen to make the
catalogs coordinates based, with the designation LGGS J001956.99+591707.5 referring to the IC10 star whose coordinates are $\alpha_{\rm 2000}=00^h19^m56.99^s$ and
$\delta_{\rm 2000}=+59^\circ17'07.5"$.
Shortly after Paper~I  was published we discovered a minor problem in this M31 and
M33 catalogs: in regions of extreme crowding there are sometimes two stars within
0.1" of each other, and thus two stars might have the same designation.
There were 62 cases of this in M31 (out of 371,781 stars), and 23 cases in M33 (out of 146,622 stars).  There were no cases where 3 or more stars were involved.  We have avoided this problem here by using an ``A" for the brightest star, and a ``B" for the fainter star.  Since the photometry of stars this crowded is quite dubious, these designations
serve primarily as flags that multiple components were identified.  Rather than allow this confusion to persist in the M31 and M33 catalogs, we are making revised versions
available here in on-line form, as Tables~\ref{tab:m31} and \ref{tab:m33}.  

\begin{figure}
\epsscale{1.0}
\plotone{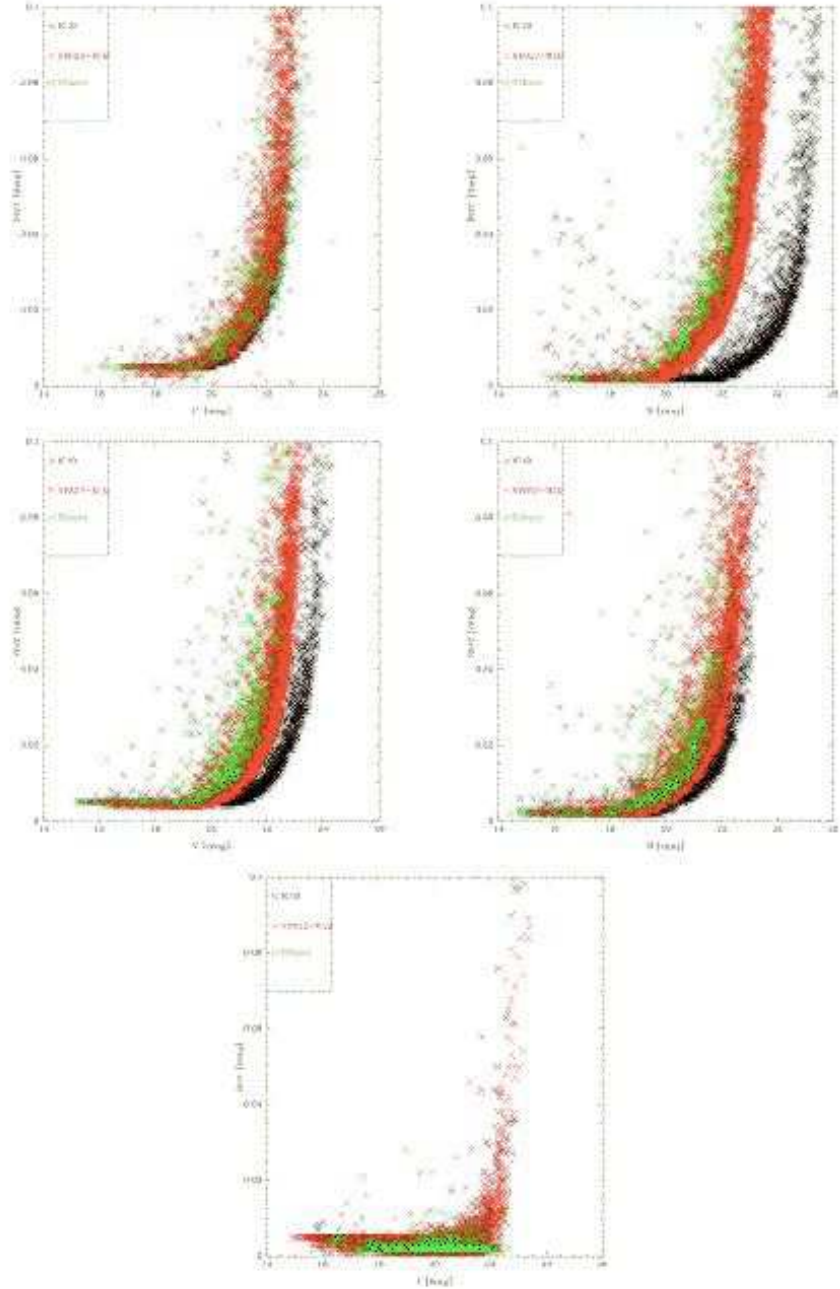}
\vskip -50pt
\caption{\label{fig:errors} Errors as a function of magnitude.  The black points 
are IC10, the red points are the combination of NGC 6822 and WLM,
and the green points come from the other 4 galaxies.  For clarity, we have plotted
only every tenth point.}
\end{figure}

In Table~\ref{tab:errors} we show the typical (internal) errors of
this photometry.  Consistent with the exposure times and conditions given
in Table~\ref{tab:journal}, we found that the errors as a function of magnitude
were comparable for Phoenix, Sex B, Sex A, and Pegasus.  The galaxies
NGC 6822 and WLM had somewhat smaller errors at a given magnitude.
Given our efforts to compensate for the large reddening in IC10, the errors
for that galaxy are smaller still.  We also show the spread in errors in 
Fig.~\ref{fig:errors}.

Our stated goal was to achieve 1-2\% photometry for massive ($>20M_\odot$) 
stars.  Did
we achieve this?   In Table~\ref{tab:bright} we list the expected brightness
of a 20$\cal M_\odot$ star in {\it UBVRI}, where we have adopted the distances
and reddenings from Table~\ref{tab:galaxies}, and intrinsic colors following
Bessell et al.\ (1998) and FitzGerald (1970).  We adopt the reddening
excesses given by Schlegel et al.\ (1998), using their values for the
``CTIO UBVRI system", as the filter responses and 
CCD quantum efficiency is similar to ours.  (We include values for M31,
M33, the LMC, and SMC to aid in interpreting the CMDs later.) 

First, let us consider the most heavily reddened galaxy in our sample,
IC10.  A 20$M_\odot$ star on the zero-age main sequence (ZAMS) 
will be quite faint at $U$ (23.2), with
correspondingly large errors (12\%).  Still, our efforts to compensate for
the reddening by going quite deep at $B$, $V$, and $R$ in IC10 were
largely successful, and although we did not achieve 1-2\% photometry, we
did do better than 5\% for such a star.  At an age of 8~Myr, when the star
is at the end of its main-sequence life 
(i.e., on the terminal age main sequence, TAMS), and the star is an early B-type
supergiant, the photometric errors would be quite small, even at $U$.
During its He-burning red supergiant (RSG) phase, the star would be invisible
at $U$, but have small photometric errors ($<1$\%) in the other bandpasses,
even $B$.

For a 20$M_\odot$ star in the other galaxies a comparison of Tables~\ref{tab:errors}
and \ref{tab:bright} show that we achieved our goals, except for the $U$-band 
during the RSG phase.  Since we require that a star be detected in $B$, $V$, and $R$
to be included in the catalog, but not $U$ or $I$, this should not have any effect on
whether or not a RSG is included in our catalogs.

\begin{figure}
\epsscale{0.8}
\plotone{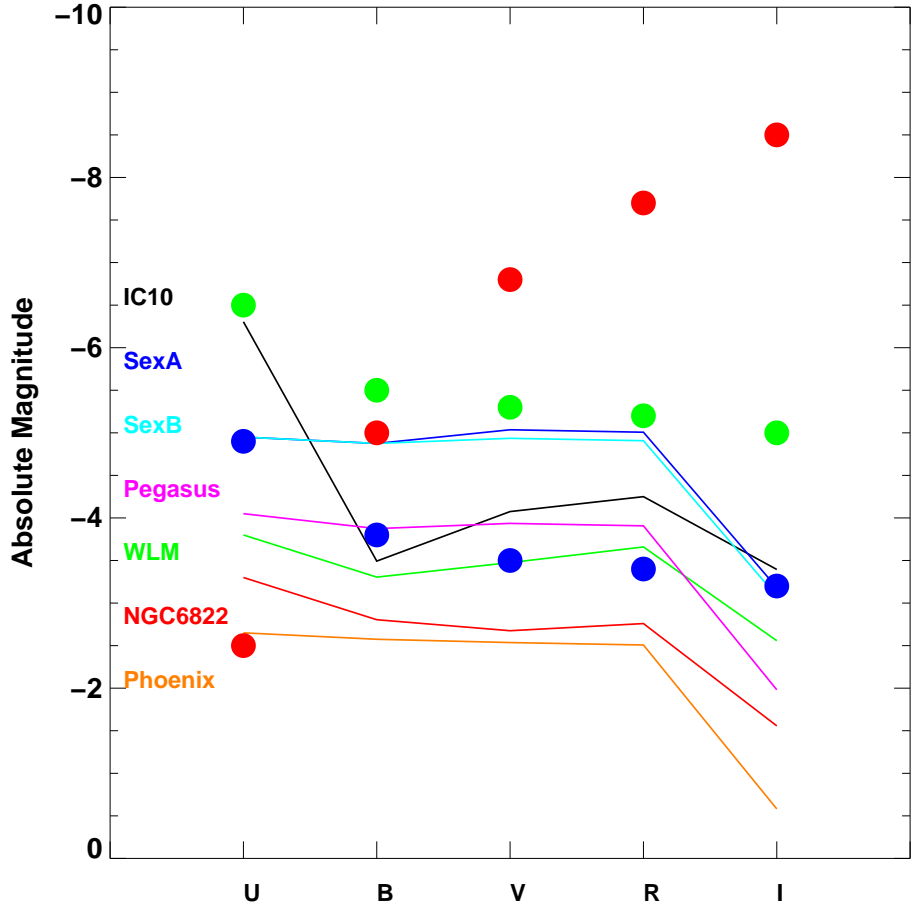}
\caption{\label{fig:knut} Photometric error as a function of absolute magnitude.
The colored dots show the expected absolute magnitude of a 20$M_\odot$ star
on the ZAMS (blue dots), TAMS (green dots), and as a RSG (red dots), for each
of the bandpasses.  The solid color curves show the 2\% photometric errors actually
achieved (Table~\ref{tab:errors}) for
each galaxy.}
\end{figure}

We provide a graphical representation of what we achieved in Fig.~\ref{fig:knut}. 
The large colored dots show the
absolute magnitude in each bandpass (i.e., $M_U$, $M_B$, etc.)  for a 20$M_\odot$
star on the
ZAMS (blue dots), on the TAMS (green dots), and as a RSG (red dots).
For each galaxy, we use a colored line to show where we achieved 2\% photometric
errors in general.  Thus we can see that we readily achieved 2\% or better photometry
for the TAMS, but that for IC10, Sextans A, and Sextans B we didn't quite achieve
this for the ZAMS.  This underscores the point made by Rousseau et al.\ (1978)
and re-emphasized by Massey et al.\ (1995b): massive stars evolve at essentially
constant bolometric luminosity, but that means that a star near the ZAMS will be much
fainter optically than when the star is ``middle-aged" (half way between the ZAMS and
the TAMS) simply owing to the fact that the high temperatures result in very large
bolometric corrections.   Our photometry came close to meeting our original goals,
and the analysis presented here allow us to understand its actual limitations.

\section{Results and Analysis: What the Color-Magnitude Diagrams Tell Us}
\label{Sec-results}

Perhaps the clearest and most useful way to understand what our considerable
photometry is telling us about these galaxies is by the most classical approach
possible: comparing the color-magnitude diagrams (CMDs) of these systems.  
In Figs.~\ref{fig:m31cmd}
through \ref{fig:phxcmd} we present $V$ vs $B-V$ plots for all of the galaxies
in our sample.  To facilitate comparisons, we have also included M31 and M33 
(from Paper~I) and the LMC and SMC (from Massey 2002).  In order to show the
outstanding bright members, we have plotted the individual stars (rather than 
using Hess diagrams) but for illustrating the more densely populated
regions of the CMDs we have included contours of equal star numbers.  (We are
indebted to Lynne Hillenbrand for suggesting this approach.)  In general, we can
distinguish four general regions of these diagrams: the blue supergiants on the
left, the red supergiants (RSGs)
on the right, and two sequences of (mainly) foreground stars near 
$B-V\sim0.6$ and $B-V\sim1.6$ for the
galaxies at modest galactic latitudes.  In order to minimize the effect these foreground stars have
on the CMDs, we have restricted these diagrams to a subset of the field centered on the galaxy,
as described in each figure caption.
Of course, these last two regions will also contain a few
bona-fide supergiants, but these are rare, and their identification will
require a great deal of spectroscopy.  The less luminous galaxies (those with $M_V>-13$ in Table 1) 
have no discernible population of RSGs, owing to their
scant number of massive stars.

\subsection{Degree of Foreground Contamination}
\label{Sec-foreground}

In interpreting the CMDs it is necessary to understand the degree of foreground
contamination.  For this reason, we have constructed CMDs of neighboring
regions for each of the dwarf galaxy fields.   We also use an
updated version of the Bahcall \& Soniera (1980) model of the Milky Way, kindly
provided by Heather Morrison, to estimate the constituents of this 
contamination, and provide an estimate of the correction for our comparison
data (M31, M33, the LMC, and the SMC).

By way of illustration,  
let us begin by considering the CMD
of NGC~6822, which is located 
near to the plane of the Milky Way ($b=-18.4^\circ$).  In Fig.~\ref{fig:n6822cmd} (left)
we see three sequences of bright stars ($V<20$), which we have labeled ``blue supergiants", ``foreground and yellow supergiants", and ``red supergiants", in describe their
major constituents.  In Fig.~\ref{fig:n6822cmd} (middle) we show the CMDs of
the neighboring foregrounds fields.  In the foreground field (which covers the
same area) there are two primary
sequences: one of intermediate color, which (by eye) seems to contain the same
number of stars as in the NGC 6822 CMD, merging into a sequence of redder,
fainter stars.

What fraction of the stars in this region of intermediate color are expected to be
foreground, and which will be yellow supergiants?  By eye it would appear that
foreground objects will dominate overwhelmingly.  Consider only the stars
with $16<V<20$ and $0.5<B-V<1.4$.  
We count 1052 stars in this region of the CMD for NGC 6822, which includes an
area of 0.052 deg$^2$.  We count stars in neighboring regions (whose areas
sum to that used for NGC~6822), and find 991 stars, suggesting that the degree
of contamination is about 94\%.   (Recall that we would
expect variations of $\sqrt{N}$ in either group, so really this should be viewed as 
94$\pm4$\%.)  The Bahcall \& Soneira (1980) model predicts 
a similar but somewhat smaller number of stars (852) in this region of the CMD.
Of these, the model shows that 68\% are disk dwarfs, 12\% are halo dwarfs, and 
20\% are halo giants.  

Next, let us consider the other extreme, the CMD of WLM,
which is located far
from the plane of the Milky Way, at a Galactic latitude $b=-73.6^\circ$.  In the
CMD (Fig.~\ref{fig:wlmcmd}(left)) we find only a few bright
$16<V<20$ stars with intermediate colors ($0.4<B-V<1.0$), where again we have
restricted the CMD to only those stars near the galaxy, a region of 0.023 deg$^2$.
We count only 20 stars in this region of the CMD.   In the foreground field
(Fig.~\ref{fig:wlmcmd} (middle) we count 19 stars within this region of the CMD,
suggesting 95\% contmination by foreground objects.  The Bahcall \& Soniera (1980)
model again slightly underestimates the degree of contamination, predicting only
13 stars, of which only 10\% will be disk dwarfs, while 55\% will be halo dwarfs and 
34\% will be halo giants.  
So, even in this galaxy the foreground contamination 
dominates for bright stars at intermediate colors\footnote{Bresolin et al.\ (2006) estimate
the degree of foreground contamination in this region to be negligible.  We are unable
to reproduce their result.}.

It is of course coincidental that in both cases we find a contamination of 
$\sim$95\%: for NGC 6822
the low galactic latitude leads to a very substantial foreground component, but the galaxy is also
quite rich in stars.  There are far fewer foreground stars in the field of WLM, but the galaxy is also
considerably less rich.  

For our comparison CMDs of M31, M33, the SMC, and the LMC, we do not have neighboring foreground
fields to use with any of these four galaxies, but instead rely upon the Bahcall \& Soniera (1980) models scaled
to the appropriate areas.   For each galaxy, foreground contamination is quite significant for intermediate
colors.  For M31 (Fig.~\ref{fig:m31cmd}) roughly 50\%  of the bright  ($15<V<20$) stars of intermediate
color ($0.4<B-V<1.1$) are expected to be foreground.  According to the model, 86\% will be disk dwarfs, 7\%
will be halo dwarfs, and 7\% halo giants.  For the redder ($1.2<B-V<1.8$) bright stars
($16<V<20$) we expect about 85\% to be foreground, all of which should be disk 
dwarfs\footnote{Note that the various foreground contributions were 
mislabeled in the CMDs of Paper~I as ``foreground dwarfs" and ``foreground giants" for the intermediate
and red colors, respectively.  The same is true for the CMDs in Massey (2002).}.  
For M33, about 40\% of the
bright ($15<V<20$) stars of intermediate color ($0.4<B-V<1.1$) will be foreground stars, of which
73\% will be disk dwarfs, 15\% halo dwarfs, and 12\% halo giants.  For the
redder ($1.2<B-V<1.8$) bright stars ($16<V<20$) we expect about 70\% to
be foreground, made up nearly exclusively of disk dwarfs.  

For the LMC and SMC we consider brighter
stars ($11<V<15$) but find similar percentages.  For the LMC CMD stars of intermediate color
($0.4<B-V<1.1$), about 85\% will be disk dwarfs, 9\% will be disk giants, and 6\% will be halo giants.
For the SMC CMD bright stars of intermediate color
($0.4<B-V<1.1$) about 45\% will be foreground stars,of which 82\% will be disk dwarfs,
6\% disk giants, and 12\% halo giants. 

In our tests, we did find that the Bahcall \& Soniera (1980) model usually underestimated
the number of stars.  For red stars, this can be by as much as a factor of two.
For instance, for WLM we see from Fig.~\ref{fig:wlmcmd} (middle)
that there is also significant contamination in the region of the CMD occupied by RSGs.
We count 27 stars in the WLM field with
$16<V<20$ and $1.1<B-V<1.9$.  The Bahcall \& Soniera (1980)
model predicts only 8.5 stars (32\% contamination, all of them disk dwarfs)
while in our actual background fields we count 20 (74\% contamination).    

It is true, however,
 that in recent years there is an increased appreciation of the large angular
extent of these galaxies, due to (for instances) tidal tails.  The best known of these
is the extended tail of  M31 (Ibata et al.\ 2001).
It may be that even our Mosaic fields do not extend sufficiently far away to provide
a clean foreground sample for the fainter stars.  For instance, Komiyama et al.\ (2003)
have identified NGC~6822 stars that have formed far (25') from the main body
of the galaxy, tracing the H~I distribution. A few of these
stars may be visible as the clump of blue stars at $V\sim22$ visible in the
CMD of our ``foreground" field in Fig.~\ref{fig:n6822cmd}(middle).  Nevertheless,
our foreground fields are useful for distinguishing the bona-fide massive star
populations.

We summarize the degree of foreground contamination in Table~\ref{tab:fore}, where we have chosen
the regions of the CMD to examine by eye, and used the Bahcall \& Soniera (1980) model to estimate
the fractional contribution by different components of the Milky Way.  Since the model occasionally
underestimates the degree of contamination, these percentages are primarily illustrative.

\subsection{Reddenings}
\label{Sec-reddenings}
We can use these CMDs to redetermine the reddening of these galaxies, by
using the location of the plume of blue supergiants, following Massey \& Armandroff
(1995).  We use as our references the CMDs of the LMC and SMC
(Figs.~\ref{fig:lmccmd} and ~\ref{fig:smccmd}), and adopt $E(B-V)=0.13$ and
$E(B-V)=0.09$ for the two galaxies, respectively, following 
Massey et al.\ (1995a)\footnote{Van den Bergh (2000) suggests that the Massey et al.\ (1995b) result
of $E(B-V)=0.13$ for the LMC is possibly too low, noting that Harris et al.\ (1997) 
derive $E(B-V)=0.20$ from a larger sample.  Massey et al.\ (1995b) used a sample
of 414 LMC OB stars with known spectral types to derive a median reddening,
comparing the observed $B-V$ color with the intrinsic color expected for each
spectral type.  By contrast, Harris et al.\ (1997) used two-color diagrams to
estimate the reddenings for 2069 (presumed OB) stars, and report a
mean value.  While there are certainly stars with higher reddenings in the Clouds,
we retain the median values given by Massey et al.\ (1995b), as the method should
be more accurate, despite the smaller sample size.  We note that the agreement
between the results of comparing the blue plumes of the various galaxies in our
sample to that of the LMC and SMC give consistent results (at the +/- 0.02~mag level)
if we adopt the Massey et al.\ (1995b) values.}.  We give the results of this analysis
in Table~\ref{tab:reddenings}.

The values in Table~\ref{tab:reddenings} apply to the typical OB star in each of
these galaxies, although clearly there are regions that are more heavily reddened,
particularly in M31.  For comparison, we note that Massey et al.\ (1986) found
reddenings for several OB associations in M31; i.e., $E(B-V)=0.12$ for OB78 (NGC 206),
$E(B-V)$=0.24 for OB 48 and the OB 8-10 region, and $E(B-V)=0.08$ for OB 102.
These values are consistent with our finding here that the typical reddening is 
$E(B-V)=0.13$.  Similarly, Massey et al.\ (1995a) used spectroscopic samples in 
NGC~6822 and M33 to determine reddenings of $E(B-V)=0.39$ and $E(B-V)=0.13$,
respectively.  The value for NGC~6822 determined here is considerably lower (0.25), and 
we suggest that the few OB associations studied by Massey et al.\ (1995a) were
higher in reddening than typical.  This interpretation is consistent with the study by 
Bianchi et al.\ (2001) which finds $E(B-V)$ varying from 0.25 to 0.45.
Our value for M33's reddening  is in very good agreement with that of
Massey et al.\ (1995a). For IC10 Massey \& Armandroff (1995)
estimated $E(B-V)=0.75-0.80$ using their photometry of the blue plume; ours goes
considerably deeper, allowing a more accurate determination, but one that is in
good accord with this earlier finding.

How much of this reddening is attributable to foreground, and how much is internal
to these galaxies themselves?  We have used the 100$\mu$m dust maps of
Schlegel et al.\ (1998) to estimate the foreground $E(B-V)$, and we
list these values in Table~\ref{tab:reddenings}. (The low Galactic latitude of
IC10 prevents an accurate determination.) The difference between these
values and the total reddenings should give a good indication of how much is
internal to these galaxies.  We see that the median value is 0.05~mag, in 
good accord with what is commonly assumed for irregular galaxies (see 
Hunter \& Elmegreen 2006).

We were initially skeptical of the relatively high reddenings found in
Pegasus and Phoenix, the two lowest luminosity galaxies in our sample.
However, Gallagher et al. (1998) found an identical value for the total
$E(B-V)$ in their two-color {\it HST} study of Pegasus. These galaxies are
currently quite quiescent in forming stars (see Table~\ref{tab:galaxies}) but
dust is primarily contributed to the ISM via low-mass asymptotic giant branch (AGB)
stars (see Whittet 2003).  Both galaxies do show modest amounts of H~I
(Young et al.\ 2003, St-Germain et al.\ 1999).

\subsection{The Massive Star Populations in Individual Galaxies}

\begin{figure}
\plotone{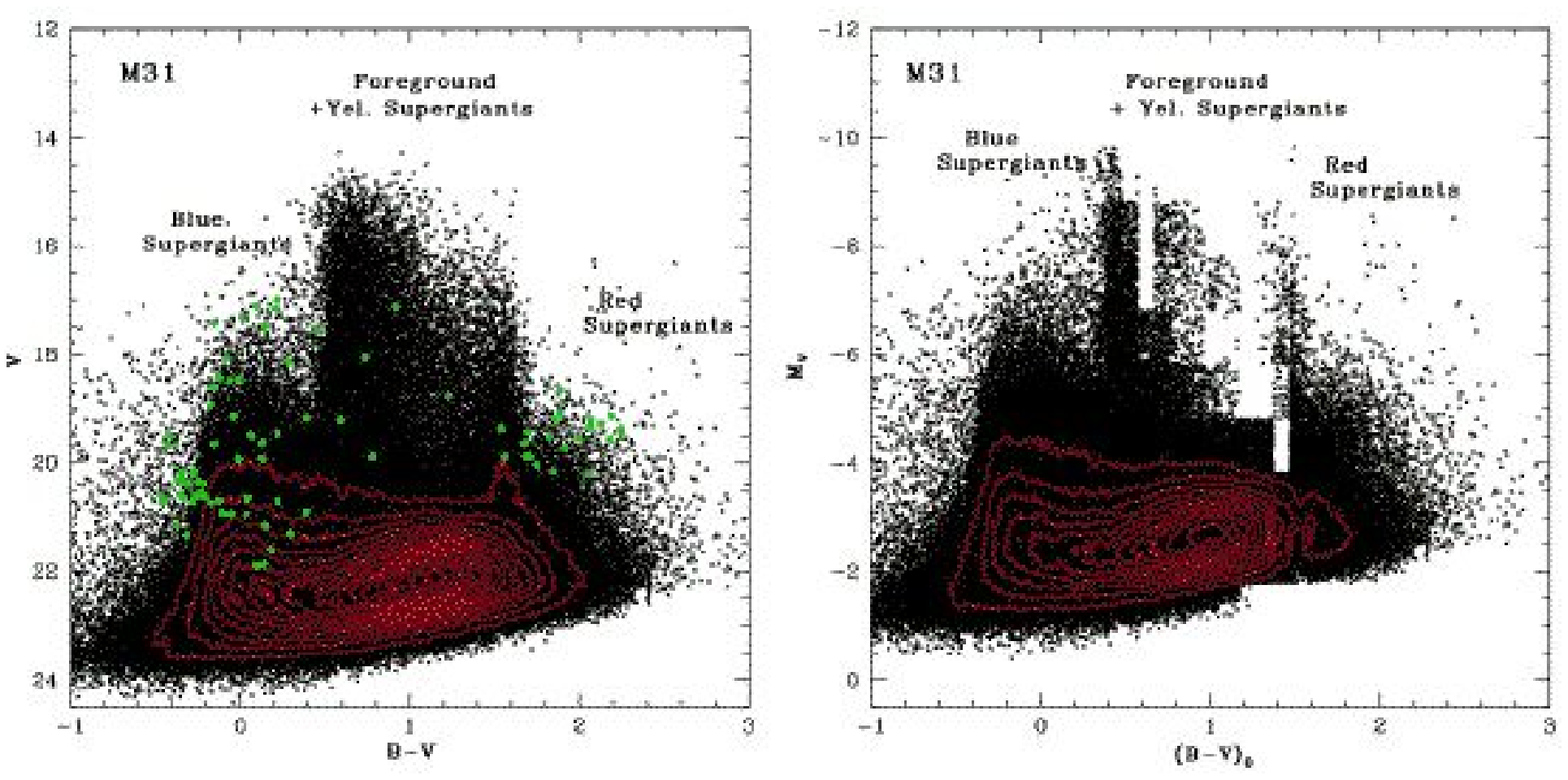}
\vskip -200pt
\caption{\label{fig:m31cmd} The color-magnitude diagram for M31. The
data are taken from Paper~I.  Left: Green symbols denote spectroscopically
confirmed members, from Table~8 of Paper~I.  
Right: We show the CMD approximately
``cleaned" of foreground stars (in a statistical sense)
using the Bahcall \& Soniera (1980) model, 
and converted to intrinsic color $(B-V)_0$ and absolute visual magnitude $M_V$.}
\end{figure}

\begin{figure}
\plotone{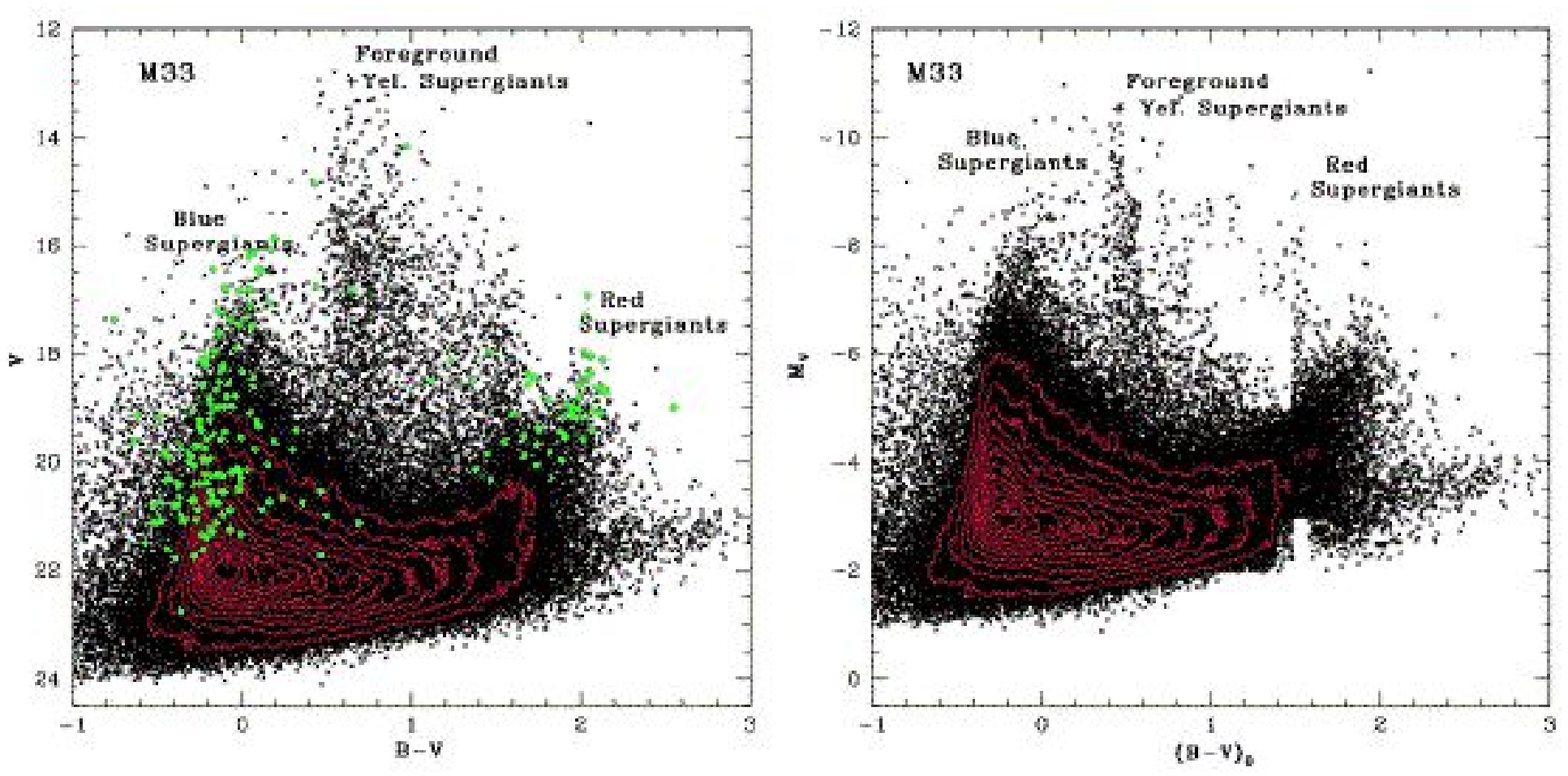}
\vskip -200pt
\caption{\label{fig:m33cmd} The color-magnitude diagram for M33. The
data are taken from Paper~I. Left:
Green symbols denote spectroscopically
confirmed members, from Table~9 of Paper~I.
Right: We show the CMD approximately
``cleaned" of foreground stars (in a statistical sense)
using the Bahcall \& Soniera (1980) model, 
and converted to intrinsic color $(B-V)_0$ and absolute visual magnitude $M_V$.}
\end{figure}

\begin{figure}
\plotone{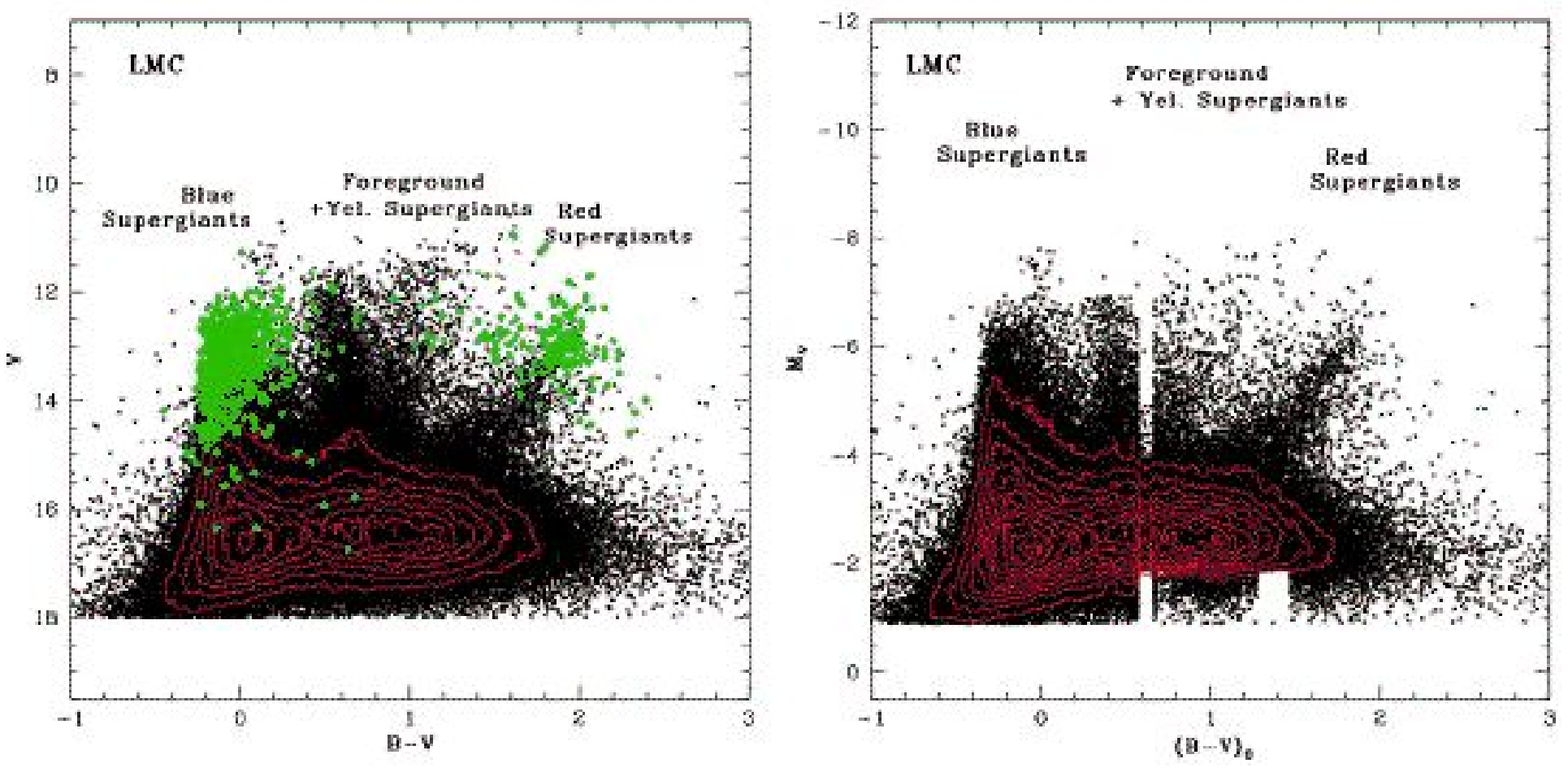}
\vskip -200pt
\caption{\label{fig:lmccmd} The color-magnitude diagram for the LMC.  The
data are taken from Massey (2002). Left:
Green symbols denote spectroscopically
confirmed members, taken from Table~4 of Massey (2002) and selected
from Table~2 of Massey \& Olsen (2003). Right:
We show the CMD approximately
``cleaned" of foreground stars (in a statistical sense)
using the Bahcall \& Soniera (1980) model, 
and converted to intrinsic color $(B-V)_0$ and absolute visual magnitude $M_V$.}
\end{figure}

\begin{figure}
\plotone{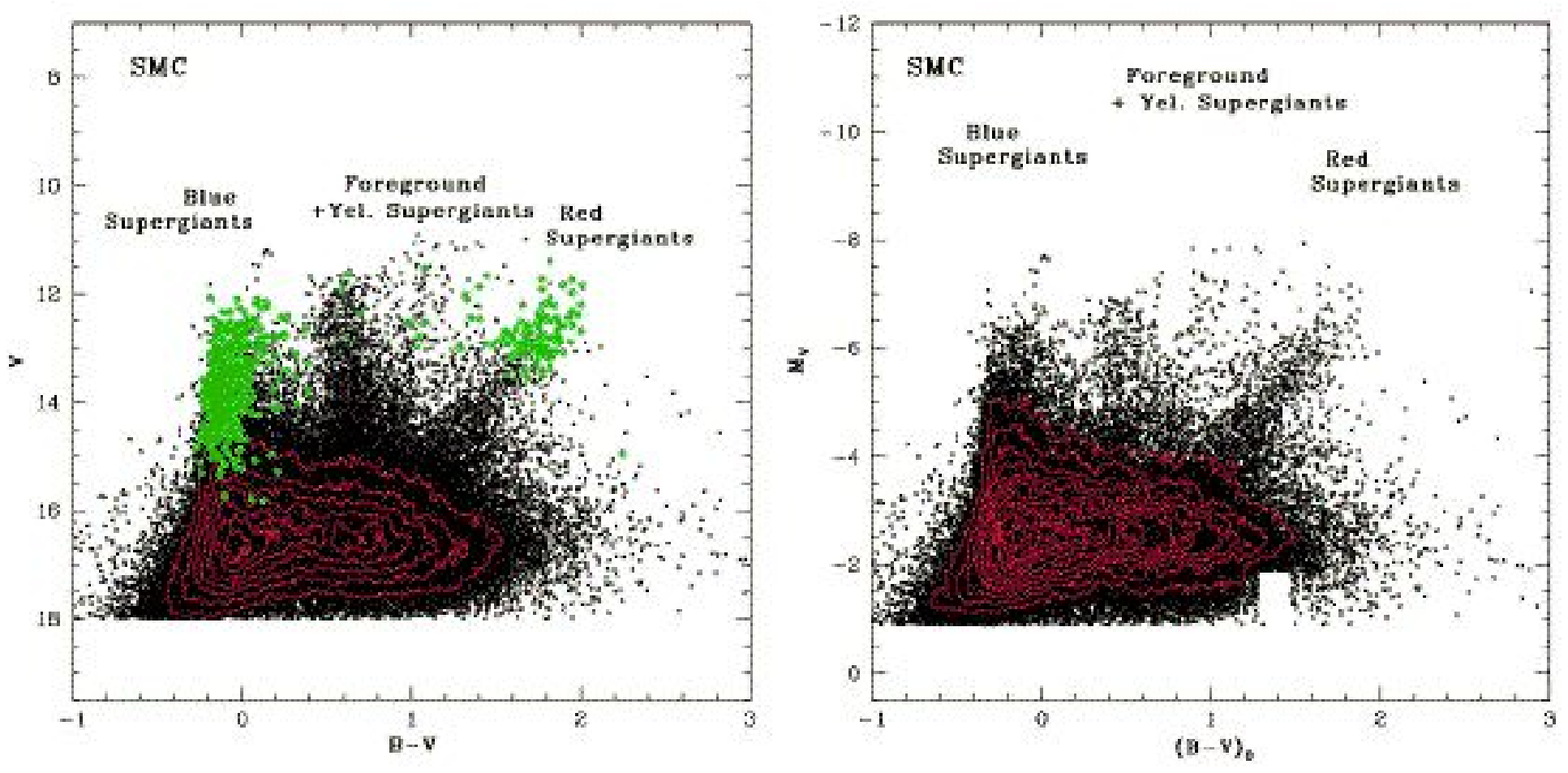}
\vskip -200pt
\caption{\label{fig:smccmd} The color-magnitude diagram for SMC. The
data are taken from Massey (2002). Left:
Green symbols denote spectroscopically
confirmed members, taken from Table~6 of Massey (2002) and selected from
Table~1 of Massey \& Olsen (2003). We show the CMD approximately
``cleaned" of foreground stars (in a statistical sense)
using the Bahcall \& Soniera (1980) model, 
and converted to intrinsic color $(B-V)_0$ and absolute visual magnitude $M_V$.}
\end{figure}

Using the results of the previous two sections, we have constructed 
CMDs which have been converted to absolute magnitude 
(using the distances and reddenings in Tables~\ref{tab:galaxies} and \ref{tab:reddenings}).   These are included at the right of 
In Figs.~\ref{fig:m31cmd} through \ref{fig:phxcmd}.    We have also cleaned out
the foreground contribution, at least in a statistical sense, by using the 
foreground fields for seven dwarfs.  For each star in the foreground CMD
(middle panel of Figs.~\ref{fig:IC10cmd} through \ref{fig:phxcmd}) we have
removed a star near that position in the galaxy CMD.  The process is only
approximate; if no star was within 0.2~mag in $B-V$ and 0.8~mag in $V$ of
a foreground star, we did not use it.  For the CMDs of galaxies lacking foreground
fields (Figs.~\ref{fig:m31cmd}-\ref{fig:smccmd}), we use the Bahcall \& Soneria (1980)
models to predict the number of foreground stars in each 0.1~mag $B-V$ bin
and each 1~mag $V$ bin.  Of course, this is considerably cruder and less certain,
and many objects which are doubtless foreground are left, 
but the resulting CMDs are still quite illustrative.   

\subsubsection{IC10}

IC10 is the most luminous galaxy in our sample of dwarfs 
(see Table~\ref{tab:galaxies}), and it is also the most heavily 
reddened due to its low Galactic latitude.  
Massey \& Armandroff (1995) argued that the galaxy is in a starburst phase, based
upon the current high star-formation rate, as judged by the known high
$H\alpha$ luminosity compared to either its H~I mass or blue-light
luminosity (Hunter \& Gallagher 1986; Thronson et al.\ 1990; Hunter et al.\ 1993),
and in accord with its extremely large number of Wolf-Rayet (WR) stars detected
at the time
(Massey et al.\ 1992).  Massey \& Armandroff (1995) found that surface density
of WR stars, averaged over {\it the entire galaxy},  is 
roughly
comparable to the most WR-rich OB associations in M33.   Since that time additional WRs have been found
(Massey\& Holmes 2002, Crowther et al.\ 2003), and as discussed by Massey \& Holmes (2002), a new survey suggests 
that the actual number of WRs in IC10 may be 
considerably higher than previously thought.

In Table~\ref{tab:ic10mem} we give the cross-references between our photometry
catalog and the stars with known spectral types from the literature.
In the case of IC10 these are all from spectroscopy of WR candidates.
The identifications have
been checked by eye against the  finding charts published by Crowther
et al.\ (2003); this was necessitated by crowding and the ambiguity of the ``system"
on which some of the published coordinates were based (i.e., Guide Star
Catalog vs USNO-B).
However, in some cases we found that stars previously identified as single were
in fact identified as multiple on our PSF-fitting; i.e., 
[MAC 92] 2, RSV9, and [MAC 92] 7. 
Others were recognized as multiple by Crowther et al.\ (2003); e.g., 
[MAC 92] 17 and [MAC92] 24.  For these cases the blending is sufficiently bad, and the
photometric difference between components sufficiently slight, that we cannot
reliably determine which component is the WR star.
Following Crowther
et al.\ (2003) we examined the {\it HST} images of these stars using in particular
the newer ACS data available since the time of the Crowther et al.\ (2003) study.
While these images in general confirmed our identification of multiple sources,
they did not shed any useful light on which component was the WR star.
Therefore we list the multiple components in Table~\ref{tab:ic10mem} where appropriate.  We have adopted
the revised spectral types of the WRs given by Crowther et al. (2003) based upon
their new spectroscopy of WRs previously confirmed spectroscopically
(Massey et al.\  1992; Massey \& Armandroff 1995; 
Massey \& Holmes 2003),  plus a few of their own candidates, although it was
not always clear if their data were of higher quality or if they were simply bolder
 in assigning exact spectral subtypes (i.e., ``WC4" vs ``WC").  The 
exceptions are the instances where they chose to reclassify a star based solely
upon their interpretation of the illustrations of spectra published by others.  
Only the WR star RSMV 13 was too faint to be included in our catalog.

\begin{figure}
\epsscale{0.30}
\plotone{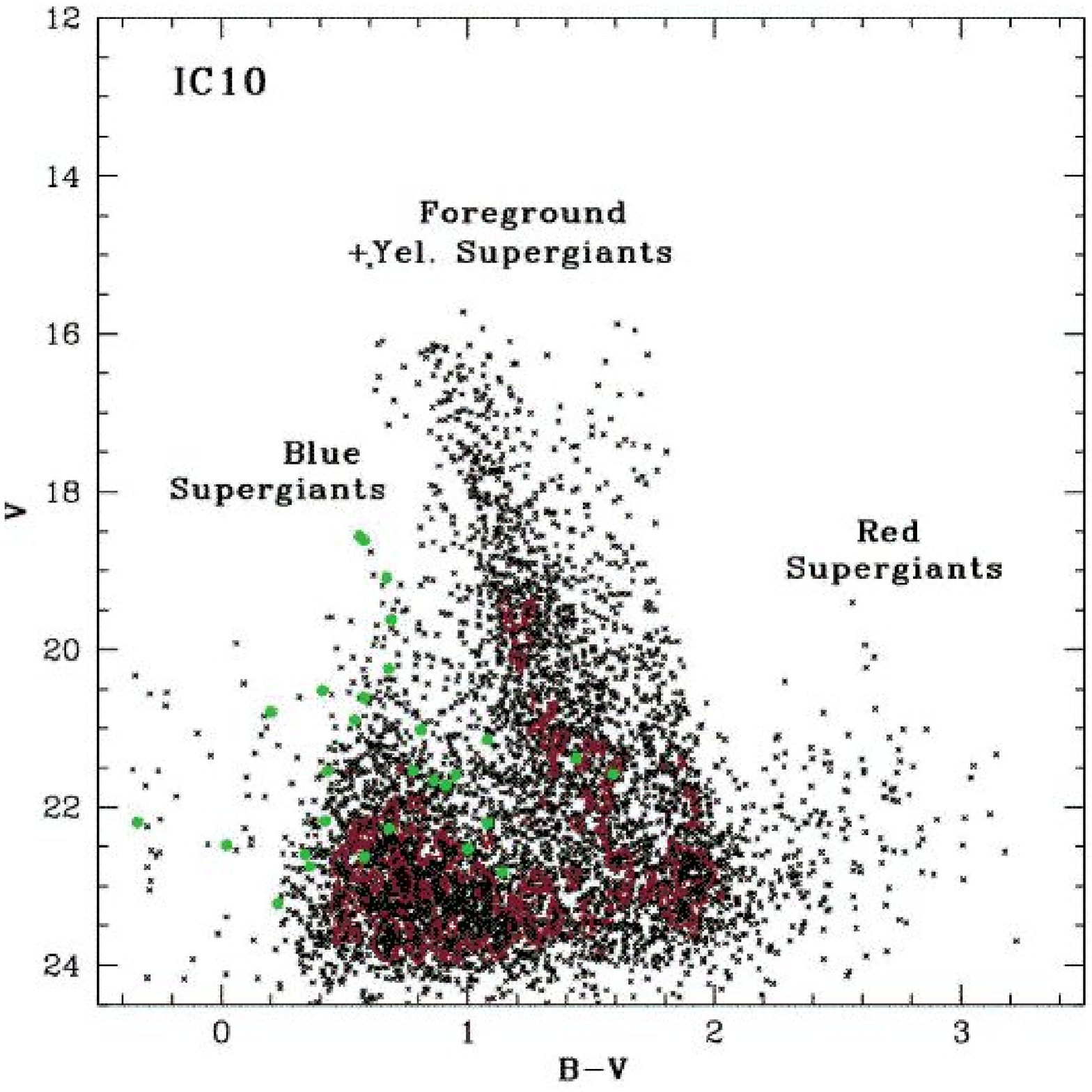}
\plotone{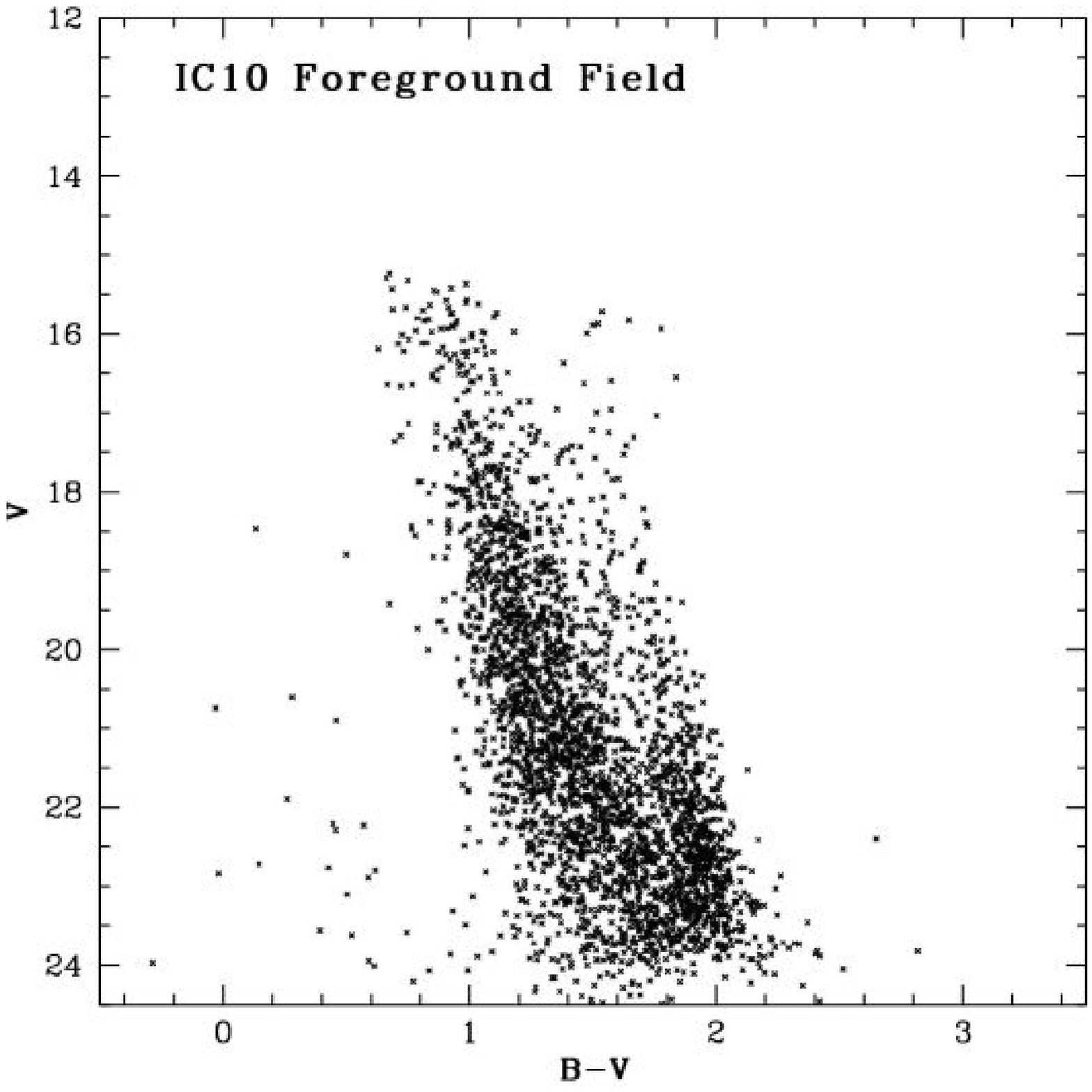}
\plotone{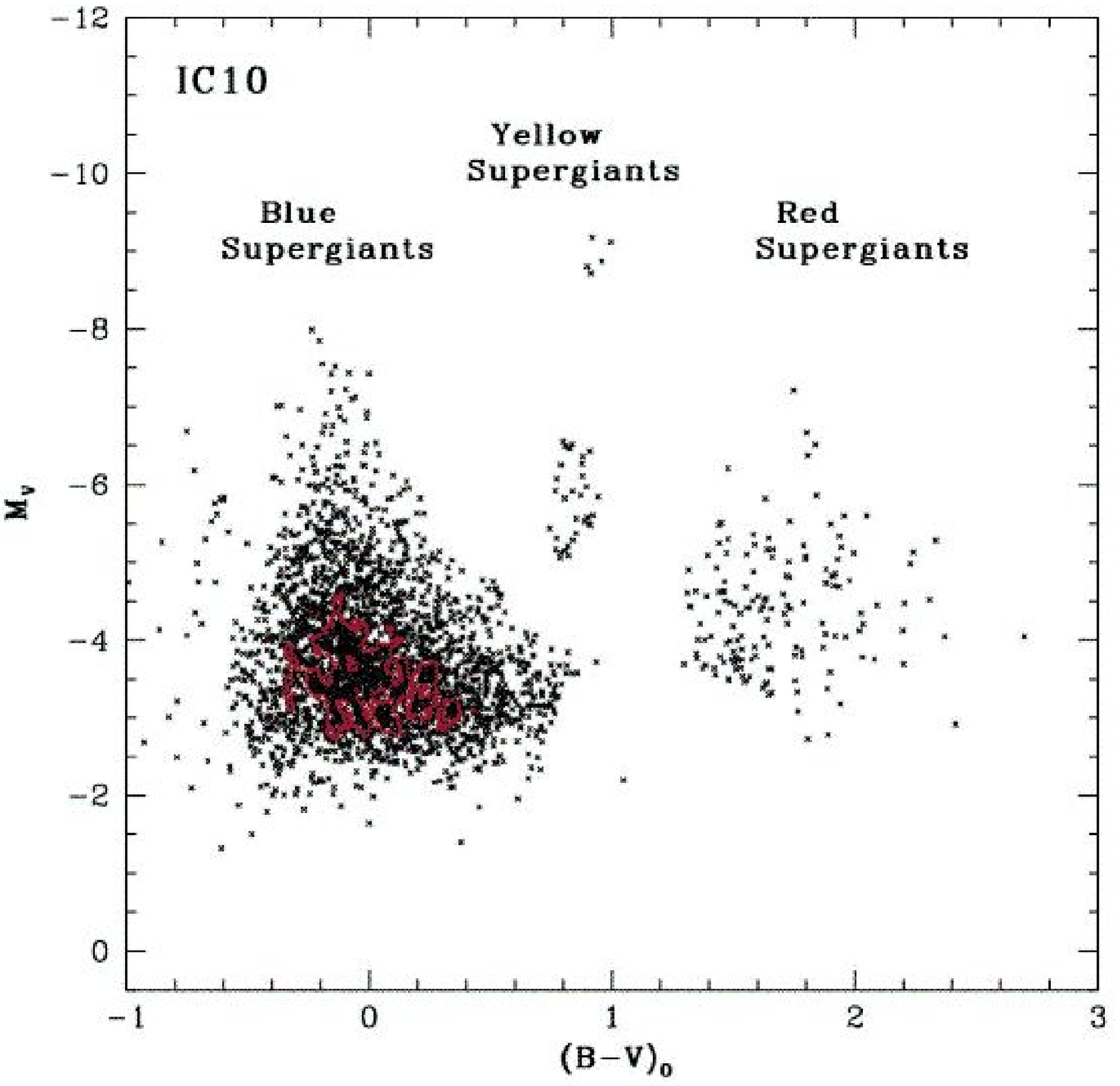}
\caption{\label{fig:IC10cmd} The color-magnitude diagram for IC10.
Left: The
CMD of the galaxy reveals a strong population of blue supergiants, as
well as some red ones, but
high extinction, $E(B-V)=0.81$, has shifted the sequences
to considerably redder colors and fainter magnitudes.  To decrease the effects of foreground contamination
we have restricted the sample to a region from $\alpha_{\rm J2000}=00^h19^m42^s$ to $00^h20^m55^s$, and
$\delta_{\rm J2000}=+59^\circ13'$ to $+59^\circ23'$, an area of 0.026 deg$^2$.  Green symbols
show spectroscopically confirmed members (Table~\ref{tab:ic10mem}). Middle:
We show the CMD the combination of two neighboring foreground
 fields with the same area, chosen from the periphery of the IC10.  
 Right: We show the CMD ``cleaned" of foreground stars (in a statistical sense)
 and converted to intrinsic color $(B-V)_0$ and absolute visual magnitude $M_V$.}
 \end{figure}

\begin{figure}
\epsscale{0.3}
\plotone{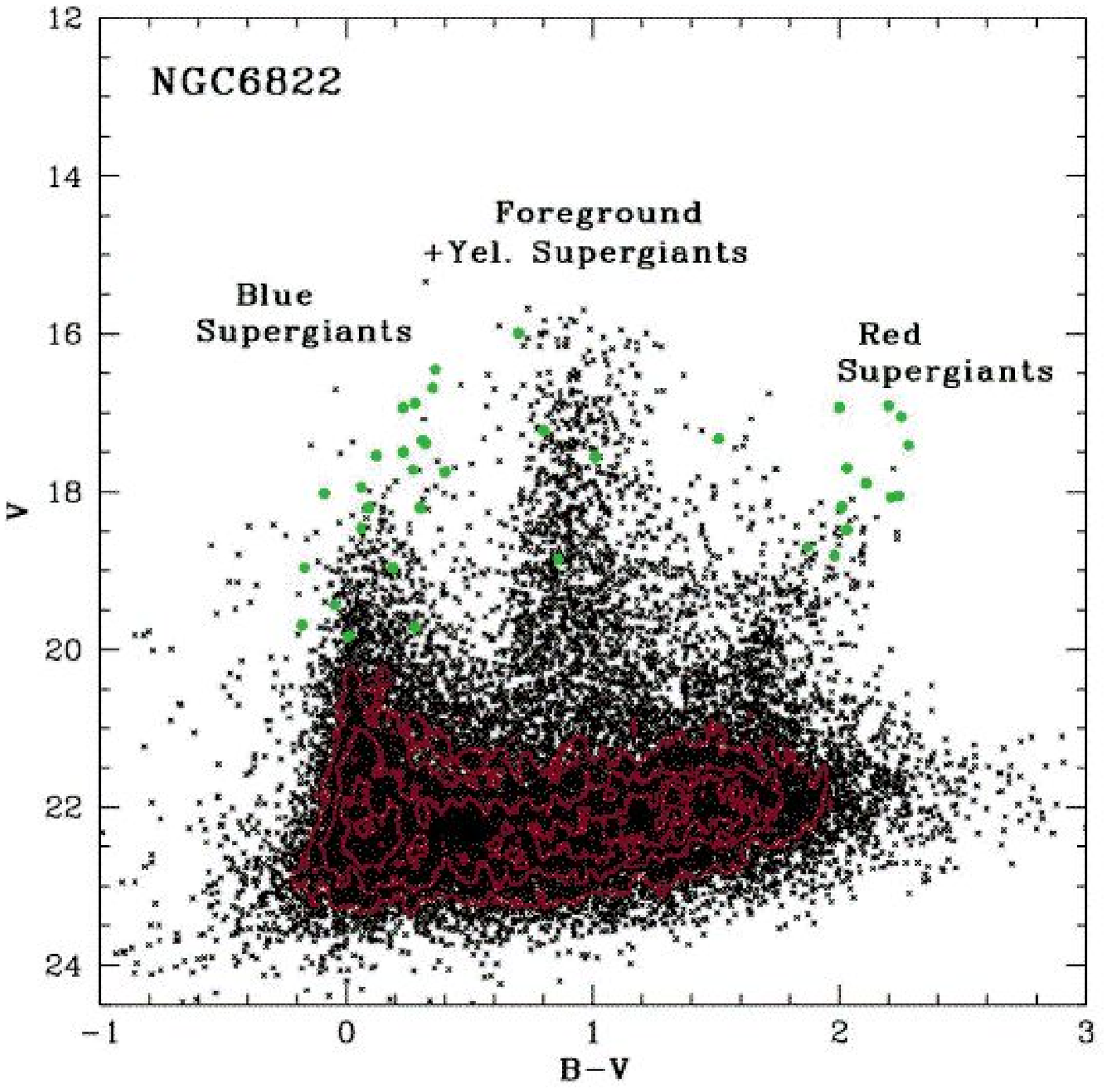}
\plotone{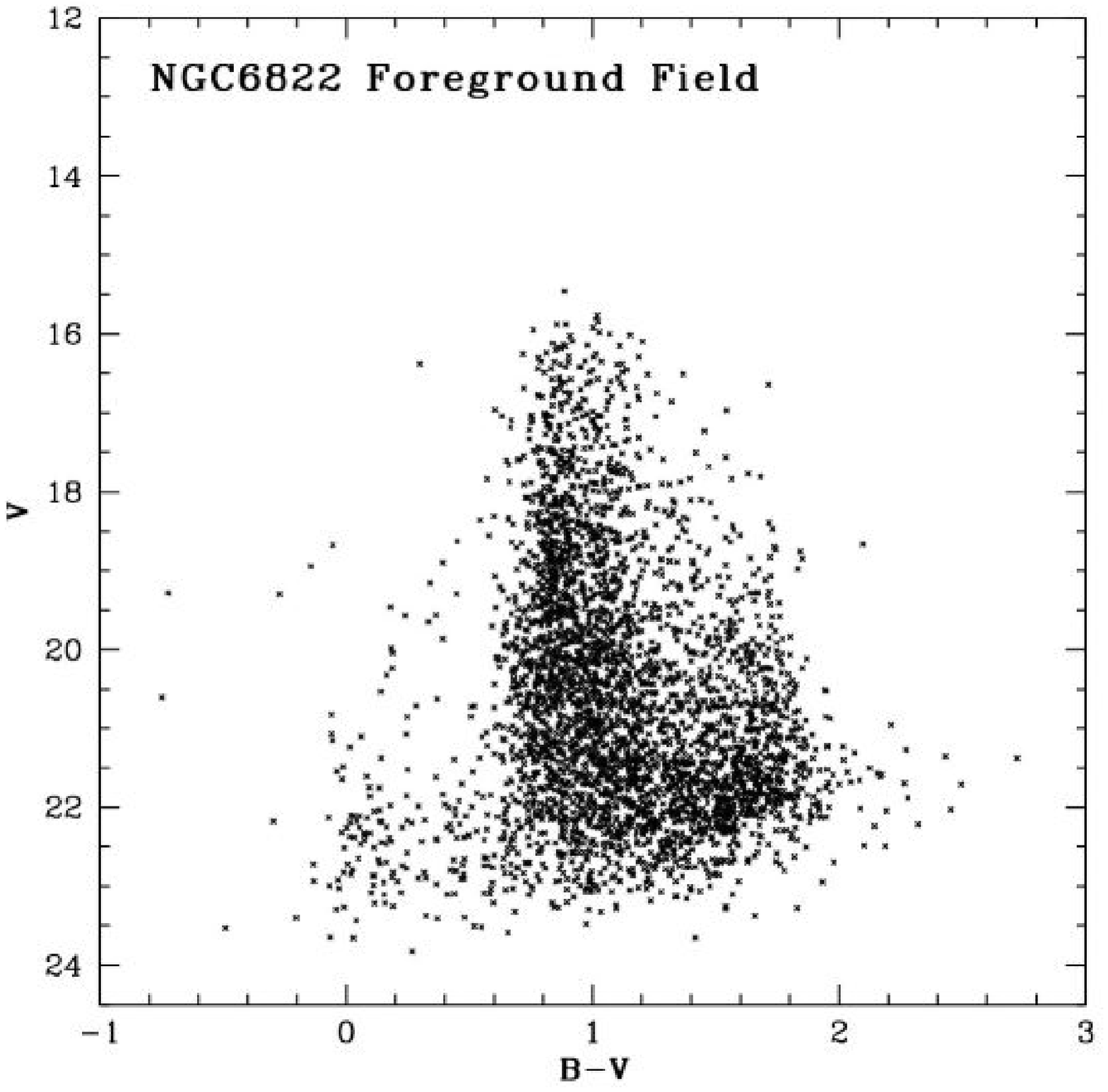}
\plotone{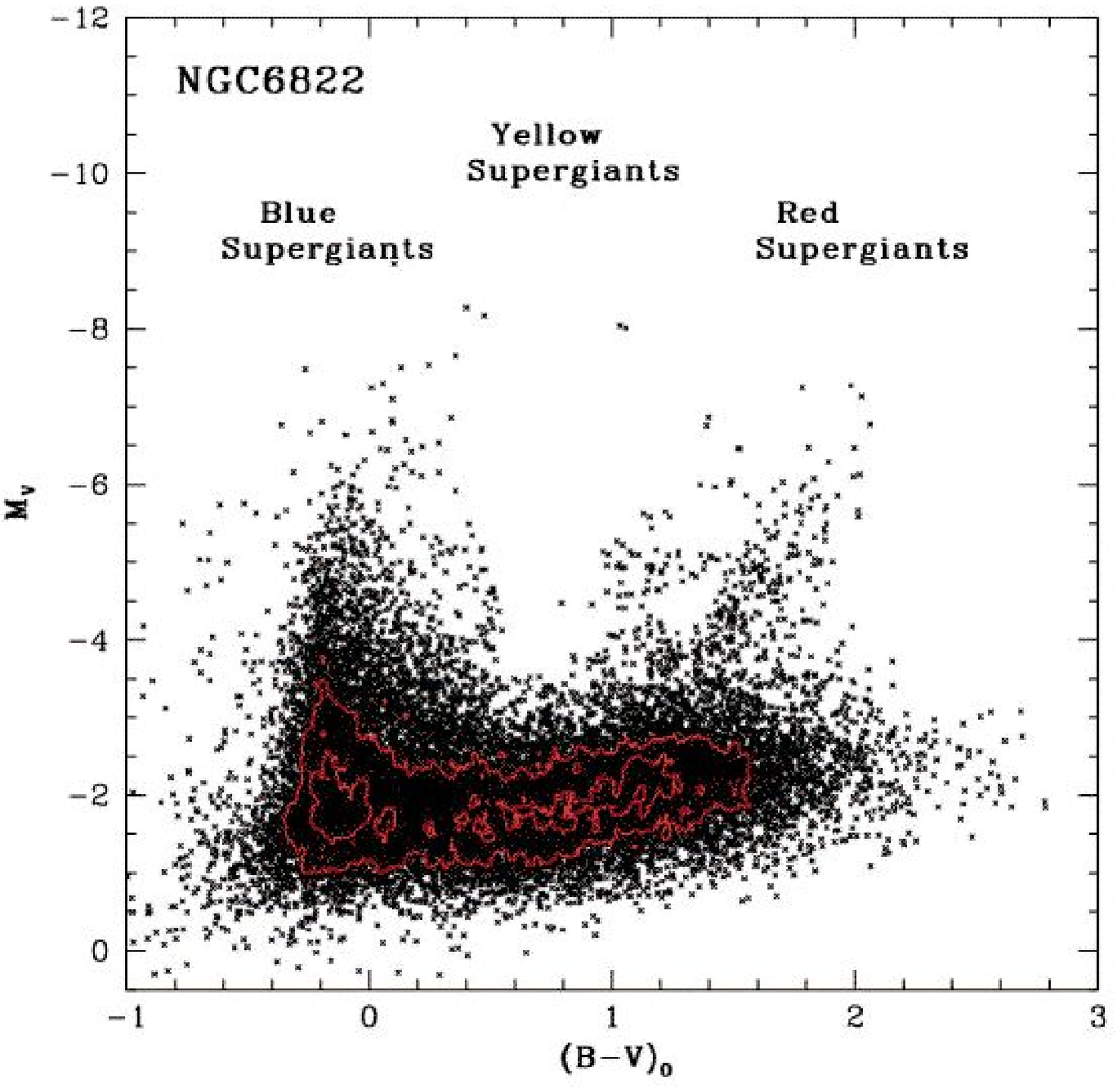}
\caption{\label{fig:n6822cmd} The color-magnitude diagram for NGC 6822. Left: The CMD of the galaxy reveals a strong population of blue and red supergiants, plus
bright stars of intermediate color  which are
 dominated by foreground stars. To decrease the effects of this contamination, we have restricted the
sample to a region from $\alpha_{\rm J2000}=19^h44^m34^s$ to $19^h45^m22^s$, and
$\delta_{\rm J2000}=-14^\circ56'$ to $-14^\circ40'$, an area of 0.052 deg$^2$.  Green
symbols show spectroscopically confirmed members (Table~\ref{tab:n6822mem}). Middle:
We show the CMD of the combination of three neighboring
foreground fields with the same area, 
chosen from the periphery of the NGC 6822.  Right: We show the CMD ``cleaned" of foreground stars (in a statistical sense)
 and converted to intrinsic color $(B-V)_0$ and absolute visual magnitude $M_V$.}
\end{figure}

\begin{figure}
\epsscale{0.3}
\plotone{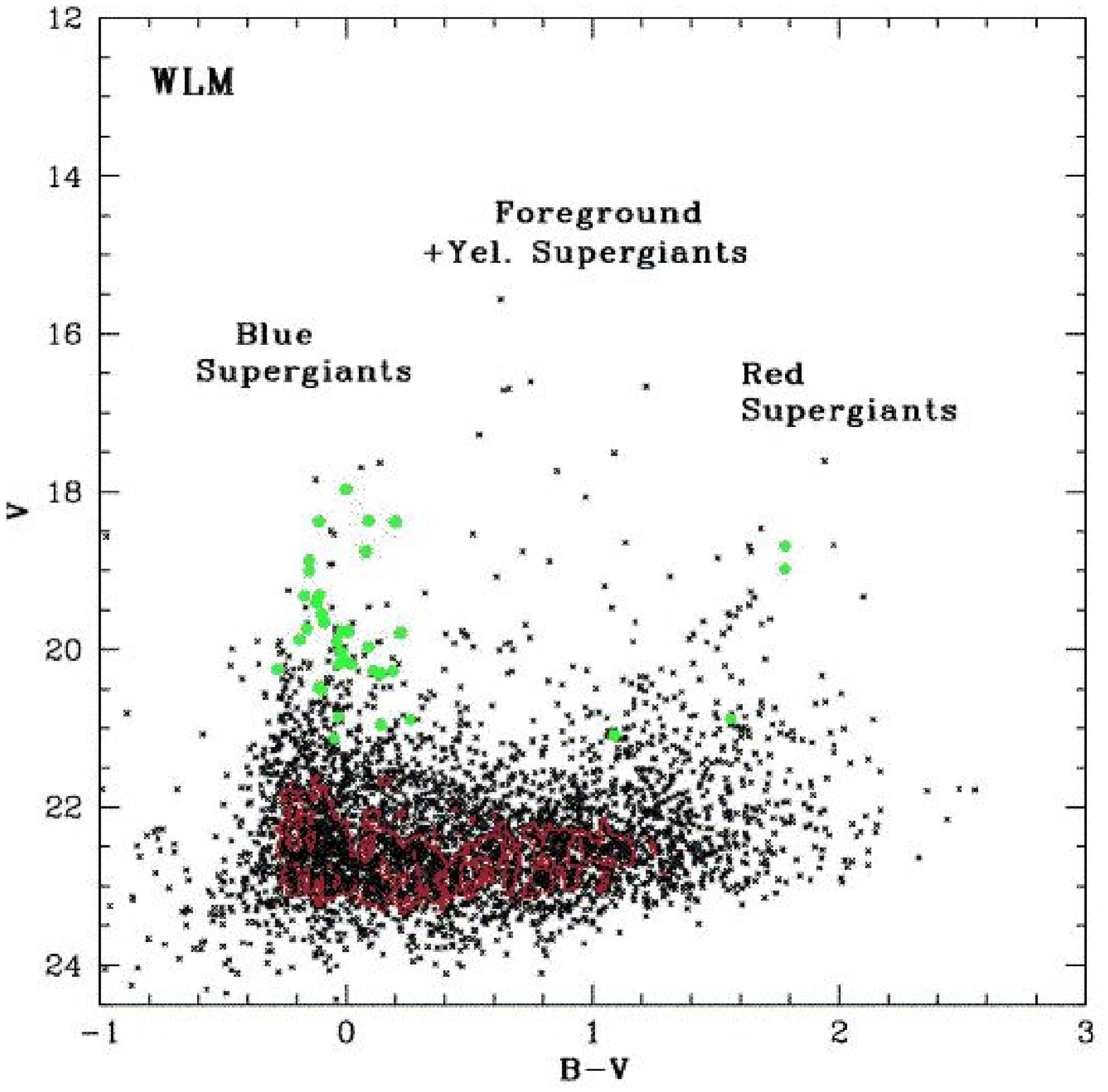}
\plotone{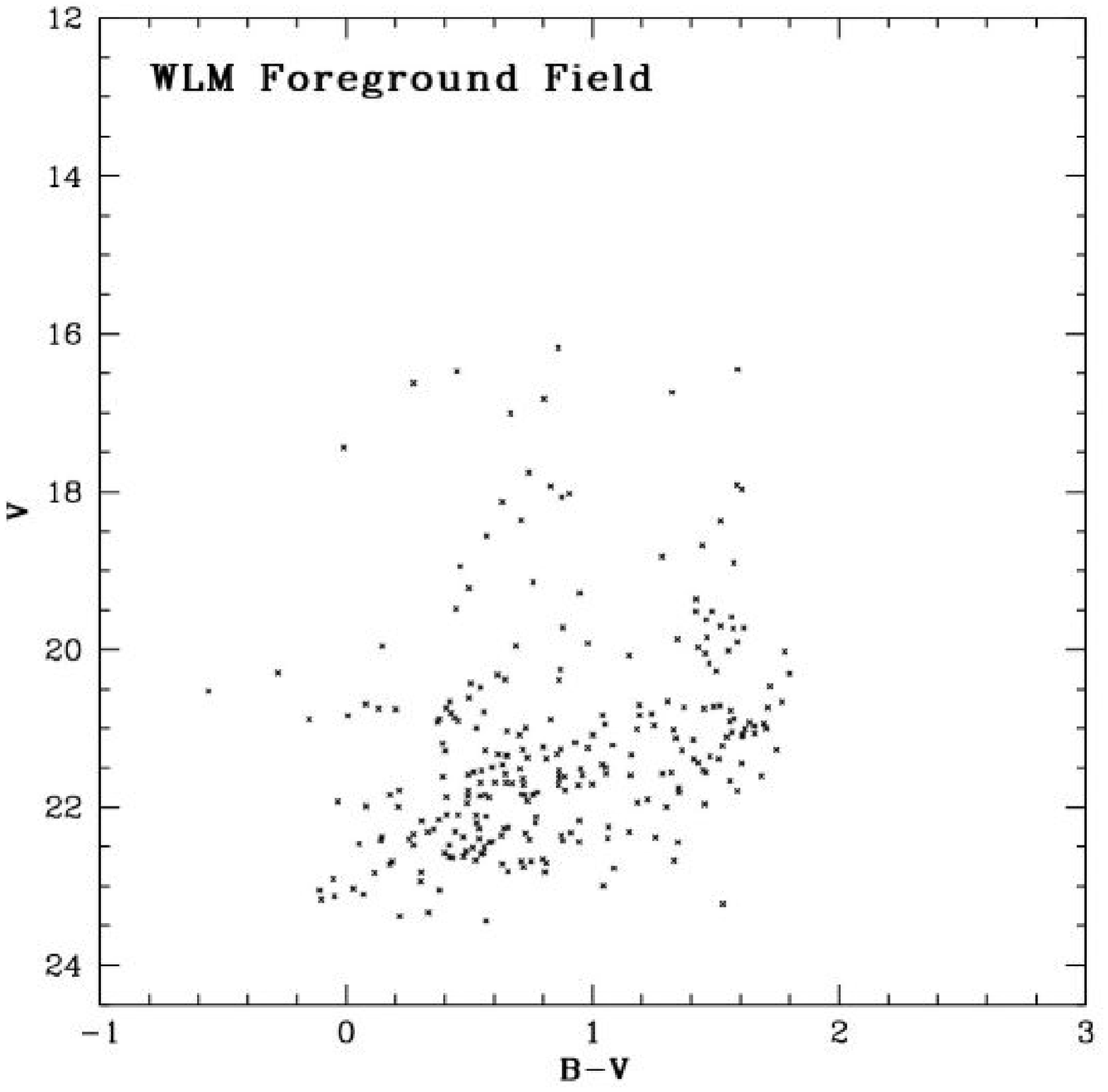}
\plotone{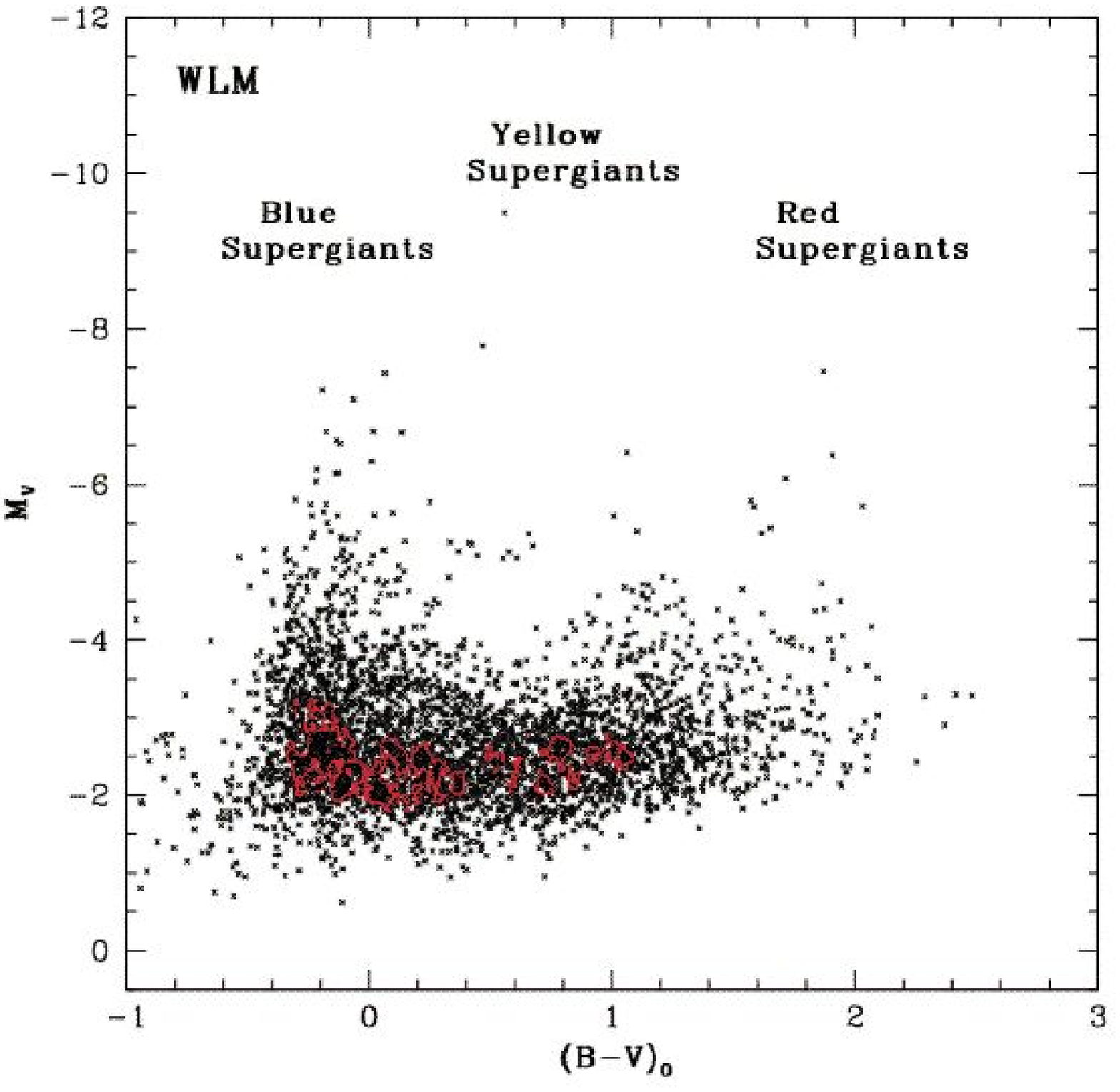}
\caption{
\label{fig:wlmcmd} 
The color-magnitude diagram for WLM and and neighboring foreground
field.
Left: The CMD of the galaxy reveals blue and red supergiants, plus a handful of
bright stars of intermediate color which is dominated by foreground stars.
To decrease the effects of this contamination, we have restricted the
sample to a region from $\alpha_{\rm J2000}=00^h01^m46^s$ to $00^h02^m12^s$, and
$\delta_{\rm J2000}=-15^\circ34'$ to $-15^\circ21'$, an area of 0.026 deg$^2$. Green
symbols show spectroscopically confirmed members (Table~\ref{tab:wlmmem}).
Middle:
We show the CMD of the combination of two neighboring foreground
fields with the same area, chosen from the periphery of the Sextans B.
Right: We show the CMD ``cleaned" of foreground stars (in a statistical sense)
 and converted to intrinsic color $(B-V)_0$ and absolute visual magnitude $M_V$.}
\end{figure}

\begin{figure}
\epsscale{0.3}
\plotone{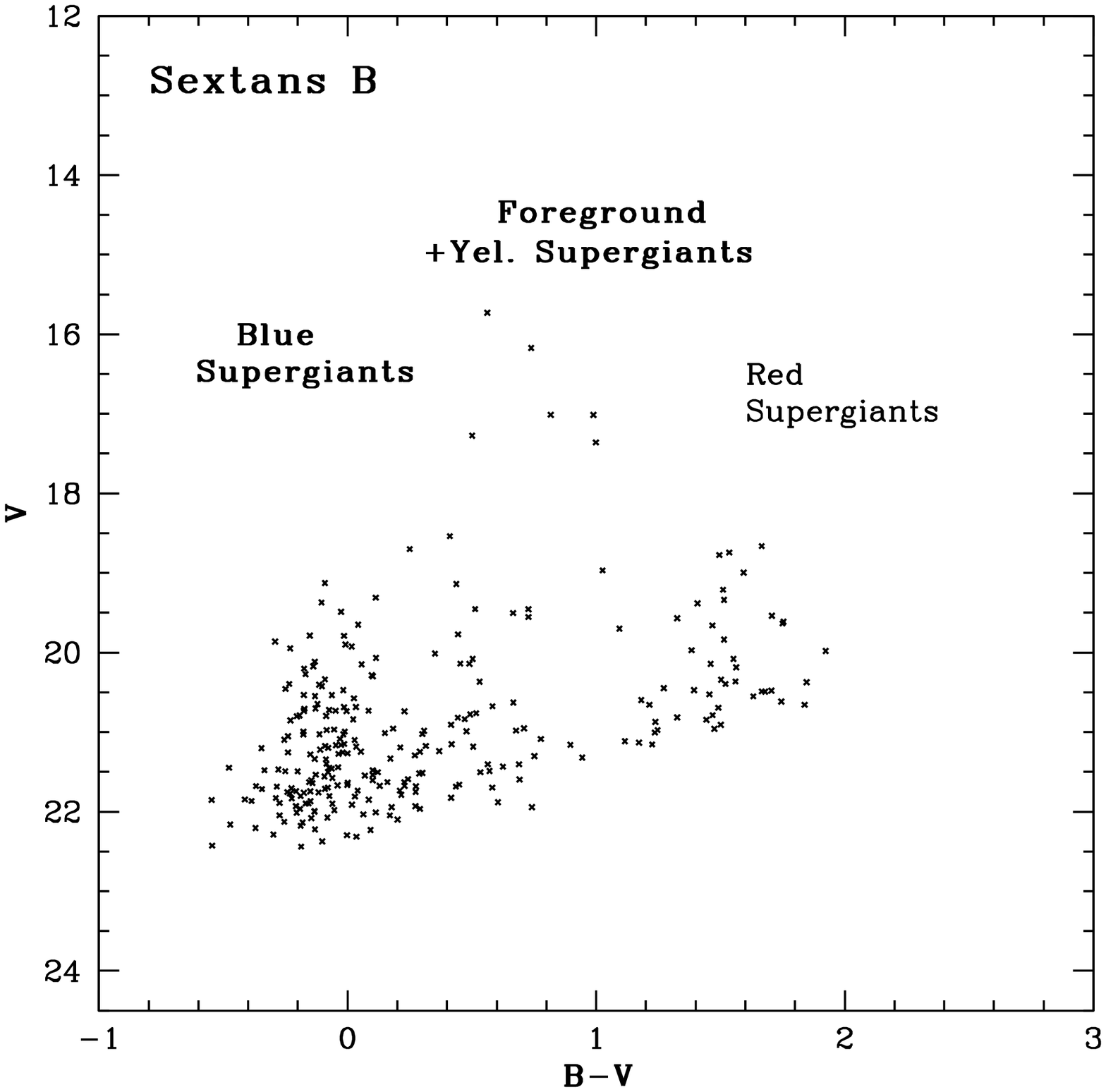}
\plotone{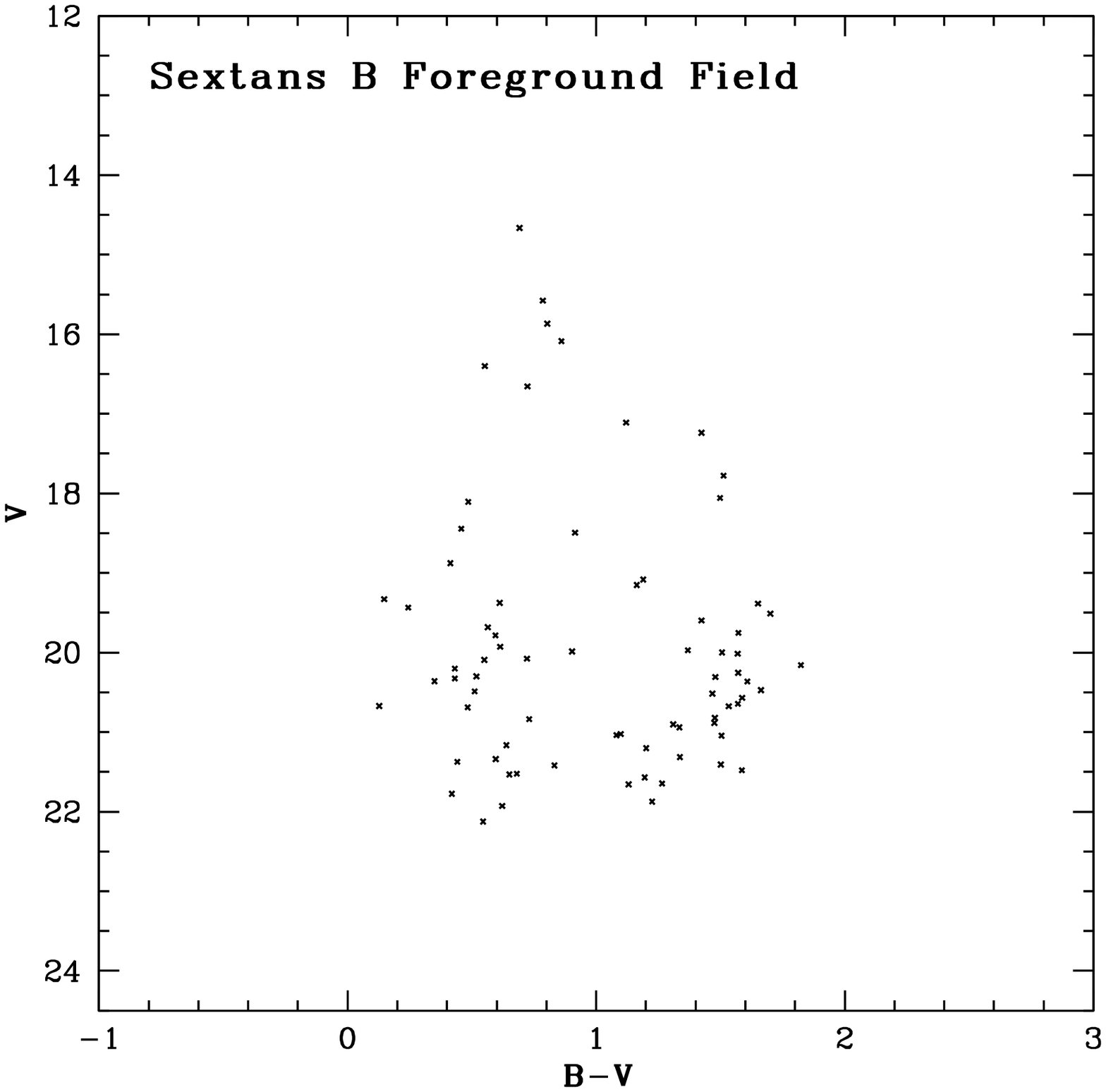}
\plotone{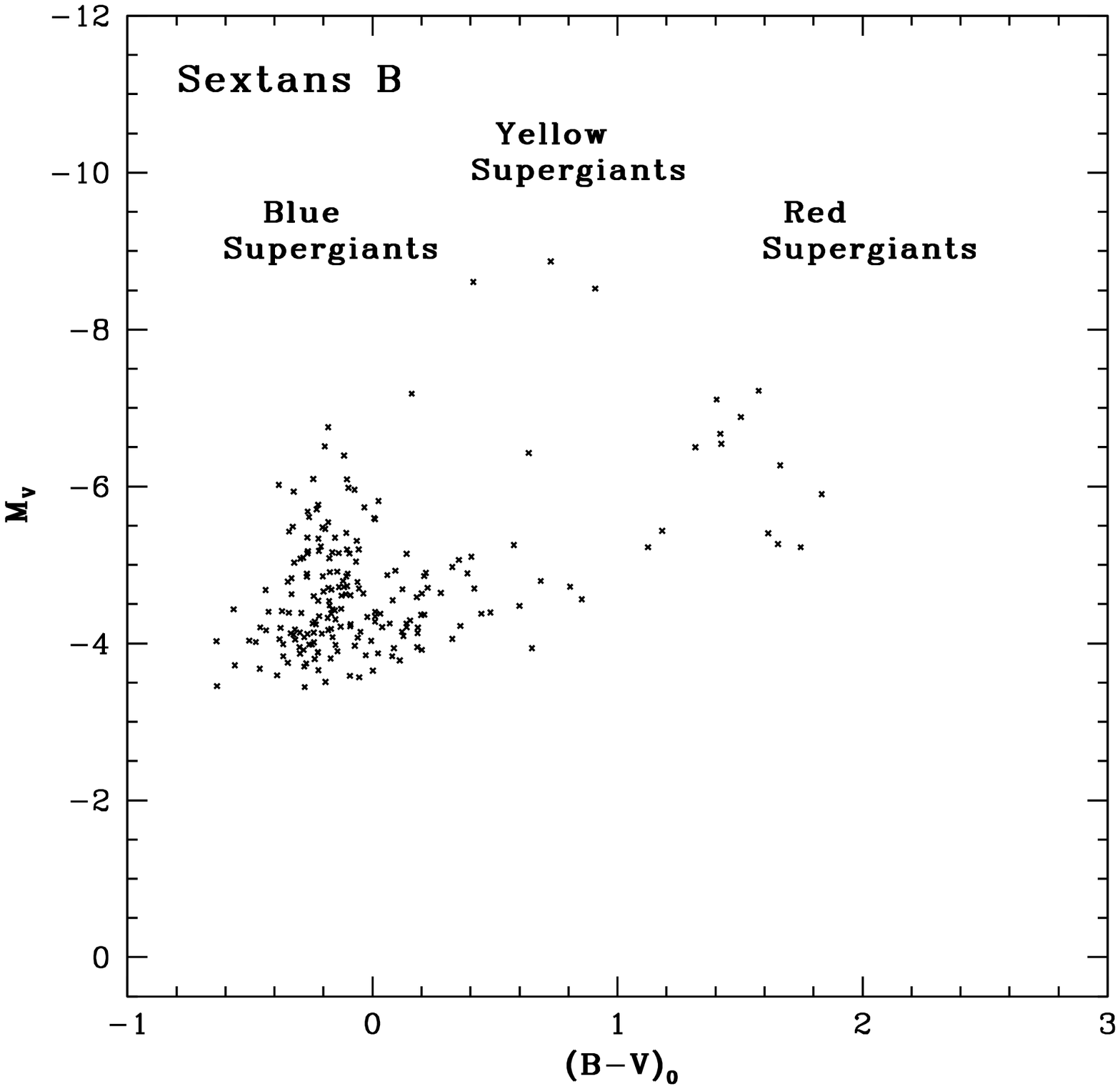}
\caption{\label{fig:sexBcmd} The color-magnitude diagram for Sextans B and neighboring
foreground field.  Left: The CMD of the galaxy reveals blue supergiants, plus a few stars of intermediate
color dominated by foreground stars, plus a smattering of red stars, some of which are
native to Sextans B.  To decrease the effect of foreground contamination, we have
restricted the same to a region from $\alpha_{\rm J2000}=09^h59^m40^s$ to $10^h00^m19^s$, and
$\delta_{\rm J2000}=+5^\circ17'$ to $+5^\circ23'$, an area of 0.016 deg$^2$.  Middle:
We show the CMD of the combination of two neighboring foreground
fields with the same area,
chosen from the periphery of the Sextans B.
Right: We show the CMD ``cleaned" of foreground stars (in a statistical sense)
 and converted to intrinsic color $(B-V)_0$ and absolute visual magnitude $M_V$.}
\end{figure}

\begin{figure}
\epsscale{0.3}
\plotone{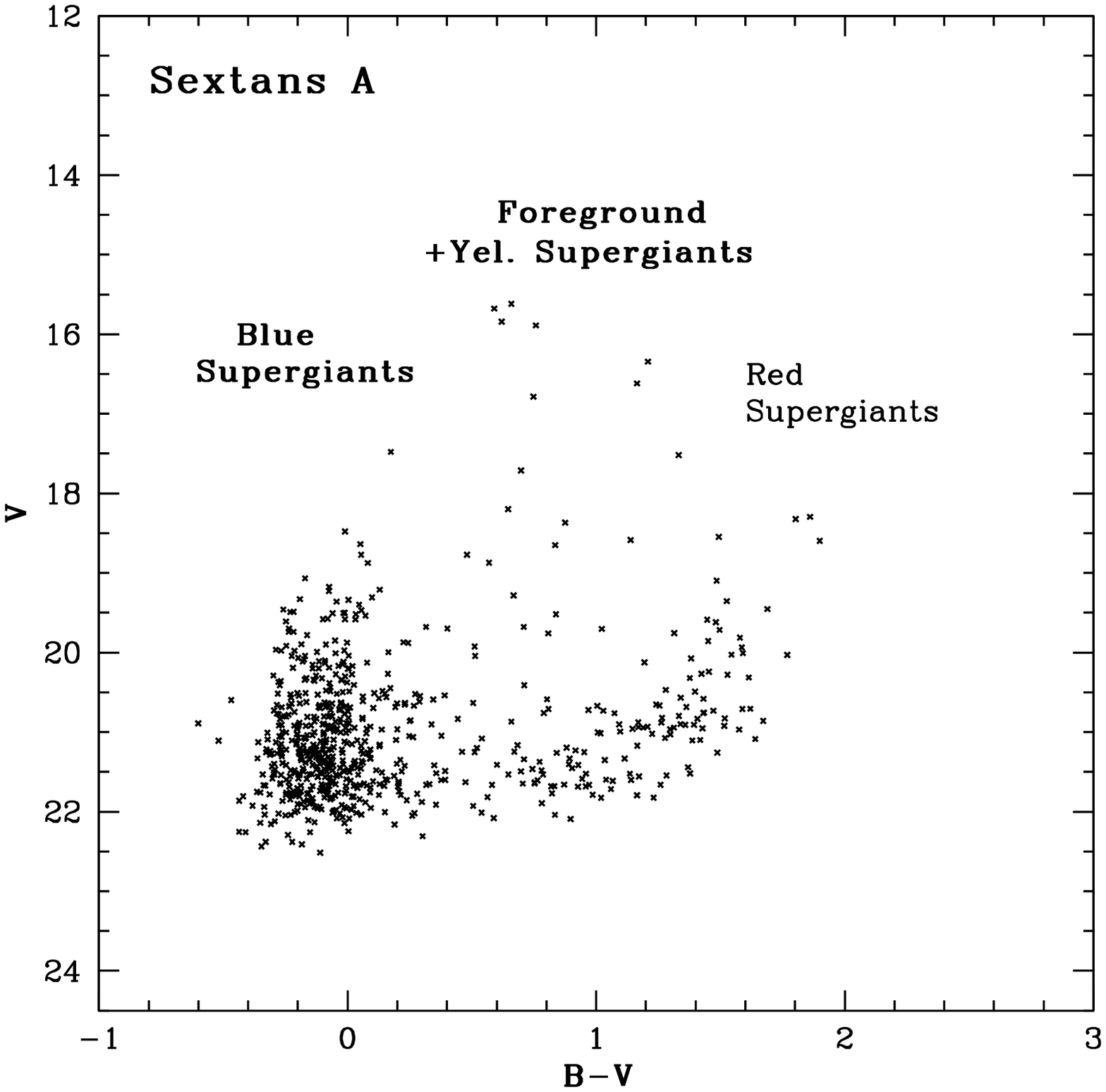}
\plotone{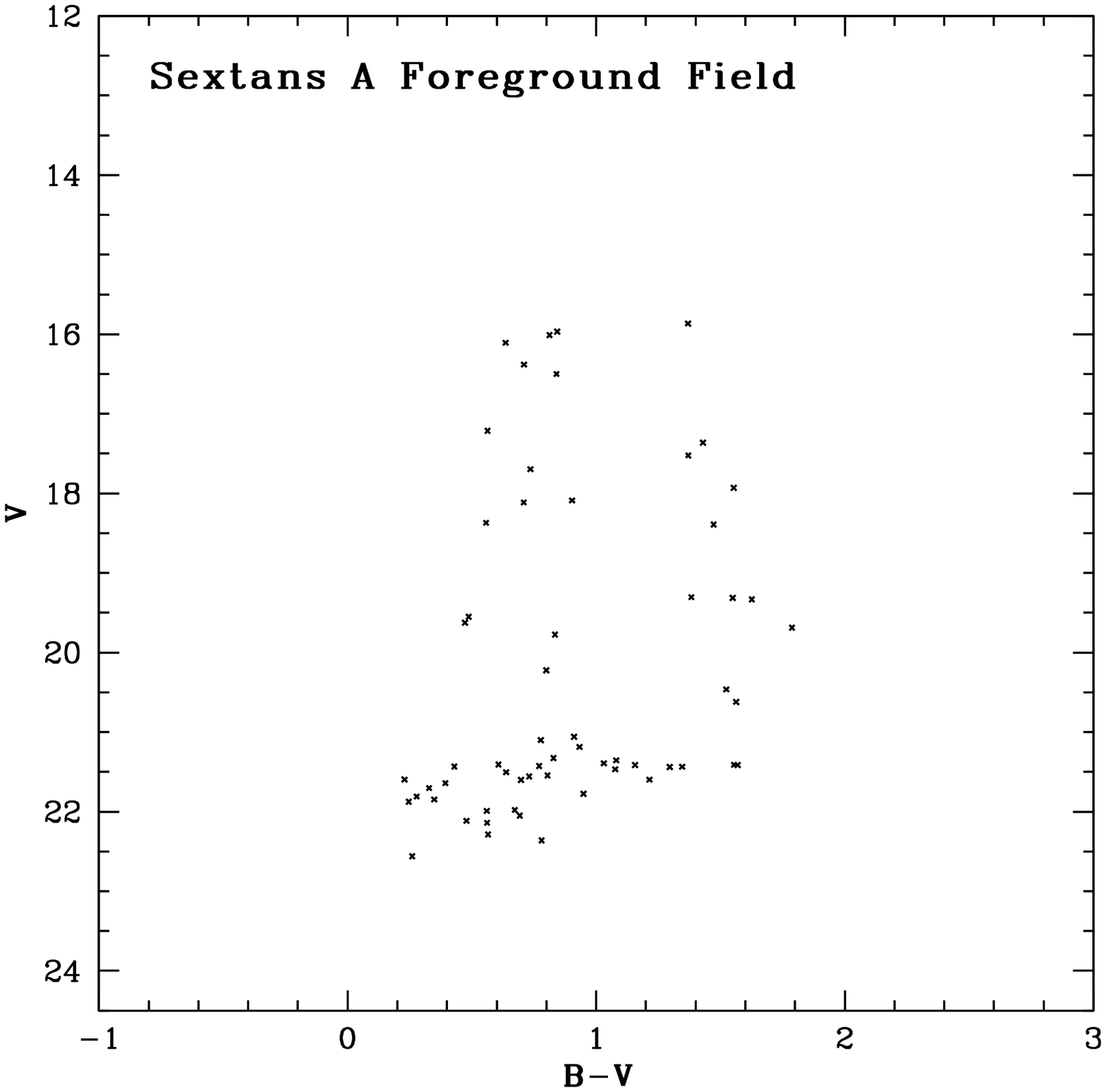}
\plotone{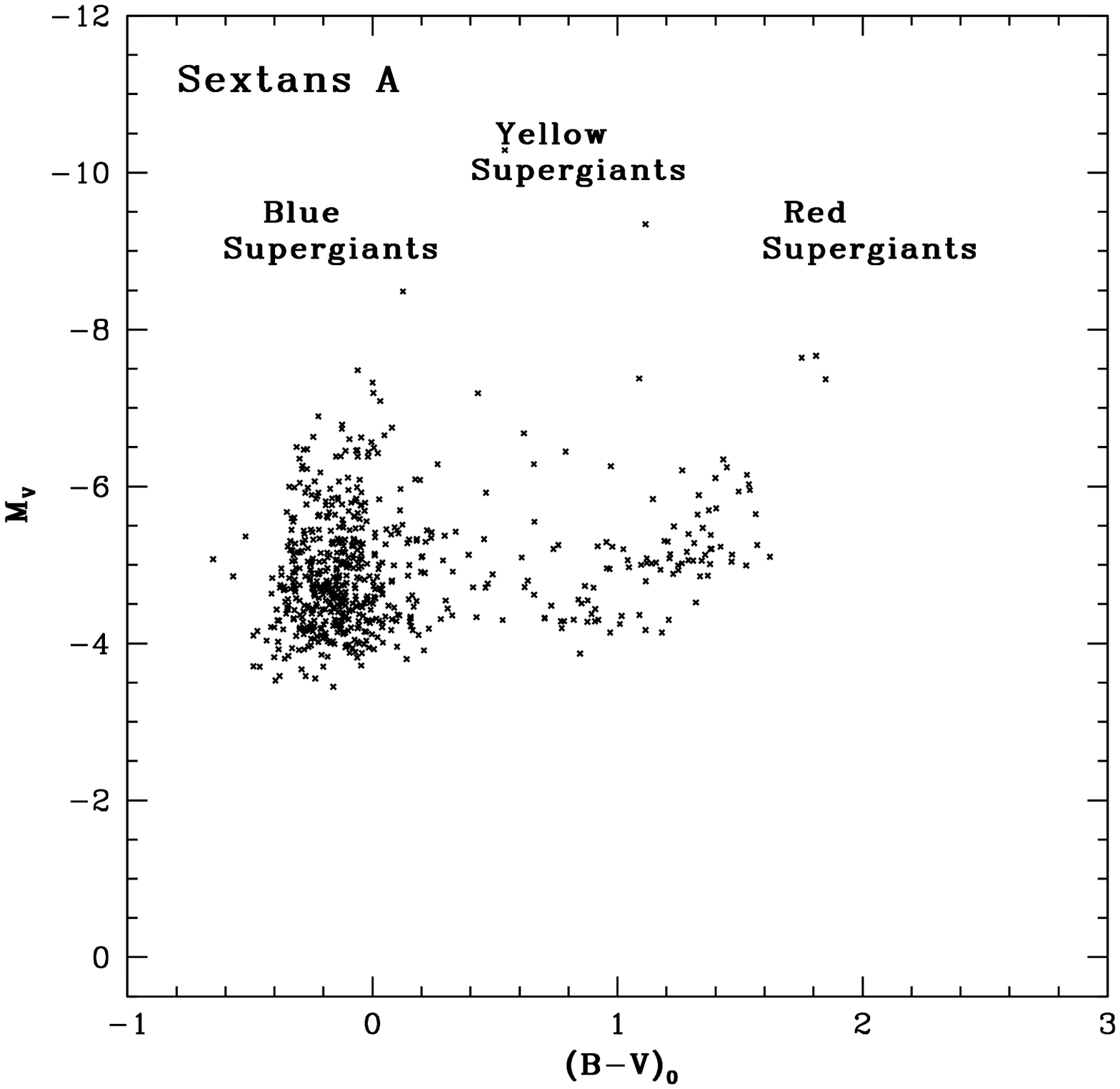}
\caption{\label{fig:sexAcmd} The color-magnitude diagram for Sextans A. Left: The CMD of the galaxy reveals blue supergiants, plus a few stars of intermediate
color dominated by foreground stars, plus a smattering of red stars, some of which are
native to Sextans A.  To decrease the effect of foreground contamination, we have
restricted the same to a region from $\alpha_{\rm J2000}=00^h01^m46^s$ to $00^h02^m12^s$, and
$\delta_{\rm J2000}=-15^\circ34'$ to $-15^\circ21'$, an area of 0.016 deg$^2$.  Middle:
We show the CMD of the combination of two neighboring foreground
fields with the same area, chosen from the periphery of the Sextans A.
Right: We show the CMD ``cleaned" of foreground stars (in a statistical sense)
 and converted to intrinsic color $(B-V)_0$ and absolute visual magnitude $M_V$.}
\end{figure}

In Fig.~\ref{fig:IC10cmd} (left) we show the unfiltered
CMD for IC10.  In comparing this to that
of the unfiltered SMC CMD, we are struck first by the large amount
of reddening.  The foreground sequence is very tilted, as expected very close
to the plane of the Milky Way, as more distant stars will be more heavily
reddened, and the line of slight goes through a substantial part of the Galaxy.
(In the other galaxies, the foreground CMDs are nearly vertical, as one
quickly runs pas the edge of the dust plane.)
The blue supergiants have been shifted from their peak near 0 in the SMC and
LMC to roughly 0.7 in IC10.   In Fig.~\ref{fig:IC10cmd} (right) we see the dereddened
CMD converted to absolute magnitude with the foreground contamination largely removed.

The second most noticeable fact about the CMD of IC10 given 
in Fig.~\ref{fig:IC10cmd}
is the relative dearth of RSGs.   Massey (2002) found that the relative number of
WRs and RSGs was extremely well correlated with metallicity
in the Local Group, at least for the SMC, NGC 6822, the LMC, M33, and M31.
The metallicity of IC10 is intermediate to that of the SMC and LMC, and
has an intermediate number of WR stars (26 confirmed, to the SMC's 12
and the LMC's 130; see Crowther et al.\ 2003, Massey \& Holmes 2002,
Massey 2003, and Massey \& Duffy 2002.)  So, one should expect
a  RSG branch that is similar in absolute numbers intermediate to 
the SMC and LMC, and a comparison of Figs.~\ref{fig:IC10cmd} with
\ref{fig:smccmd} and \ref{fig:lmccmd} shows that we are clearly {\it not} seeing
that.  (Massey \& Holmes 2003 estimate that the true number of WRs
in IC10 is about 90, which should imply even more RSGs.)  Where, then,
are the missing RSGs?

We don't know how long the burst of star-formation has been going on in IC10,
only that the amount of gas cannot sustain the present amount of star formation
for a significant fraction of a Hubble time (see discussion in Massey 
\& Armandroff 1995).  The typical age of a star in the WR phase is 3-4~Myr, while
the ages of RSGs will be 10-20~Myr.  Thus, if IC10's burst is extremely young, say
10 Myr, that would explain the lack of RSGs while allowing the large population of
WRs.  As we note in discussing NGC 6822, this lack of RSGs is consistent with
the luminosities of the visually brightest blue supergiants.

\subsubsection{NGC 6822}

The first comprehensive study of the resolved stellar population of NGC 6822 was
that of Kayser (1966, 1967), whose photographic photometry identified blue
supergiants, and a sequence of RSGs.  Some of the latter were shown to be
variable.  It is instructive to compare her CMD (Kayser 1967 Fig 3) to 
ours (Fig.~\ref{fig:n6822cmd} left).  Although ours goes deeper, of course, the
general characteristics are similar.
In our figure we find that the blue supergiants are well
distinguished from the
foreground dwarfs, despite the large number of the latter, due to NGC 6822's
relatively low Galactic latitude ($b=-18.4^\circ$).   In its CMD we see a strong
RSG component starting at  $B-V=1.8$ and $V=19.5$ and extending to $B-V>2$
and possibly $V=17.5$, much as Kayser (1966, 1967) found.  Massey (1998b)
obtained spectra of several of these RSGs.

By contrast to IC10, NGC 6822 contains only 4 WRs, and it is likely this number
is complete (Massey \& Johnson 1998).    A comparison of the 
CMDs of NGC 6822 (Fig.~\ref{fig:n6822cmd} left) and IC10 (Fig.~\ref{fig:IC10cmd}  left)
is quite instructive: we expect stars to be shifted about 1.7 mag fainter in IC10 and
to 0.55 redder colors.   The tip of the ``blue plume" in NGC 6822 is at about
$V=18$, and we see that in IC10 this is more like $V=21$---although IC10 does
have a large number of WRs, the (visually) brightest blue supergiants are not
visible in similar numbers.  Recall, though, that the brightest blue supergiants
in $V$ are the evolved A-type supergiants from intermediate high mass stars
(10-25$M_\odot$) with an age similar to RSGs; thus this comparison 
shows consistency between the strong blue and red supergiant populations of
NGC 6822, and the relatively weak population of
{\it visually} bright blue supergiants and lack of RSGs in IC10.

Only 38  of NGC 6822's massive stars have been observed
spectroscopically, and we list those in Table~\ref{tab:n6822mem}\footnote{Bianchi et al.\ (2001) and 
Catanzaro et al.\ (2003) observed several additional
stars in the NGC 6822, but no spectral types are given, apparently due to the
poor signal-to-noise of their spectra.}.  The WRs are identified
by the designations given in Massey \& Johnson (1998), while the OB stars are primarily
identified by the OB association designations given in Massey et al.\ (1995a).  Red
supergiants were identified from spectroscopic photometry by Massey (1998b); in
some cases these lack spectral types, but the strengths of the CaII triplet and radial
velocities have been used to show these are members.  In addition, there are a few
early- and late-type supergiants that were observed spectroscopically by Humphreys (1980a).
These are refereed to only by the designations given in the photographic reproductions
in the PhD thesis of Kayser (1966), with no coordinates provided.
The reproduction of these plates are usually not readable in
the copies of this thesis that are available commercially.  However, an excellent
copy is now available electronically\footnote{See http://etd.caltech.edu/etd/available/etd-09232002-112325.  We are indebted to Cathy Slesnick for tracking this down.}, and we used this to cross reference the stars with spectroscopy
by Humphreys (1980a) to our survey. 

NGC 6822 is one of the few galaxies in our sample for which previous ground-based photometry 
has been published covering most of the galaxy.  We provide a comparison
to the work of Bianchi et al.\ (2001) in Fig.~\ref{fig:n6822comp}. Out of the 3232 stars in 
Bianchi et al.\ (2001)'s catalog, we
found 3135 matches to our Mosaic catalog, where we have restricted the latter
sample to only those stars with errors less than 0.1~mag. 
We had to adjust the Bianchi et al.\ (2001) coordinates
by -0.05$^s$ and -0.17",  due
to the need to transform from the Guide Star Catalog system used  by
Bianchi et al.\ (2001) to that of USNO-B, which we use here. 
The photometry agrees extremely well in that the median differences are 
0.03~mag in $V$, 0.002~mag in $B-V$, and
0.020~mag in $U-B$, all in the sense of our values minus those of Bianchi
et al.\ (2001).   However, it is clear from Fig.~\ref{fig:n6822comp}b that there
is a significant color term in $B-V$: the bluest stars in our catalog appear slightly
less blue in Bianchi et al.\ (2001), while the red stars in our sample are redder
than in Bianchi et al.\ (2001).   No independent {\it UBV} calibration was obtained
for the ground-based data of Bianchi et al.\ (2001), but rather their photometry
was tied to the Massey et al. (1995) study of OB stars in several small regions
in NGC 6822, which covered a more modest color range than here\footnote{
The first author of the present paper was responsible for the reduction of the
ground-based photometry of Binachi et al.\ (2001).}.  For $U-B$ (Fig.~\ref{fig:n6822comp}c) there is a
modest color term for the reddest stars, and in addition there is a sequence
of stars that are much redder in the $U-B$ of Bianchi et al.\ (2001) than in
the present study.  The size of the color term is not unexpected, as Bianchi et al.\ (2001)
warn that the reddest stars might be systematically affected by errors at the 0.1~mag
level.  The sequence of much redder stars is harder to understand.  To investigate
this further, we selected only the stars with differences $>0.5$ mag in $U-B$ and
show these stars plotted against $B-V$ in Fig.~\ref{fig:n6822comp}d.  It is clear that
the problem occurs primarily around a $B-V$ of 0, where stars in Bianchi et al. (2001)
may show {\it much} redder $U-B$ values than their $B-V$ warrants.  Massey (2002)
has reported similar problems in $U-B$ CCD photometry near $B-V \sim 0$, owing
to the usual mismatches of the intrinsic filter plus detector bandpass compared to the
standard bandpass, and the size of the Balmer jump.  

 \subsubsection{WLM}
 
It is clear from Fig.~\ref{fig:wlmcmd} that WLM is still relatively rich in massive stars,
at least compared to the lower luminosity galaxies (Figs.~\ref{fig:sexBcmd}-\ref{fig:phxcmd}),
as shown by the strength of the ``plume" of  blue stars with $B-V\le 0.2$.  The galaxy is far from
the Galactic plane ($b=-73.6^\circ$), 
and has minimal reddening (Minniti \& Zijlstra 1997, and our Table~\ref{tab:reddenings}).  Nevertheless,
as discussed in \S~\ref{Sec-foreground}, the vast majority of bright ($16<V<20$) stars
of intermediate color ($0.4<B-V<1.1$) are foreground.

Five WLM stars were confirmed spectroscopically as early-type supergiants by Venn et al.\ (2003), and these plus an additional
33 stars were observed spectroscopically by Bresolin et al.\ (2006).   The fact that all 38 of these stars could be observed in just
a little over two hours of integration with the VLT emphasizes the power of modern large telescope with multi-object spectroscopic
capabilities; this number is the same as the number of stars observed spectroscopically over the past 27 years 
in the considerably better studied galaxy NGC 6822 (Table~\ref{tab:n6822mem}).  Bresolin et al.\ (2006) identified primarily B and
early A supergiants, as well as two O stars.  They also identified four G-type supergiants.   The fact that all four of these stars
of intermediate color were supergiants is rather surprising, as these
stars are found in a region of the CMD where we expect a significant foreground contribution (Fig.~\ref{fig:wlmcmd} (middle)).
Using the Bahcall \& Soniera (1980) model we compute that for 
$18<V<21$ and $1.0<B-V<1.8$ (the range covered by their G supergiants) we expect 35 foreground stars; in good agreement
with the 43 stars we count from our neighboring foreground fields.  There are 84 stars in this part of the CMD, so we would
expect about 40-50\% to be foreground. (In contrast, Bresolin et al.\ 2006 suggest that the foreground contribution is negligible, a conclusion with which
we do not agree.)
The luminosity criteria used by Bresolin et al.\ (2006) is based on an
empirical relation between the equivalent width of H$\gamma$ and the luminosity class found for somewhat earlier
type (O9-F8) stars, as their dispersion was relatively low (5\AA).   Further work may be needed to
clarify  the exact nature of these four stars, as halo giants would have radial velocities similar to that of the WLM.

Several groups have obtained and discussed global, ground-based photometry of this galaxy
(Minniti \& Zulstra 1997, McConnachie et al. 2005) but not published the individual values.
Bresolin et al.\ (2006) list $V$ and $V-I$ photometry of their 38 spectroscopic candidates, based
upon some as yet unpublished photometric study.  Excluding the one blended  object
(J000158.73-153001.5) and one outlier in color (J000156.62-152501.5) we find very good
agreement in the photometry, with a median difference
$V$(Mosaic)-$V$(Bresolin)=-0.03 and $V-I$(Mosaic)-$V-I$(Bresolin)=-0.01.  Deep {\it HST}
photometry has been analyzed by Dolphin (2000) and Rejkuba et al.\ (2000).

\subsection{The Other Galaxies}

Spectroscopy has not yet been carried out for the stars in the other four galaxies, and
so we just discuss them briefly here.

Although Sextans B is more luminous than Sextans A (Table~\ref{tab:galaxies}) it is clear
from comparing the CMDs (Figs.~\ref{fig:sexBcmd} and \ref{fig:sexAcmd}) that there are considerably
more massive stars present in Sextans A.  Both show some enhancement of bright red stars over
the adjacent foreground fields, and so it is likely that these galaxies contain some RSGs.

Analysis of deep CCD photometry of Sextans B has been published by Sakai et al.\ (1997), but
individual measurements were not published other than for a hand full of stars identified only on
the finding charts of Sandage \& Carlson (1985a), and we have not done the matching that would
be needed to compare our results. (See also Piotto et al.\ 1994).  Deep {\it HST} photometry of
a part of Sextans~A has been analyzed by Dohm-Palmer et al.\ (1997a,b).

\begin{figure}
\epsscale{0.3}
\plotone{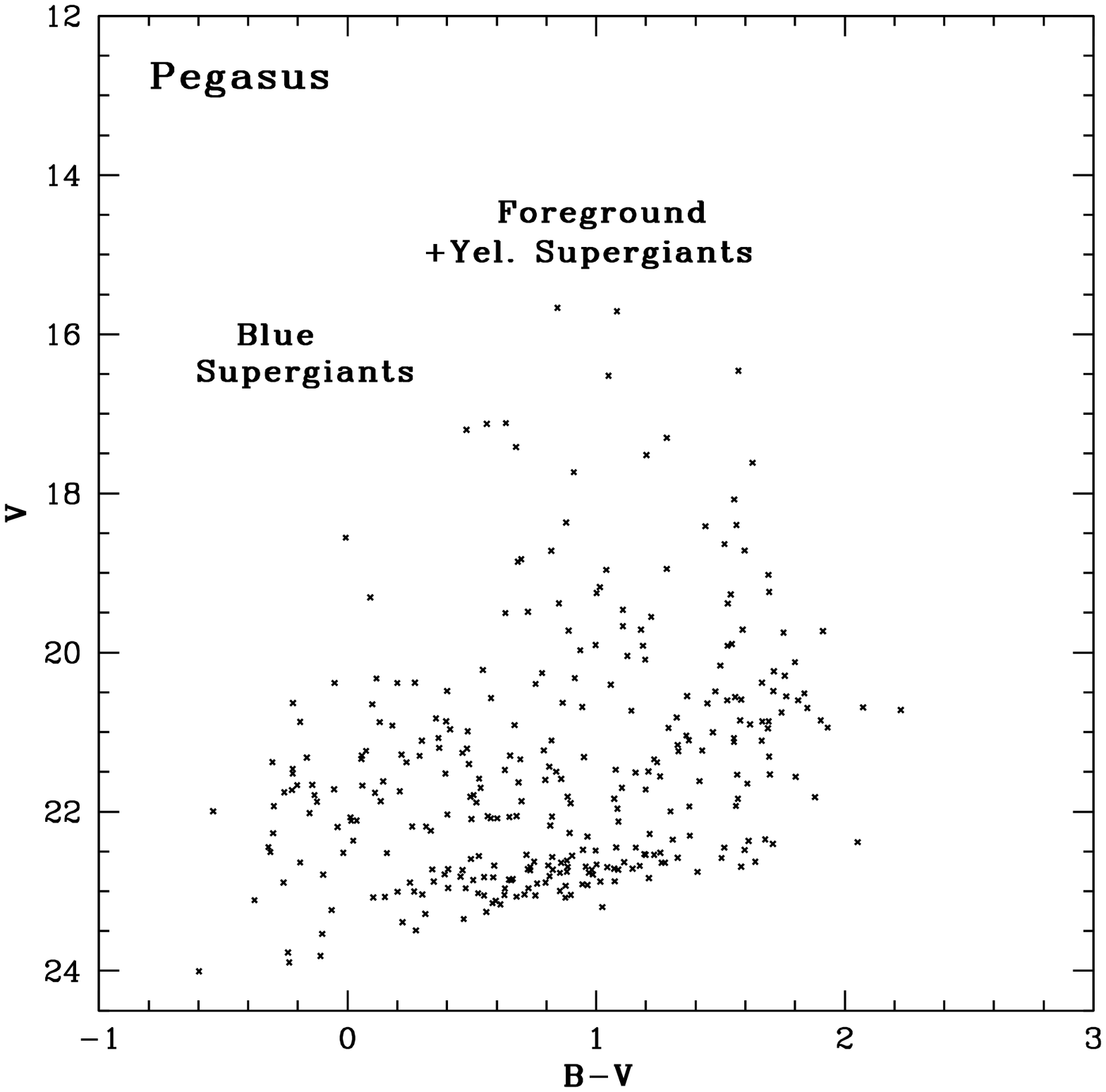}
\plotone{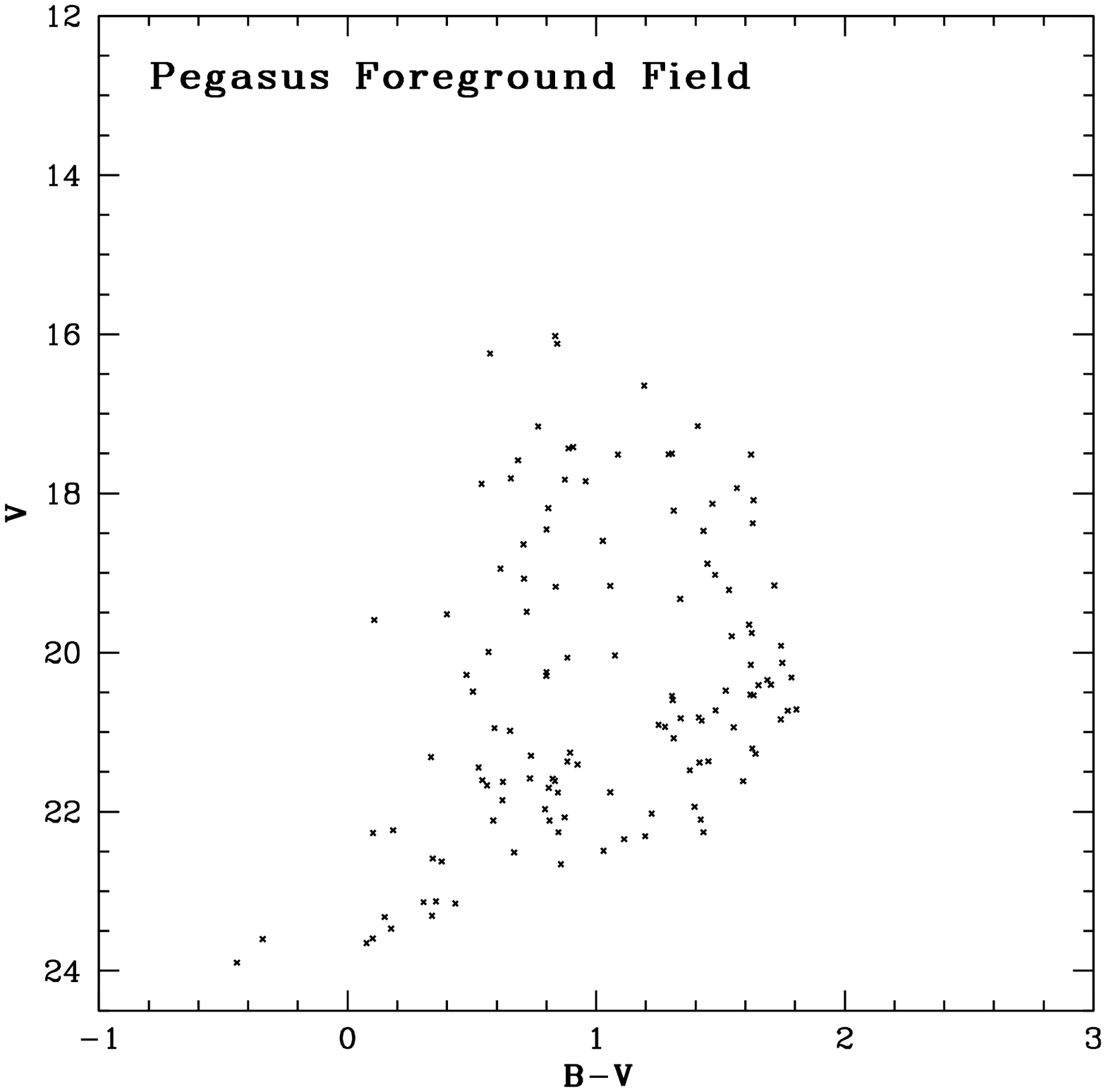}
\plotone{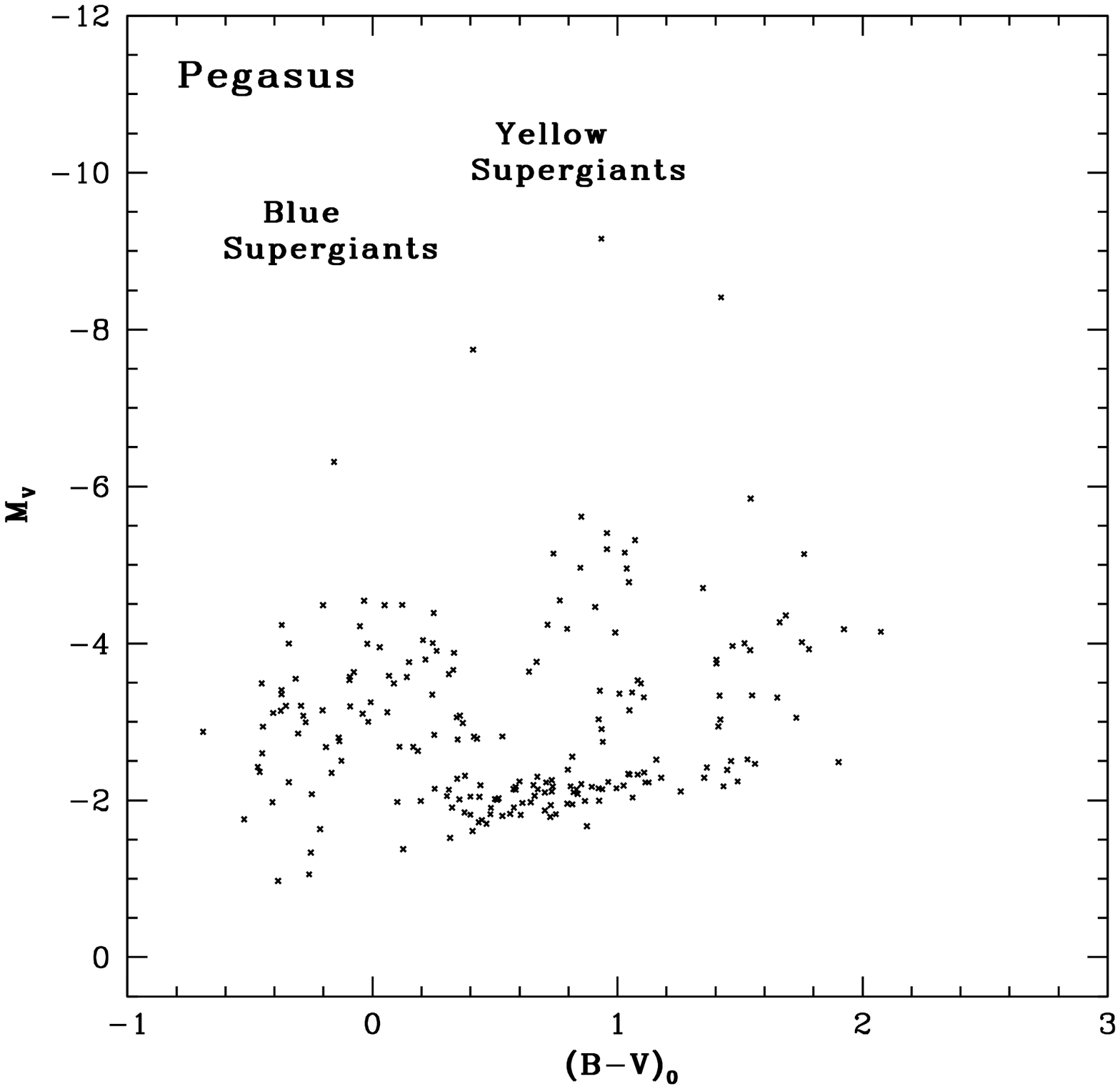}
\caption{\label{fig:pegcmd} The color-magnitude diagram for Pegasus. 
Left: The CMD of the galaxy reveals blue supergiants, plus a few redder
stars dominated by foreground stars. To decrease the effect of foreground contamination, we have
restricted the same to a region from $\alpha_{\rm J2000}=23^h28^m15^s$ to $23^h28^m53^s$, and
$\delta_{\rm J2000}=+14^\circ41'$ to $+14^\circ49'$, an area of 0.020 deg$^2$.  Middle:
We show the CMD of the combination of two neighboring foreground
fields with the same area, chosen from the periphery of the Pegasus dwarf.
Right: We show the CMD ``cleaned" of foreground stars (in a statistical sense)
 and converted to intrinsic color $(B-V)_0$ and absolute visual magnitude $M_V$.}
\end{figure}

\begin{figure}
\epsscale{0.3}
\plotone{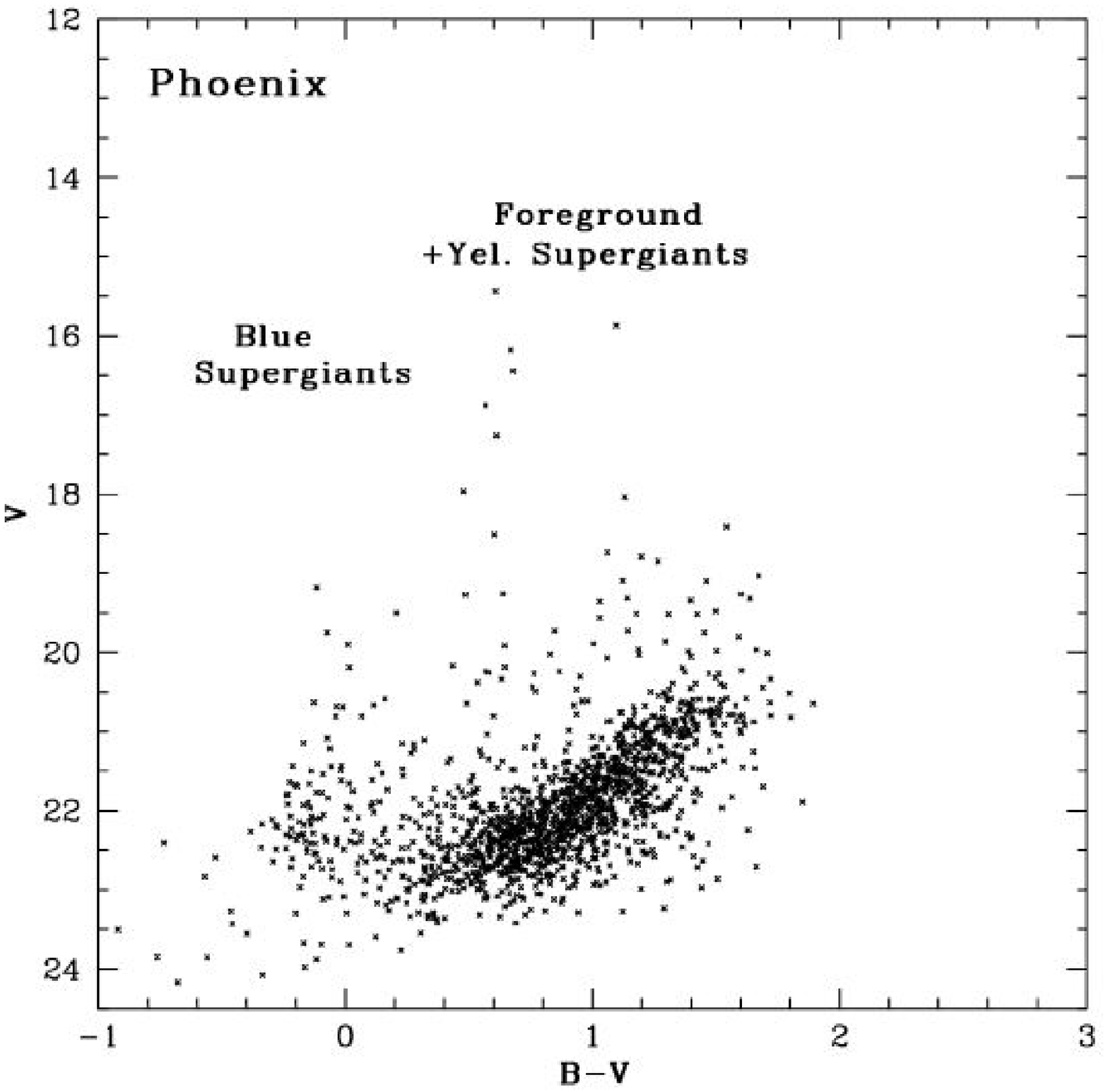}
\plotone{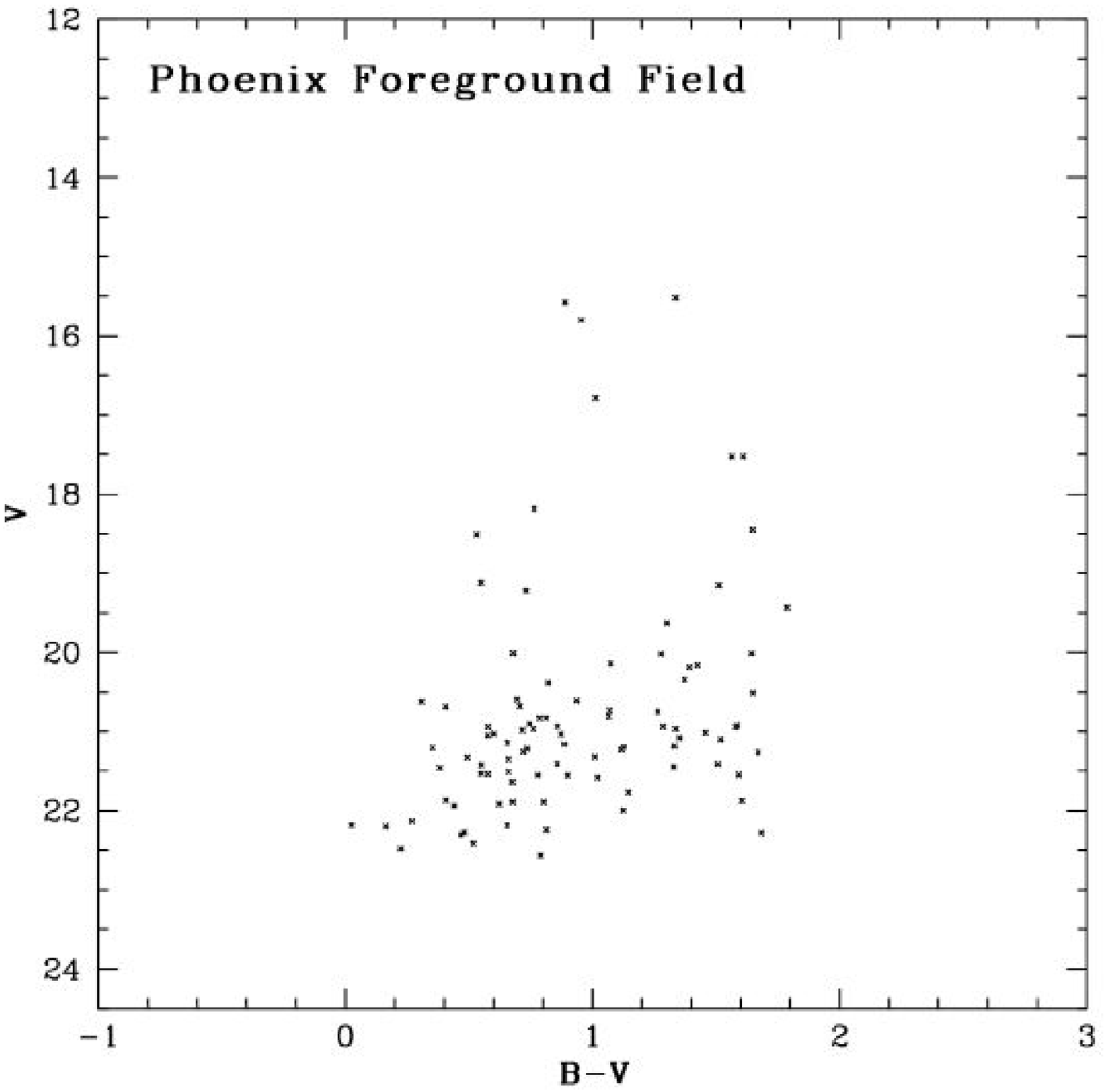}
\plotone{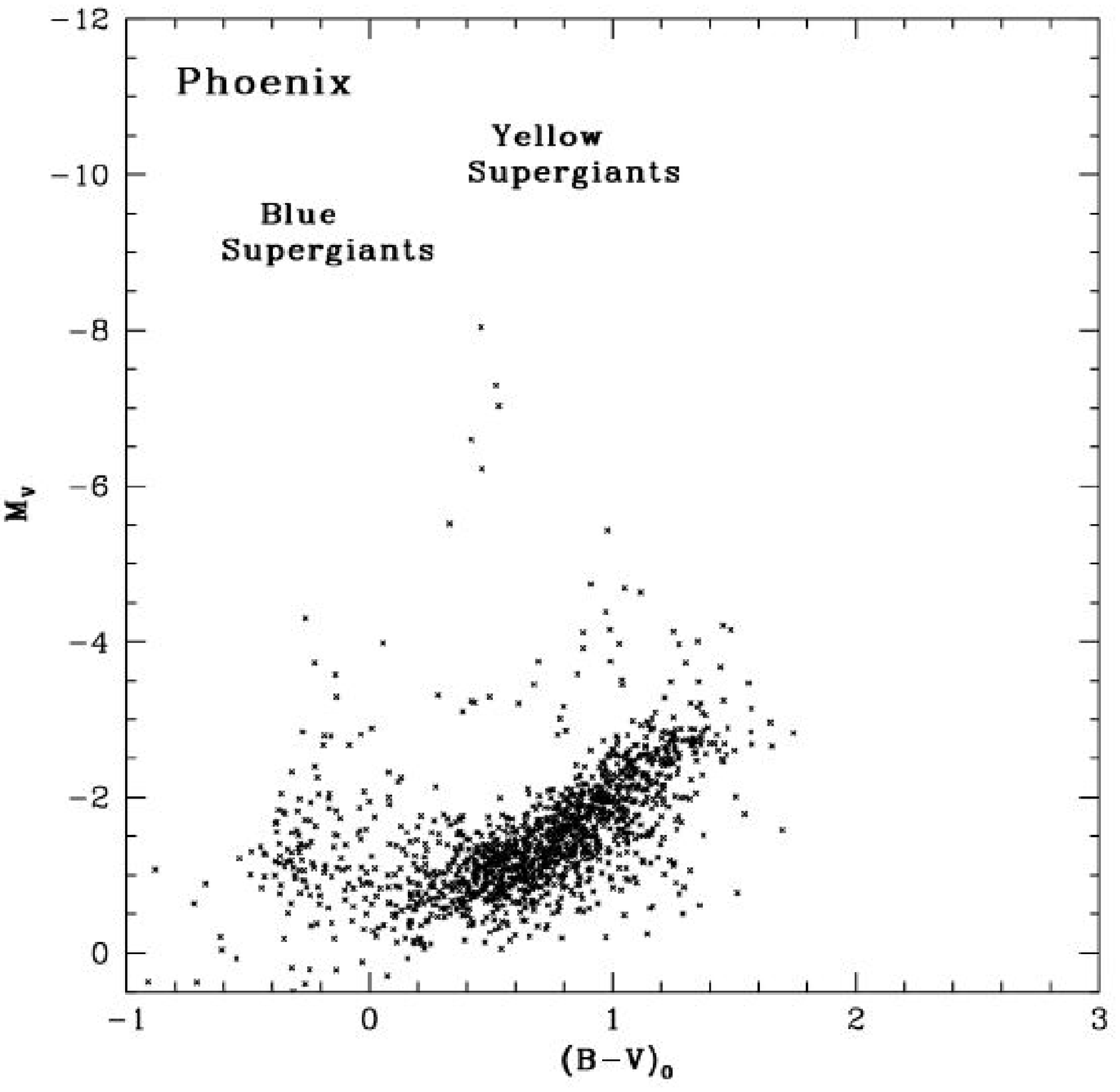}
\caption{\label{fig:phxcmd} The color-magnitude diagram for Phoenix. Left: The CMD of the galaxy reveals a wealth of faint stars of intermediate
and red color, and a few blue supergiants.
To decrease the effect of foreground contamination, we have
restricted the same to a region from $\alpha_{\rm J2000}=01^h50^m42^s$ to $01^h51^m25^s$, and
$\delta_{\rm J2000}=-44^\circ31'$ to $-44^\circ23'$, an area of 0.017 deg$^2$.  Middle:
We show the CMD of the combination of two neighboring foreground
fields with the same area, chosen from the periphery of the Phoenix.
Right: We show the CMD ``cleaned" of foreground stars (in a statistical sense)
 and converted to intrinsic color $(B-V)_0$ and absolute visual magnitude $M_V$.}
\end{figure}

\begin{figure}
\epsscale{0.4}
\plotone{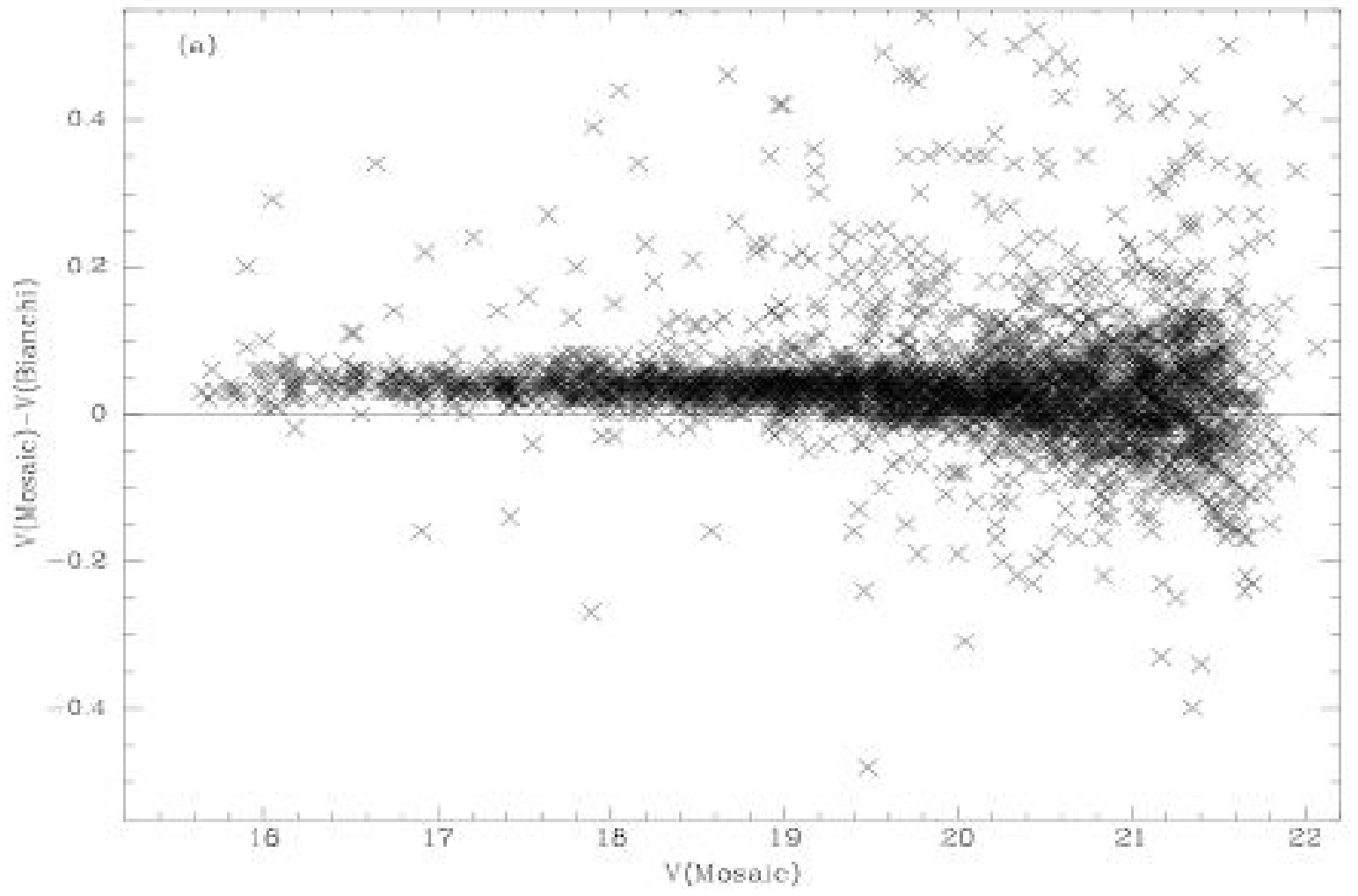}
\plotone{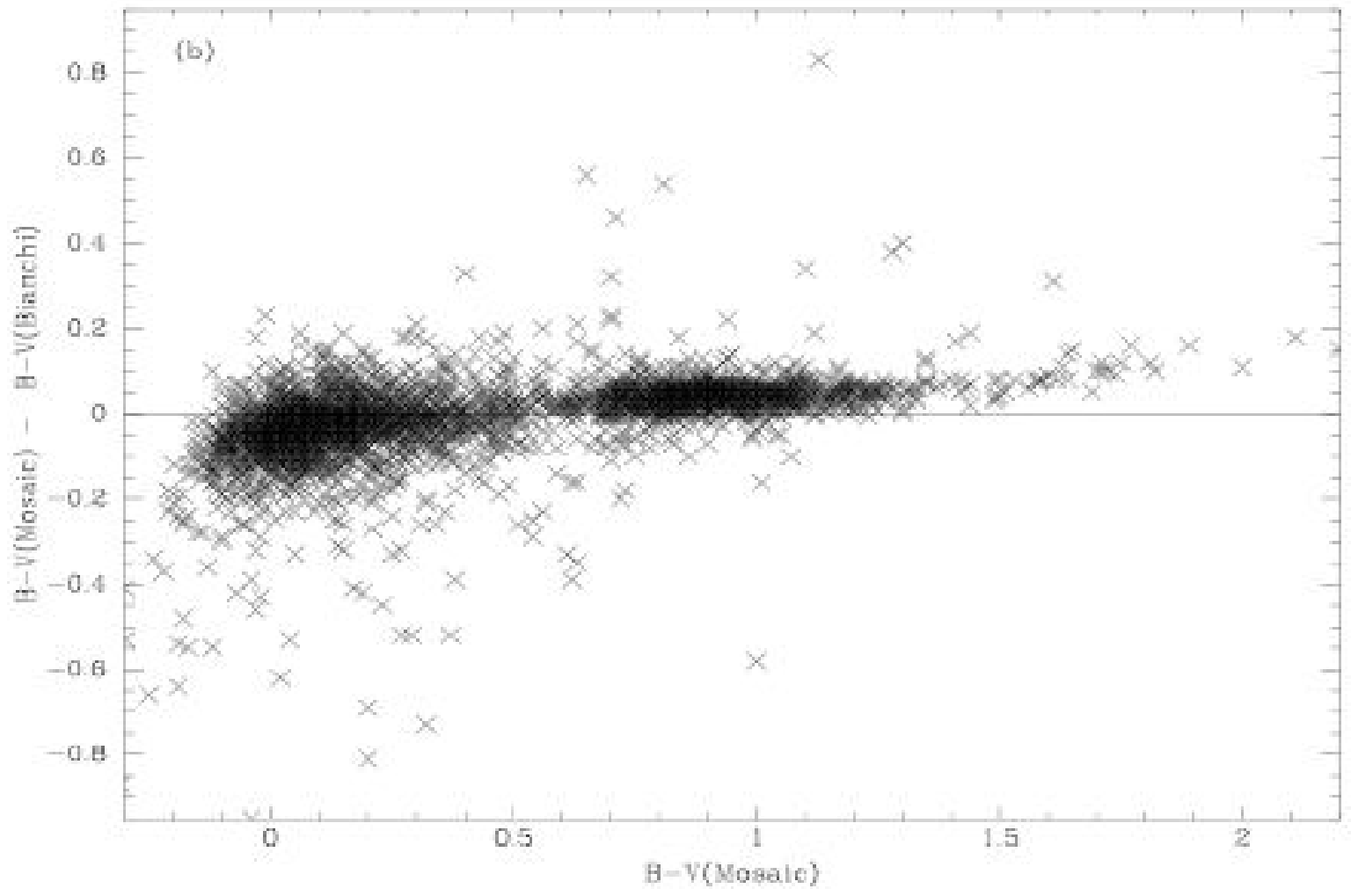}
\plotone{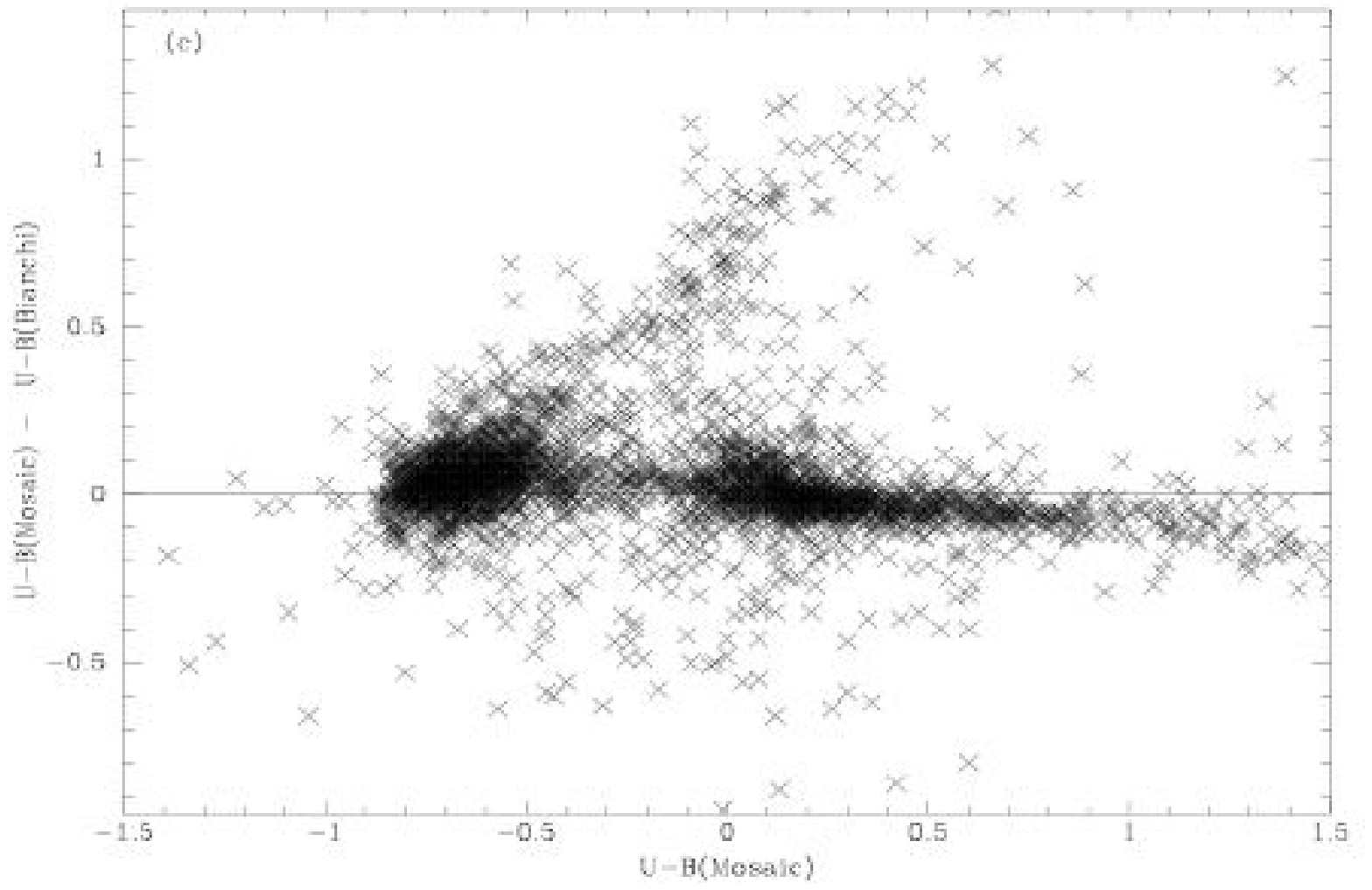}
\plotone{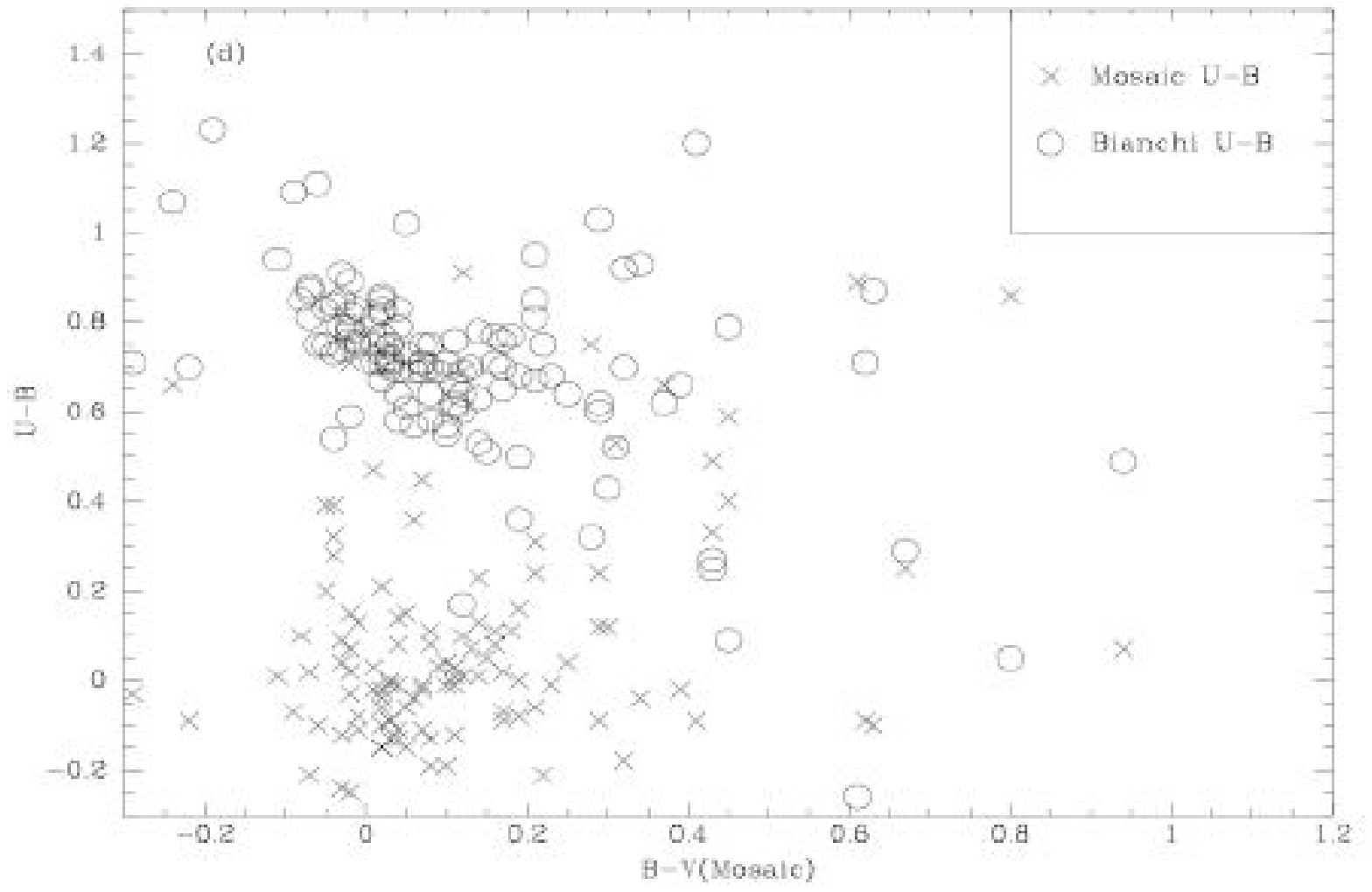}
\caption{\label{fig:n6822comp} Comparison of our photometry with that of
Bianchi et al.\ (2001).  (a) There is a slight offset (0.03) mag in the $V$ photometry. 
 (b) The average $B-V$ agrees, but there is a significant color term. (c) The average $U-B$ agrees, but there is not only a color term, but a significant sequence of stars
 that are considerably redder Bianchi et al.\ (2001)'s $U-B$ than in ours.  (d) We show only those stars with $U-B$ differences $>$0.5~mag from (c).  For
 these stars the $U-B$ photometry
 from Bianchi et al.\ (2001) (open circles) is considerably greater than that expected
 for their $B-V$ colors, while that from our survey (x's) are more in keeping with
 what one would expect. }
\end{figure}

Pegasus contains only a smattering of blue supergiants (compare the
CMD of the galaxy field (Fig.~\ref{fig:pegcmd} (left)) with
that of the neighboring foreground field (Fig.~\ref{fig:pegcmd}
(middle)).  These are apparent in Fig.~\ref{fig:pegcmd} 
(right), in which we have subtracted the foreground, statistically.
Even the lowest luminosity galaxy in our sample, Phoenix, contains
a stronger blue supergiant population (Fig.~\ref{fig:phxcmd}).  The
primary difference is that Phoenix contains a very strong population
of intermediate age stars, in accord with the findings of Canterna
\& Flower (1977). As such, Phoenix appears to be intermediate between
an irregular galaxy and a dwarf spheroidal (van den Bergh 2000).
Neither galaxy appears to have a significant number of RSGs.

Deep {\it BVR} photometry of Pegasus was discussed by Lee (1995a,b)
but individual measurements were not apparently ever published.
Lee (1995b) mentions the presence of bright RSGs in his CMD, but a
comparison of our CMD of the galaxy (Fig.~\ref{fig:pegcmd} 
(left) with the CMD of neighboring foreground fields
(Fig.~\ref{fig:pegcmd} (middle) makes this unlikely.  Gallagher
et al.\ (1998) have provided a very comprehensive study of the
stellar populations of Phoenix based upon deep {\it HST} and
ground-based imaging.

\section{Summary and Future Work}
\label{Sec-summary}

We have obtained broad-band photometry of 88,144 stars 
(identified in {\it B,V, and R}) in
IC10 (20,663 stars), NGC 6822 (51,877 stars), WLM (7,656 stars), Sextans B (800 stars),
Sextans A (1,516 stars), Pegasus (1,390 stars), and Phoenix (4,242 stars).  Combined with our
earlier work in Paper I on M31 (371,781 stars) and M33 (145,522 stars), our survey consists
of over half a million stars with accurate coordinates and photometry.  The corresponding images
are available from our ftp site\footnote{ftp://ftp.lowell.edu/pub/massey/lgsurvey/datarelease/} or
through NOAO\footnote{http://archive.noao.edu/nsa/}. Although our survey lacks the high angular
resolution afforded by {\it HST}, it provides full areal coverage of these galaxies, including neighboring
regions that can be used for evaluating the contribution of foreground stars to the CMDs; the photometry
is also on a well-defined standard system that can be used for monitoring long-term variability.  It is our
hope that these data will serve both as the impetus and ``finding charts" for further space-based imaging,
and for many spectroscopic programs at large aperture.

We have inter-compared the CMDs of these systems and those of the Magellanic Clouds.  This has led
to improved estimates of the typical reddening of an OB star in these systems, based upon the colors
of the plume of blue supergiants, and adopting the reddening values for the LMC and SMC based upon
spectroscopy of hundreds of stars.  The CMDs reveal strong or modest blue supergiants in all of these
galaxies.     All but Pegasus and Phoenix also show red supergiants in their CMDs, although we find
a curious deficiency of RSGs in the starburst galaxy IC10.   Without exception, the bright stars of intermediate
color in the CMDs are strongly (30-100\%) contaminated by foreground Galactic stars, a combination of
disk and halo dwarfs, and halo giants. 

Follow-up projects we plan in the near future include:
\begin{enumerate} 
\item Determination of the IMF for massive stars at differing metallicities.  
\item Identification of yellow supergiants (F-G~I).  Given the large foreground contamination, this will require
spectroscopy
of hundreds, and possibly thousands, of stars, but such studies can be readily accomplished by multi-object
spectroscopy on 6.5-m class telescopes.  Knowledge of the numbers of yellow supergiants would provide
critical test of massive stellar evolutionary theories.  Such stars would also provide a good astrometric reference,
as they are relatively bright ($V\sim 16$) and have negligible proper motion.
\item Spectroscopic confirmation of the RSGs in these systems, both to provide statistics on the relative number
of red and blue massive stars at differing metallicities (for comparison with evolutionary models), and to
provide targets for spectroscopic determinations of physical properties.
\item Determination of physical properties of blue supergiants at differing metallicities.
\end{enumerate}.

\acknowledgments
We are grateful for the encouragement of our colleagues in this project; help with the initial proposal was also provided
by N. King and A. Saha.  R.T.M. was supported through the NSF's REU program, and we gratefully acknowledge
AST-0453611.

\clearpage

\clearpage

\clearpage

\clearpage

\begin{deluxetable}{l l l r r r c c c c c c}
\tabletypesize{\tiny}
\tablewidth{0pc}
\tablenum{1}
\tablecolumns{11}
\tablecaption{\label{tab:galaxies}Properties of Dwarf Galaxies in Our Sample\tablenotemark{a}}
\tablehead{
\colhead{Galaxy}
&\colhead{Alias}
&\colhead{Type}
&\colhead{$l$ [deg]}
&\colhead{$b$ [deg]}
&\colhead{$M_V$}
&\colhead{$E(B-V)$\tablenotemark{b}}
&\colhead{Dist.~[Mpc]}
&\colhead{log O/H+12} 
&\colhead{$\log{\dot{M}}$\tablenotemark{c}  }
&\colhead{$\log{\dot{M_D}}\!$\tablenotemark{c}   }
}
\startdata
IC10   & UGC 192   & Ir IV:    &118.97&-3.34& -16.3 & 0.81& 0.66 & 8.2 &  -1.3 & -1.3 \\
NGC 6822 & \nodata & Ir IV-V &25.34&-18.39& -16.0 & 0.25& 0.50 & 8.1&  -2.0 & -2.0 \\
WLM    & DDO 221   & Ir IV-V  &75.85&-73.63& -14.4 & 0.07 & 0.93 & 7.7 &  -2.8& -2.9 \\
Sextans B&DDO 70   & IrIV-V  &233.20 & +43.78& -14.3 & 0.09 & 1.32 & 7.6 &-3.0& -2.9  \\ 
Sextans A&DDO 75   & Ir V     &246.17 & +39.86&  -14.2 & 0.05  & 1.45 & 7.5   &-2.2&  -1.4 \\
Pegasus & DDO 216  & Ir V     &94.77 & -43.55&  -12.3 & 0.15 & 0.76 & 7.9  &-4.4& -4.2  \\
Phoenix  & \nodata & dIr/dSph &272.19 & -68.95&  -9.8 & 0.15 & 0.40 & \nodata  & \nodata  \\
\cutinhead{Other LG Galaxies Given for Comparison}
M31  &  NGC224 & Sb I-II & 121.17 & -21.57 & -21.2 & 0.13 & 0.76 & 9.0 & -1.3\tablenotemark{d} & -3.3\tablenotemark{d} \\
M33  &  NGC598  & Sc II-III &133.61 & -31.33 & -18.9 & 0.12 & 0.83 & 8.4 & -1.0\tablenotemark{d} & -2.0\tablenotemark{d} \\
LMC & \nodata & Ir III-IV  & 280.19 & -33.29& -18.5 & 0.13 & 0.050 & 8.4 & -0.8\tablenotemark{d}& -1.8\tablenotemark{d} \\
SMC & \nodata & Ir IV-V  & 302.81 &-44.33 & -17.1  & 0.09&0.059 & 8.0 &-1.4\tablenotemark{d}& -2.3\tablenotemark{d} \\
\enddata
\tablenotetext{a}{Data from van den Bergh 2000, except as noted.}
\tablenotetext{b}{Total $E(B-V)$ determined in Section~\ref{Sec-reddenings}.}
\tablenotetext{c}{Star formation rates (SFRs) are given in terms of $M_\odot$ yr$^{-1}$
integrated over the entire galaxy ($log\dot M$), and also normalized by the 
area of the $V$-band scale-length (i.e.,
$\dot{M_D}$, units of$M_\odot$ yr$^{-1}$ kpc$^{-2}$);
from Hunter \& Elmegreen 2004, except as
noted.}
\tablenotetext{d}{Calculated here, using the H$\alpha$ luminosities from 
Kennicutt \& Hodge (1986) and Kennicutt et al. 1989,
corrected by an additional $A_{H\alpha}=0.2$ mag, and adopting disk scale lengths of 5.6~kpc (M 31,
Walterbos \& Kennicutt 1988),
1.7~kpc (M33, Sparke \& Gallagher 2000), 
1.5~kpc  and 0.9~kpc (LMC and SMC, Bothun \& Thompson 1988).}
\end{deluxetable}

\begin{deluxetable}{l l l c c c c c c c c c c c c c c}
\rotate
\setlength{\tabcolsep}{0.04in}
\tabletypesize{\tiny}
\tablewidth{0pc}
\tablenum{2}
\tablecolumns{17}
\tablecaption{\label{tab:journal}Mosaic Observations}
\tablehead{
\colhead{Field}
&\colhead{Obs.}
&\colhead{Date}
& \multicolumn{2}{c}{$U$}
& \colhead{}
& \multicolumn{2}{c}{$B$}
& \colhead{}
& \multicolumn{2}{c}{$V$}
& \colhead{}
& \multicolumn{2}{c}{$R$}
& \colhead{}
& \multicolumn{2}{c}{$I$}  \\ \cline{4-5} \cline{7-8} \cline{10-11} \cline{13-14} \cline{16-17} 
\colhead{}  & & & \colhead{Exps.} & \colhead{DIQ(")} && \colhead{Exps.} & \colhead{DIQ(")} && \colhead{Exps.} & \colhead{DIQ(")} && \colhead{Exps.} & \colhead{DIQ(")}
&& \colhead{Exps.} & \colhead{DIQ(")}

}

\startdata

IC10\tablenotemark{a} & KPNO &
    2001 Sep 20-22 & 
     5x600s &  1.1    && 
     5x60s, 2x900s, 1x1800s, 5x2400s & 1.0 && 
     5x60s, 3x600s &                              0.9 &&
     5x50s, 3x200s                                  & 0.9&& 
      5x150s                                             & 0.9 \\

NGC 6822 & CTIO& 2000 Sep 1-2 &
   6x600s , 1x180s & 1.2 &&
   5x120s, 1x180s  & 1.4 &&
   5x120s, 1x180s    & 1.4 &&
   5x100s, 1x180s   & 1.4 &&
   5x120s, 1x180s    & 1.0 \\

WLM  &  CTIO & 2000 Sep 1-2 &
    5x600s, 1x60s  & 1.3 &&
    5x90s, 1x60s  & 1.2 &&
    5x70s, 1x60s  & 1.1 &&
    5x70s, 1x60s   & 1.0 &&
    5x200s, 1x60s & 1.0 \\

Phoenix &  CTIO & 2000 Sep 1&
    5x600s & 1.3 &&
    5x60s   & 1.3 &&
    5x60s & 1.2 &&
    5x50s & 1.2 &&
    5x200s &  1.0 \\

Sextans B\tablenotemark{a}  & KPNO & 2002 Feb 13 &
   5x600s & 1.2 &&
   5x60s & 1.2 &&
   5x60s & 1.2 &&
   5x50s & 1.1 &&
   5x150s  & 0.9 \\

Sextans A\tablenotemark{a} &KPNO&
   2002 Feb 13 &
   5x600s &1.4 &&
   5x60s & 1.2 &&
   5x60s & 1.2 &&
   5x50s & 1.1 &&
   5x150s & 1.1 \\

Pegasus\tablenotemark{b} & KPNO &
    2000 Oct 03 & 
    5x600s & 1.1 && 
    5x60s & 1.0 && 
    5x60s & 1.0 && 
    5x50s & 1.0 &&
    5x150s & 0.8 \\

\enddata

\tablenotetext{a}{Centered on chip 2; FOV of calibrated data is 20'x30'.}
\tablenotetext{b}{Centered in the middle, but FOV of calibrated data is 20'x30'.}

\end{deluxetable}

\begin{deluxetable}{l c c c c c c c c c}
\tablewidth{0pc}
\tablenum{3}
\tablecolumns{9}
\tablecaption{\label{tab:colorterms}Color Terms for the CTIO 4-m Mosaic Camera\tablenotemark{a}}
\tablehead{
\colhead{Color}
&\multicolumn{8}{c}{Chip\tablenotemark{c}} \\ \cline{2-9}
\colhead{Term\tablenotemark{b}}
&\colhead{1}
&\colhead{2}
&\colhead{3}
&\colhead{4}
&\colhead{5}
&\colhead{6}
&\colhead{7}
&\colhead{8}
}
\startdata
$K_{U1}$  & $-0.092$  &  $-0.047$   &  $-0.075$ & $ -0.098$    &$-0.068$   &$-0.044$ &  $-0.015$  & $-0.043$\\
$K_{U2}$  & $ -0.272$ &   $-0.216$  &   $-0.203$ & $ -0.229$   & $-0.210$  &$ -0.228$ & $ -0.225$ & $ -0.251$\\
$K_B$      &  $-0.164$  & $ -0.160$   &  $-0.130$ & $ -0.130$   & $-0.156$  &$ -0.151$ & $ -0.171$ & $ -0.154$\\
$K_V$     &  $+0.006$  &  $-0.014$  &  $+0.000$ &  $+0.015$   &$ -0.009$ &  $-0.002$ & $ -0.015$  & $+0.011$\\
$K_R$     &  $ +0.012$  &  $-0.016$   &  $ -0.008$ &  $+0.006$  &  $-0.021$  & $+0.007$  & $-0.034$  &$ +0.009$\\
$K_I$       & $+0.027$  &  $+0.079$   & $+0.101$  & $+0.124$   & $+0.054$  & $+0.054$ & $ +0.085$  & $+0.080$\\
\enddata
\tablenotetext{a}{Typical uncertainties in the color terms are 0.020.}
\tablenotetext{b}{The color terms are defined as follows:
$$u_{\rm Mosaic} = K_{U1} (U-B)_{\rm std} + C_U, (U-B)_{\rm std}>0$$
$$u_{\rm Mosaic} = K_{U2} (U-B)_{\rm std} + C_U, (U-B)_{\rm std}<0$$
$$b_{\rm Mosaic} = K_{B}   (B-V)_{\rm std} + C_B$$
$$v_{\rm Mosaic} = K_{V}    (B-V)_{\rm std} + C_V$$
$$r_{\rm Mosaic} = K_{R}    (V-R)_{\rm std} + C_R$$
$$i_{\rm Mosaic} = K_{I}   (R-I)_{\rm std} + C_I,$$
\noindent
where the $K$'s are the color terms, and $C$'s are the zero-points.
}
\tablenotetext{c}{Numbered as in the Mosaic II manual (Schommer et al.\ 2000), starting with the 
south-western chip
and continuing north along the western set of four, and then north along the eastern four.}

\end{deluxetable}

\begin{deluxetable}{l r r r r r r r r r r r}
\tablewidth{0pc}
\tablenum{4}
\tablecolumns{12}
\tablecaption{\label{tab:agree}Average Agreement Between Different Calibrations\tablenotemark{a}}
\tablehead{
\colhead{}
&\multicolumn{5}{c}{NGC 6822} 
&
&\multicolumn{5}{c}{WLM}   \\ \cline{2-6} \cline{8-12}
& \multicolumn{2}{c}{Lowell 1.1-m}
&
&\multicolumn{2}{c}{CTIO 0.9-m}
&
&\multicolumn{2}{c}{Lowell 1.1-m}
&
&\multicolumn{2}{c}{CTIO 0.9-m} \\ \cline{2-3} \cline{5-6}  \cline{8-9} \cline{11-12}
\colhead{Index}
&\colhead{\#}
&\colhead{Diff}
&\colhead{}
&\colhead{\#}
&\colhead{Diff}
&\colhead{}
&\colhead{\# }
&\colhead{Diff }
&\colhead{}
&\colhead{\#}
&\colhead{Diff.}
}
\startdata
V       &133 &+0.016   &&          165& $-$0.005 &&      20& $-$0.005 &&     35& $-$0.002\\
B$-$V   & 133 &+0.020   &&          165 &$-$0.021  &&    20 &$-$0.005  &&     35& $-$0.001\\
U$-$B   &130& $-$0.037  &&          164& $-$0.004  &&    20& $-$0.028   &&    34& +0.018\\
V$-$R   &132& +0.008  &&         164  & 0.000  &&    20&  $-$0.008   &&   35& $-$0.010\\
R$-$I     &  69& +0.016  &&          95& $-$0.003  &&    12&  +0.014 &&     27& $-$0.001\\
\enddata
\tablenotetext{a}{We give the median differences between the final calibrated
photometry and that of the Lowell 1.1-m and CTIO 0.9-m calibration data
for NGC 6822 and WLM.}
\end{deluxetable}

\begin{deluxetable}{l r r r r r r r r r r r r r r r r r l l}
\pagestyle{empty}
\rotate
\tabletypesize{\tiny}
\tablewidth{0pc}
\tablenum{5}
\tablecolumns{20}
\tablecaption{\label{tab:ic10}IC10 Catalog}
\tablehead{
\colhead{LGGS}
& \colhead{$\alpha_{\rm 2000}$}
& \colhead{$\delta_{\rm 2000}$}
& \colhead{$V$}
& \colhead{$\sigma_V$}
& \colhead{$B-V$}
& \colhead{$\sigma_{B-V}$}
& \colhead{$U-B$}
& \colhead{$\sigma_{U-B}$}
& \colhead{$V-R$}
& \colhead{$\sigma_{V-R}$}
& \colhead{$R-I$}
& \colhead{$\sigma_{R-I}$}
& \colhead{$N_V$}
& \colhead{$N_B$}
& \colhead{$N_U$}
& \colhead{$N_R$}
& \colhead{$N_I$}
&\colhead{Sp.\ Type}
&\colhead{Ref.}
}
\startdata
J001908.25+592855.5&00 19 08.25&+59 28 55.5&20.550& 0.021& 1.375& 0.049& 0.547& 0.104& 0.864& 0.024& 0.976& 0.012& 1& 1& 1& 1& 1&         & \\                  
J001908.27+593126.8&00 19 08.27&+59 31 26.8&18.556& 0.008& 1.358& 0.015& 0.750& 0.024& 0.827& 0.009& 0.882& 0.005& 1& 1& 1& 1& 1&         & \\                  
J001908.30+592050.7&00 19 08.30&+59 20 50.7&18.509& 0.009& 1.224& 0.014& 0.683& 0.022& 0.766& 0.010& 0.752& 0.005& 1& 1& 1& 1& 1&         & \\                  
J001908.32+590945.5&00 19 08.32&+59 09 45.5&18.531& 0.008& 1.289& 0.014& 0.567& 0.017& 0.858& 0.010& 0.841& 0.006& 1& 1& 1& 1& 1&         & \\                  
J001908.33+591514.9&00 19 08.33&+59 15 14.9&19.151& 0.008& 1.042& 0.015& 0.435& 0.022& 0.730& 0.010& 0.715& 0.006& 1& 1& 1& 1& 1&         & \\                  
\enddata
\tablecomments{Notes---Units of right ascension are hours, minutes,
and seconds, and units of declination are degrees, arcminutes, and
arcseconds. Note that an entry of ``99.999" denotes no measurement.
Table 5 is published in its entirely in the electronic edition of
the {\it Astronomical Journal}.  A portion is shown here for
guidance regarding its form and content.}
\tablerefs{For spectral types:
(1) Massey et al.\ 1992;
(2) Massey \& Armandroff 1995;
(3) Massey \& Holmes 2002;
(4) Crowther et al.\ 2003.
}
\end{deluxetable}

\begin{deluxetable}{l r r r r r r r r r r r r r r r r r l l}
\pagestyle{empty}
\rotate
\tabletypesize{\tiny}
\tablewidth{0pc}
\tablenum{6}
\tablecolumns{20}
\tablecaption{\label{tab:n6822}NGC 6822 Catalog}
\tablehead{
\colhead{LGGS}
& \colhead{$\alpha_{\rm 2000}$}
& \colhead{$\delta_{\rm 2000}$}
& \colhead{$V$}
& \colhead{$\sigma_V$}
& \colhead{$B-V$}
& \colhead{$\sigma_{B-V}$}
& \colhead{$U-B$}
& \colhead{$\sigma_{U-B}$}
& \colhead{$V-R$}
& \colhead{$\sigma_{V-R}$}
& \colhead{$R-I$}
& \colhead{$\sigma_{R-I}$}
& \colhead{$N_V$}
& \colhead{$N_B$}
& \colhead{$N_U$}
& \colhead{$N_R$}
& \colhead{$N_I$}
&\colhead{Sp.\ Type}
&\colhead{Ref.}
}
\startdata
J194335.94-145605.5&19 43 35.94&-14 56 05.5&21.023& 0.028& 1.247& 0.056&99.999&99.999& 0.827& 0.041& 0.818& 0.030& 1& 1& 0& 1& 1&         & \\                  
J194335.95-145043.5&19 43 35.95&-14 50 43.5&21.733& 0.072& 1.126& 0.106&99.999&99.999& 0.578& 0.091& 0.779& 0.056& 1& 1& 0& 1& 1&         & \\                  
J194336.01-150039.3&19 43 36.01&-15 00 39.3&21.683& 0.038& 1.023& 0.081&99.999&99.999& 0.544& 0.064& 0.552& 0.052& 1& 1& 0& 1& 1&         & \\                  
J194336.03-145851.2&19 43 36.03&-14 58 51.2&19.199& 0.008& 1.544& 0.020&99.999&99.999& 0.995& 0.010& 1.030& 0.006& 1& 1& 0& 1& 1&         & \\                  
J194336.04-150013.7&19 43 36.04&-15 00 13.7&21.102& 0.030& 1.264& 0.047&99.999&99.999& 0.809& 0.039& 0.683& 0.025& 1& 1& 0& 1& 1&         & \\                  
\enddata
\tablecomments{Notes---Units of right ascension are hours, minutes,
and seconds, and units of declination are degrees, arcminutes, and
arcseconds. Note that an entry of ``99.999" denotes no measurement.
Table 6 is published in its entirely in the electronic edition of
the {\it Astronomical Journal}.  A portion is shown here for
guidance regarding its form and content.}
\tablerefs{For spectral types:
(1) Massey et al.\ 1995a;
(2) Massey \& Johnson 1998; 
(3) Muschielok et al.\ 1999;
(4) Venn et al.\ 2001;
(5) Humphreys 1980a;
(6) Massey 1998b.}
\end{deluxetable}

\begin{deluxetable}{l r r r r r r r r r r r r r r r r r l l}
\pagestyle{empty}
\rotate
\tabletypesize{\tiny}
\tablewidth{0pc}
\tablenum{7}
\tablecolumns{20}
\tablecaption{\label{tab:wlm}WLM Catalog}
\tablehead{
\colhead{LGGS}
& \colhead{$\alpha_{\rm 2000}$}
& \colhead{$\delta_{\rm 2000}$}
& \colhead{$V$}
& \colhead{$\sigma_V$}
& \colhead{$B-V$}
& \colhead{$\sigma_{B-V}$}
& \colhead{$U-B$}
& \colhead{$\sigma_{U-B}$}
& \colhead{$V-R$}
& \colhead{$\sigma_{V-R}$}
& \colhead{$R-I$}
& \colhead{$\sigma_{R-I}$}
& \colhead{$N_V$}
& \colhead{$N_B$}
& \colhead{$N_U$}
& \colhead{$N_R$}
& \colhead{$N_I$}
&\colhead{Sp.\ Type}
&\colhead{Ref.}
}
\startdata
J000044.71-153324.4 &00 00 44.71&-15 33 24.4&21.254& 0.032& 1.780& 0.114&99.999&99.999& 0.962& 0.042&99.999&99.999& 1& 1& 0& 1& 0&         & \\                 
J000044.81-152808.8 &00 00 44.81&-15 28 08.8&20.696& 0.018& 1.395& 0.050&99.999&99.999& 0.904& 0.022&99.999&99.999& 1& 1& 0& 1& 0&         & \\                 
J000045.04-152842.2 &00 00 45.04&-15 28 42.2&19.982& 0.012& 1.317& 0.031& 1.210& 0.113& 0.836& 0.013&99.999&99.999& 1& 1& 1& 1& 0&         & \\                 
J000045.06-152246.1 &00 00 45.06&-15 22 46.1&21.754& 0.061& 0.529& 0.089&-0.853& 0.085& 0.555& 0.082&99.999&99.999& 1& 1& 1& 1& 0&         & \\                 
J000045.20-151944.4 &00 00 45.20&-15 19 44.4&20.223& 0.015& 1.498& 0.035&99.999&99.999& 1.208& 0.017& 1.625& 0.009& 1& 1& 0& 1& 1&         & \\        
\enddata
\tablecomments{Notes---Units of right ascension are hours, minutes,
and seconds, and units of declination are degrees, arcminutes, and
arcseconds. Note that an entry of ``99.999" denotes no measurement.
Table 7 is published in its entirely in the electronic edition of
the {\it Astronomical Journal}.  A portion is shown here for
guidance regarding its form and content.}
\tablerefs{For spectral types:
(1) Bresolin et al.\ 2006 }
\end{deluxetable}

\begin{deluxetable}{l r r r r r r r r r r r r r r r r r}
\pagestyle{empty}
\rotate
\tabletypesize{\tiny}
\tablewidth{0pc}
\tablenum{8}
\tablecolumns{18}
\tablecaption{\label{tab:sexB}Sextans B Catalog}
\tablehead{
\colhead{LGGS}
& \colhead{$\alpha_{\rm 2000}$}
& \colhead{$\delta_{\rm 2000}$}
& \colhead{$V$}
& \colhead{$\sigma_V$}
& \colhead{$B-V$}
& \colhead{$\sigma_{B-V}$}
& \colhead{$U-B$}
& \colhead{$\sigma_{U-B}$}
& \colhead{$V-R$}
& \colhead{$\sigma_{V-R}$}
& \colhead{$R-I$}
& \colhead{$\sigma_{R-I}$}
& \colhead{$N_V$}
& \colhead{$N_B$}
& \colhead{$N_U$}
& \colhead{$N_R$}
& \colhead{$N_I$}
}
\startdata
J095922.35+051211.7 &09 59 22.35&+05 12 11.7&18.184& 0.008& 0.571& 0.013&-0.111& 0.014& 0.370& 0.009& 0.352& 0.005& 1& 1& 1& 1& 1\\                 
J095922.44+051221.6 &09 59 22.44&+05 12 21.6&19.964& 0.023& 1.526& 0.053& 1.223& 0.156& 0.979& 0.025& 0.942& 0.010& 1& 1& 1& 1& 1 \\                 
J095922.83+052943.4 &09 59 22.83&+05 29 43.4&18.091& 0.008& 0.668& 0.011&-0.041& 0.010& 0.402& 0.011& 0.388& 0.007& 1& 1& 1& 1& 1 \\                 
J095922.88+053137.8 &09 59 22.88&+05 31 37.8&20.932& 0.060& 0.995& 0.104& 0.659& 0.141& 0.556& 0.070& 0.513& 0.036& 1& 1& 1& 1& 1 \\                 
J095922.92+051001.3 &09 59 22.92&+05 10 01.3&20.020& 0.031& 0.516& 0.043&-0.272& 0.035& 0.297& 0.037& 0.342& 0.020& 1& 1& 2& 1& 1 \\                
\enddata
\tablecomments{Notes---Units of right ascension are hours, minutes,
and seconds, and units of declination are degrees, arcminutes, and
arcseconds. Note that an entry of ``99.999" denotes no measurement.
Table 8 is published in its entirely in the electronic edition of
the {\it Astronomical Journal}.  A portion is shown here for
guidance regarding its form and content.}
\end{deluxetable}
\begin{deluxetable}{l r r r r r r r r r r r r r r r r r}
\pagestyle{empty}
\rotate
\tabletypesize{\tiny}
\tablewidth{0pc}
\tablenum{9}
\tablecolumns{18}
\tablecaption{\label{tab:sexA}Sextans A Catalog}
\tablehead{
\colhead{LGGS}
& \colhead{$\alpha_{\rm 2000}$}
& \colhead{$\delta_{\rm 2000}$}
& \colhead{$V$}
& \colhead{$\sigma_V$}
& \colhead{$B-V$}
& \colhead{$\sigma_{B-V}$}
& \colhead{$U-B$}
& \colhead{$\sigma_{U-B}$}
& \colhead{$V-R$}
& \colhead{$\sigma_{V-R}$}
& \colhead{$R-I$}
& \colhead{$\sigma_{R-I}$}
& \colhead{$N_V$}
& \colhead{$N_B$}
& \colhead{$N_U$}
& \colhead{$N_R$}
& \colhead{$N_I$}
}
\startdata
J101021.29-045018.6 &10 10 21.29&-04 50 18.6&16.003& 0.005& 0.848& 0.007&99.999&99.999& 0.511& 0.007&99.999&99.999& 1& 1& 0& 1& 0 \\                 
J101021.41-044609.4 &10 10 21.41&-04 46 09.4&21.256& 0.055& 1.077& 0.097&99.999&99.999& 0.857& 0.062& 0.714& 0.029& 1& 1& 0& 1& 1 \\                 
J101021.68-045246.3 &10 10 21.68&-04 52 46.3&21.051& 0.047& 1.310& 0.092&99.999&99.999& 0.640& 0.056& 0.645& 0.030& 1& 1& 0& 1& 1\\                 
J101022.48-044503.1 &10 10 22.48&-04 45 03.1&19.404& 0.014& 1.523& 0.033& 1.145& 0.082& 1.028& 0.016& 1.349& 0.008& 1& 1& 1& 1& 2 \\                 
J101022.50-044124.1 &10 10 22.50&-04 41 24.1&19.982& 0.017& 0.292& 0.025&99.999&99.999& 0.461& 0.027& 0.447& 0.021& 1& 1& 0& 1& 2 \\            
\enddata
\tablecomments{Notes---Units of right ascension are hours, minutes,
and seconds, and units of declination are degrees, arcminutes, and
arcseconds. Note that an entry of ``99.999" denotes no measurement.
Table 9 is published in its entirely in the electronic edition of
the {\it Astronomical Journal}.  A portion is shown here for
guidance regarding its form and content.}
\end{deluxetable}

\begin{deluxetable}{l r r r r r r r r r r r r r r r r r}
\pagestyle{empty}
\rotate
\tabletypesize{\tiny}
\tablewidth{0pc}
\tablenum{10}
\tablecolumns{18}
\tablecaption{\label{tab:peg}Pegasus  Catalog}
\tablehead{
\colhead{LGGS}
& \colhead{$\alpha_{\rm 2000}$}
& \colhead{$\delta_{\rm 2000}$}
& \colhead{$V$}
& \colhead{$\sigma_V$}
& \colhead{$B-V$}
& \colhead{$\sigma_{B-V}$}
& \colhead{$U-B$}
& \colhead{$\sigma_{U-B}$}
& \colhead{$V-R$}
& \colhead{$\sigma_{V-R}$}
& \colhead{$R-I$}
& \colhead{$\sigma_{R-I}$}
& \colhead{$N_V$}
& \colhead{$N_B$}
& \colhead{$N_U$}
& \colhead{$N_R$}
& \colhead{$N_I$}
}
\startdata
J232720.15+145207.4 &23 27 20.15&+14 52 07.4&16.059& 0.005& 0.761& 0.009& 0.180& 0.009& 0.415& 0.007&99.999&99.999& 1& 1& 1& 1& 0 \\                 
J232720.23+144410.7 &23 27 20.23&+14 44 10.7&22.169& 0.103& 1.101& 0.305&99.999&99.999& 1.200& 0.114& 1.626& 0.048& 1& 1& 0& 1& 1 \\                 
J232720.51+145345.6 &23 27 20.51&+14 53 45.6&19.042& 0.011& 1.504& 0.025& 1.133& 0.066& 0.929& 0.014& 0.959& 0.008& 1& 1& 1& 1& 1 \\                 
J232720.56+143751.6 &23 27 20.56&+14 37 51.6&16.467& 0.006& 0.892& 0.008& 0.423& 0.007& 0.502& 0.008&99.999&99.999& 1& 1& 1& 1& 0 \\                 
J232720.77+145120.1 &23 27 20.77&+14 51 20.1&21.109& 0.033& 1.522& 0.089&99.999&99.999& 0.896& 0.040& 0.931& 0.023& 1& 1& 0& 1& 1 \\                
\enddata
\tablecomments{Notes---Units of right ascension are hours, minutes,
and seconds, and units of declination are degrees, arcminutes, and
arcseconds. Note that an entry of ``99.999" denotes no measurement.
Table 10 is published in its entirely in the electronic edition of
the {\it Astronomical Journal}.  A portion is shown here for
guidance regarding its form and content.}
\end{deluxetable}

\begin{deluxetable}{l r r r r r r r r r r r r r r r r r}
\pagestyle{empty}
\rotate
\tabletypesize{\tiny}
\tablewidth{0pc}
\tablenum{11}
\tablecolumns{18}
\tablecaption{\label{tab:phx}Phoenix  Catalog}
\tablehead{
\colhead{LGGS}
& \colhead{$\alpha_{\rm 2000}$}
& \colhead{$\delta_{\rm 2000}$}
& \colhead{$V$}
& \colhead{$\sigma_V$}
& \colhead{$B-V$}
& \colhead{$\sigma_{B-V}$}
& \colhead{$U-B$}
& \colhead{$\sigma_{U-B}$}
& \colhead{$V-R$}
& \colhead{$\sigma_{V-R}$}
& \colhead{$R-I$}
& \colhead{$\sigma_{R-I}$}
& \colhead{$N_V$}
& \colhead{$N_B$}
& \colhead{$N_U$}
& \colhead{$N_R$}
& \colhead{$N_I$}
}
\startdata
J014928.54-443314.7 &01 49 28.54&-44 33 14.7&22.089& 0.084& 0.254& 0.109& 0.470& 0.110& 0.452& 0.107& 0.238& 0.066& 1& 1& 1& 1& 1 \\                 
J014929.04-443519.3 &01 49 29.04&-44 35 19.3&20.630& 0.021& 1.604& 0.065&99.999&99.999& 1.268& 0.022& 1.670& 0.008& 1& 1& 0& 1& 1 \\                 
J014929.69-442318.7 &01 49 29.69&-44 23 18.7&18.499& 0.007& 1.402& 0.019& 0.970& 0.026& 0.960& 0.009& 0.990& 0.006& 1& 1& 1& 1& 1 \\                 
J014929.70-443119.8 &01 49 29.70&-44 31 19.8&18.128& 0.006& 1.105& 0.014& 1.286& 0.018& 0.716& 0.008& 0.662& 0.005& 1& 1& 1& 1& 1\\                 
J014929.74-443600.8 &01 49 29.74&-44 36 00.8&18.356& 0.006& 0.687& 0.011& 0.163& 0.011& 0.376& 0.009& 0.412& 0.007& 1& 1& 1& 1& 1\\       
\enddata
\tablecomments{Notes---Units of right ascension are hours, minutes,
and seconds, and units of declination are degrees, arcminutes, and
arcseconds. Note that an entry of ``99.999" denotes no measurement.
Table 11 is published in its entirely in the electronic edition of
the {\it Astronomical Journal}.  A portion is shown here for
guidance regarding its form and content.}
\end{deluxetable}

\begin{deluxetable}{l r r r r r r r r r r r r r r r r r l l}
\pagestyle{empty}
\rotate
\tabletypesize{\tiny}
\tablewidth{0pc}
\tablenum{12}
\tablecolumns{20}
\tablecaption{\label{tab:m31}Revised M31 Catalog}
\tablehead{
\colhead{LGGS}
& \colhead{$\alpha_{\rm 2000}$}
& \colhead{$\delta_{\rm 2000}$}
& \colhead{$V$}
& \colhead{$\sigma_V$}
& \colhead{$B-V$}
& \colhead{$\sigma_{B-V}$}
& \colhead{$U-B$}
& \colhead{$\sigma_{U-B}$}
& \colhead{$V-R$}
& \colhead{$\sigma_{V-R}$}
& \colhead{$R-I$}
& \colhead{$\sigma_{R-I}$}
& \colhead{$N_V$}
& \colhead{$N_B$}
& \colhead{$N_U$}
& \colhead{$N_R$}
& \colhead{$N_I$}
&\colhead{Sp.\ Type}
&\colhead{Ref.}
}
\startdata
J003701.92+401233.2 &00 37  1.92&+40 12 33.2&19.862& 0.017&-0.021& 0.021&-0.928& 0.015& 0.204& 0.023&99.999&99.999& 1& 2& 1& 1& 0&         & \\                 
J003701.93+401218.4 &00 37  1.93&+40 12 18.4&18.739& 0.008& 1.494& 0.015& 0.945& 0.036& 0.946& 0.014&99.999&99.999& 1& 2& 1& 1& 0&         & \\                 
J003702.03+401141.4 &00 37  2.03&+40 11 41.4&21.225& 0.043& 1.362& 0.085&99.999&99.999& 0.748& 0.049& 0.694& 0.024& 1& 1& 0& 1& 1&         & \\                 
J003702.05+400633.5 &00 37  2.05&+40 06 33.5&21.091& 0.044& 0.050& 0.061&-1.110& 0.052& 0.042& 0.074&99.999&99.999& 1& 2& 1& 1& 0&         & \\                 
J003702.13+400945.6 &00 37  2.13&+40 09 45.6&16.091& 0.006& 1.287& 0.007& 0.983& 0.007& 0.792& 0.010&99.999&99.999& 1& 2& 1& 1& 0&         & \\              
\enddata
\tablecomments{Notes---Units of right ascension are hours, minutes,
and seconds, and units of declination are degrees, arcminutes, and
arcseconds. Note that an entry of ``99.999" denotes no measurement.
Table 5 is published in its entirely in the electronic edition of
the {\it Astronomical Journal}.  A portion is shown here for
guidance regarding its form and content.}
\tablerefs{For spectral types:
(1) Paper I;
(2) Trundle et al.\ 2002;
(3) Humphreys 1979;
(4) Massey et al.\ 1995a;
(5) Massey et al.\ 1986;
(6) Humphreys et al.\ 1990;
(7) Bianchi et al.\ 1994;
(8) P. Massey, 1996-2006, unpublished;
(9) Hubble \& Sandage 1953;
(10) Humphreys et al.\ 1988;
(11) Massey 1998b;
(12) Massey \& Johnson 1998 and references therein.
}
\end{deluxetable}
\begin{deluxetable}{l r r r r r r r r r r r r r r r r r l l}
\pagestyle{empty}
\rotate
\tabletypesize{\tiny}
\tablewidth{0pc}
\tablenum{13}
\tablecolumns{20}
\tablecaption{\label{tab:m33}Revised M33 Catalog}
\tablehead{
\colhead{LGGS}
& \colhead{$\alpha_{\rm 2000}$}
& \colhead{$\delta_{\rm 2000}$}
& \colhead{$V$}
& \colhead{$\sigma_V$}
& \colhead{$B-V$}
& \colhead{$\sigma_{B-V}$}
& \colhead{$U-B$}
& \colhead{$\sigma_{U-B}$}
& \colhead{$V-R$}
& \colhead{$\sigma_{V-R}$}
& \colhead{$R-I$}
& \colhead{$\sigma_{R-I}$}
& \colhead{$N_V$}
& \colhead{$N_B$}
& \colhead{$N_U$}
& \colhead{$N_R$}
& \colhead{$N_I$}
&\colhead{Sp.\ Type}
&\colhead{Ref.}
}
\startdata
J013146.16+301855.6 &01 31 46.16&+30 18 55.6&19.555& 0.013& 1.533& 0.025& 1.141& 0.047& 1.030& 0.015&99.999&99.999& 1& 1& 1& 1& 0&         & \\                 
J013146.18+302932.4 &01 31 46.18&+30 29 32.4&20.560& 0.068& 0.645& 0.117&99.999&99.999& 0.564& 0.115&99.999&99.999& 1& 1& 0& 1& 0&         & \\                 
J013146.18+302931.4 &01 31 46.18&+30 29 31.4&21.027& 0.061& 0.090& 0.113& 0.266& 0.118& 1.012& 0.111&99.999&99.999& 1& 1& 1& 1& 0&         & \\                 
J013146.20+302706.2 &01 31 46.20&+30 27  6.2&21.057& 0.032& 1.857& 0.084&99.999&99.999& 0.924& 0.036&99.999&99.999& 1& 1& 0& 1& 0&         & \\                 
J013146.21+302026.9 &01 31 46.21&+30 20 26.9&21.179& 0.038& 0.962& 0.066& 0.749& 0.096& 0.588& 0.047&99.999&99.999& 1& 1& 1& 1& 0&         & \\                 
\enddata
\tablecomments{Notes---Units of right ascension are hours, minutes,
and seconds, and units of declination are degrees, arcminutes, and
arcseconds. Note that an entry of ``99.999" denotes no measurement.
Table 5 is published in its entirely in the electronic edition of
the {\it Astronomical Journal}.  A portion is shown here for
guidance regarding its form and content.}
\tablerefs{For spectral types:
(1) Humphreys 1980b;
(2) Massey et al.\ 1996;
(3) Massey et al.\  1995a;
(4) P. Massey 1996-2006, unpublished;
(5) Monteverde et al.\ 1996;
(6) Hubble \& Sandage 1953;
(7) van den Bergh et al.\  1975;
(8) Massey et al.\ 1998b;
(9) Massey \& Johnson 1998 and references therein.
}
\end{deluxetable}

\begin{deluxetable}{l c c c c c c c c c c c c c c c c c}
\pagestyle{empty}
\tabletypesize{\tiny}
\tablewidth{0pc}
\tablenum{14}
\tablecolumns{18}
\tablecaption{\label{tab:errors}Median Errors}
\tablehead{
\colhead{}
&\multicolumn{5}{c}{IC10}
&
&\multicolumn{5}{c}{WLM+NGC 6822}
&
&\multicolumn{5}{c}{Others} \\ \cline{2-6} \cline{8-12} \cline{14-18}
\colhead{Magnitude}
&\colhead{$\sigma_U$}
&\colhead{$\sigma_B$}
&\colhead{$\sigma_V$}
&\colhead{$\sigma_R$}
&\colhead{$\sigma_I$}
&
&\colhead{$\sigma_U$}
&\colhead{$\sigma_B$}
&\colhead{$\sigma_V$}
&\colhead{$\sigma_R$}
&\colhead{$\sigma_I$}
&
&\colhead{$\sigma_U$}
&\colhead{$\sigma_B$}
&\colhead{$\sigma_V$}
&\colhead{$\sigma_R$}
&\colhead{$\sigma_I$}
}
\startdata
15.0-15.5&\nodata&0.035&0.005&0.002&0.000&&\nodata&0.022&0.005&0.002&0.005&&  0.029&0.023&0.005&0.002&0.004\\
15.5-16.0&\nodata&0.025&0.005&0.002&0.005&&  0.005&0.011&0.005&0.002&0.005&&  0.005&0.014&0.005&0.002&0.005\\
16.0-16.5& 0.005&0.009&0.005&0.002&0.005 &&  0.005&0.005&0.005&0.002&0.005&&  0.005&0.002&0.005&0.002&0.005\\
16.5-17.0& 0.005&0.003&0.005&0.002&0.005 &&  0.005&0.002&0.005&0.002&0.005&&  0.005&0.002&0.005&0.002&0.005\\
17.0-17.5& 0.005&0.002&0.005&0.002&0.005 &&  0.005&0.002&0.005&0.002&0.005&&  0.005&0.002&0.005&0.002&0.003\\
17.5-18.0& 0.005&0.002&0.005&0.002&0.005 &&  0.005&0.002&0.005&0.002&0.005&&  0.005&0.002&0.005&0.002&0.003\\
18.0-18.5& 0.005&0.002&0.005&0.002&0.005 &&  0.005&0.002&0.005&0.002&0.004&&  0.005&0.002&0.005&0.003&0.002\\
18.5-19.0& 0.005&0.002&0.005&0.002&0.004 &&  0.005&0.002&0.005&0.002&0.003&&  0.005&0.002&0.005&0.004&0.002\\
19.0-19.5& 0.005&0.002&0.005&0.002&0.002 &&  0.005&0.002&0.005&0.003&0.002&&  0.006&0.005&0.006&0.006&0.002\\
19.5-20.0& 0.005&0.002&0.005&0.003&0.002 &&  0.006&0.003&0.005&0.006&0.002&&  0.008&0.007&0.009&0.009&0.002\\
20.0-20.5& 0.006&0.002&0.005&0.005&0.002 &&  0.008&0.005&0.006&0.008&0.002&&  0.010&0.010&0.012&0.012&0.002\\
20.5-21.0& 0.009&0.002&0.005&0.008&0.002 &&  0.012&0.009&0.009&0.012&0.002&&  0.016&0.014&0.017&0.018&0.002\\
21.0-21.5& 0.013&0.002&0.007&0.011&0.002 &&  0.017&0.013&0.014&0.018&0.002&&  0.021&0.022&0.024&0.025&0.002\\
21.5-22.0& 0.019&0.002&0.010&0.016&0.004 &&  0.027&0.019&0.022&0.029&0.002&&  0.032&0.032&0.034&0.038&0.002\\
22.0-22.5& 0.030&0.004&0.014&0.026&0.016 &&  0.045&0.030&0.034&0.047&0.010&&  0.044&0.048&0.053&0.065&0.005\\
22.5-23.0& 0.061&0.007&0.022&0.045&0.052 &&  0.086&0.053&0.060&0.074&0.036&&  0.064&0.087&0.106&0.123&0.021\\
23.0-23.5& 0.124&0.010&0.035&0.086&0.138 &&  0.169&0.090&0.106&0.121&0.116&&  0.098&0.158&0.192&0.283&0.080\\
23.5-24.0& 0.228&0.015&0.058&0.155&0.260 &&  0.289&0.142&0.189&0.210&0.229&&  0.166&0.294&0.400&0.498&0.190\\
24.0-24.5& 0.348&0.022&0.120&0.306&\nodata&& 0.443&0.234&0.337&0.391&\nodata&&0.281&0.440&0.477&\nodata&\nodata\\
24.5-25.0& 0.551&0.036&0.214&\nodata&\nodata&& 0.522&0.363&0.468&0.391&\nodata&&0.433&0.492&\nodata&\nodata&\nodata\\
25.0-25.5& \nodata & 0.083 & 0.419 & \nodata & \nodata && \nodata & 0.467 & \nodata & \nodata &\nodata && \nodata & \nodata & \nodata & \nodata & \nodata \\
25.5-26.0& \nodata & 0.174 & \nodata & \nodata & \nodata && \nodata & \nodata & \nodata & \nodata & \nodata && \nodata & \nodata & \nodata & \nodata & \nodata \\  
\enddata
\end{deluxetable}

\begin{deluxetable}{l r r r r r}
\tabletypesize{\tiny}
\tablewidth{0pc}
\tablenum{15}
\tablecolumns{6}
\tablecaption{\label{tab:bright}Brightness of a 20$M_\odot$ Star}
\tablehead{
\colhead{Stage\tablenotemark{a}}
&\colhead{$U$} 
&\colhead{$B$}
&\colhead{$V$}
&\colhead{$R$}
&\colhead{$I$}
}
\startdata
\cutinhead{IC10}
ZAMS (9.5~V)   & 23.2   & 23.8  &  23.2   & 22.8  &  22.5\\
TAMS (B1~I)    & 21.6   & 22.1  &  21.4  &  21.0   & 20.7\\
RSG            & 25.6    &22.6   & 19.9  &  18.5   & 17.2\\
\cutinhead{NGC 6822}
ZAMS (O9.5~V)   & 19.8   & 20.8   & 20.8 &   20.7   & 20.8\\
TAMS (B1 ~I)     &18.2 &   19.1   & 19.0   & 18.9   & 19.0\\
RSG              &22.2    &19.6   & 17.5   & 16.4   & 15.5\\
\cutinhead{WLM}
ZAMS (O9.5~V)  &  20.3  &  21.3  &  21.6 &   21.6 &   21.8\\
TAMS (B1~I)     & 18.7   & 19.6  &  19.8   & 19.8  &  20.0\\
RSG              &22.7   & 20.1   & 18.3   & 17.3   & 16.5\\
\cutinhead{Sextans B}
ZAMS (O9.5~V)   & 21.2   & 22.2 &   22.4  &  22.4   & 22.6\\
TAMS (B1~I)      19.6  &  20.5  &  20.6  &  20.6  &  20.8\\
RSG             & 23.6 &   21.0   & 19.1  &  18.1  &  17.3\\
\cutinhead{Sextans A}
ZAMS (O9.5~V)   & 21.2   & 22.2  &  22.5 &   22.5  &  22.7\\
TAMS (B1~I)     & 19.6   & 20.5   & 20.7  &  20.7   & 20.9\\
RSG             & 23.6   & 21.0   & 19.2 &   18.2   & 17.4\\
\cutinhead{Pegasus}
ZAMS (O9.5~V)  &  20.3    &21.2  &  21.4  &  21.4   &21.5\\
TAMS (B1~I)     & 18.7   & 19.5 &   19.6 &   19.6 &   19.7\\
RSG             & 22.7    &20.0 &   18.1&    17.1&    16.2\\
\cutinhead{Phoenix}
ZAMS (O9.5~V)  & 18.9 &   19.9&    20.0  &  20.0 &   20.1\\
TAMS (B1~I)    & 17.3  &  18.2 &   18.2  &  18.2  &  18.3\\
RSG            & 21.3  &  18.7   & 16.7  &  15.7  &  14.8\\
\cutinhead{M31}
ZAMS (O9.5~V)   & 20.2  &  21.2 &   21.3 &   21.3  &  21.5\\
TAMS (B1~I)      &18.6  & 19.5   & 19.5 &   19.5 &   19.7\\
RSG             & 22.6   & 20.0   & 18.0   & 17.0  &  16.2\\
\cutinhead{M33}
ZAMS (O9.5~V)   & 20.3  &  21.3   & 21.5 &   21.5 &   21.6\\
TAMS (B1~I)     & 18.7   & 19.6   & 19.7 &   19.7  &  19.8\\
RSG           &   22.7  &  20.1 &   18.2  &  17.2    &16.3\\
\cutinhead{LMC}
ZAMS (O9.5~V)   &  9.2   & 10.3   & 10.4   & 10.4  &  10.6\\
TAMS (B1~I)     &  7.6    & 8.6    & 8.6   &  8.6  &   8.8\\
RSG            &  11.6   &  9.1   &  7.1    & 6.1    & 5.3\\
\cutinhead{SMC}
ZAMS (O9.5~V)  &   9.0  &  10.1  &  10.4   & 10.5   & 10.7\\
TAMS (B1~I)     &  7.4   &  8.4   &  8.6   &  8.7   &  8.9\\
RSG            &  11.4   &  8.9 &    7.1   &  6.2    & 5.4\\
\enddata
\tablenotetext{a}{A 20$M_\odot$ 
zero-age main sequence (ZAMS) star taken to be 09.5~V with
$M_V=-3.5$, and $(U-B)_0=-1.1$, $(B-V)_0=-0.3$, $V-R=-0.1$, and
$(R-I)_0=-0.2$.  A 20$M_\odot$ terminal-age main sequence (TAMS) star
taken to be B1~I with $M_V=-5.3$, $(U-B)_0=-1.0$, $(B-V)=-0.2$,
$(V-R)_0=-0.1$, and $(R-I)_0=-0.2$.  A 20$M_\odot$ RSG is assumed to
have $M_V=-6.8$, and $(U-B)_0=+2.5$, $(B-V)_0=+1.8$, $(V-R)_0=+0.9$,
$(R-I)_0=+0.8$.}
\end{deluxetable}

\begin{deluxetable}{l c c c c c c c}
\pagestyle{empty}
\tabletypesize{\small}
\tablewidth{0pc}
\tablenum{16}
\tablecolumns{8}
\tablecaption{\label{tab:fore}Foreground Contribution to Selected Portions of the CMDs}
\tablehead{
\colhead{Galaxy}
&\multicolumn{2}{c}{Selection Criteria}
&\colhead{Area} 
&\colhead{Foreground}
&\multicolumn{3}{c}{Relative Contributions (\%)} \\ \cline {2-3} \cline{6-8}
&\colhead{$V$ Range}&\colhead{$B-V$ Range}&\colhead{(deg$^2$)}&(\%)&
\colhead{Disk dwarfs}
&\colhead{Halo dwarfs}
&\colhead{Halo giants}
}
\startdata
M31 & 15-20 & 0.4-1.1 & 2.2 & 50 & 86 & 7 & 7 \\ 
M31 & 16-20 & 1.2-1.8 & 2.2 & 85 & 100 & 0 & 0 \\
M33 & 15-20 & 0.4-1.1 & 0.8 & 40 & 73 & 15 & 12 \\
M33 & 16-20 & 1.2-1.8 & 0.8 & 70 & 99 & 1 & 0 \\
LMC & 11-15 & 0.4-1.1 & 14.5 & 54 & 85 & 9 & 6\\
SMC & 11-15 & 0.4-1.1 & 7.2 & 45 & 82 & 6 & 12 \\
IC10 & 16-20 & 0.5-2.0 & 0.026 & 100 & 100 & 0 & 0 \\
NGC 6822 & 16-20 & 0.5-1.4 & 0.052 & 94 & 68 & 12 & 20 \\
WLM  & 16-20 & 0.4-1.0 & 0.023 & 95 & 11 & 55 & 34 \\
WLM & 16-20 & 1.1-1.9 & 0.023 & 74 & 97 & 3 & 0 \\
Sextans B & 16-20 & 0.4-1.1 & 0.016 & 85 & 42 & 34 & 24 \\
Sextans A & 16-20 & 0.4-1.1 & 0.016 & 85 & 47 & 30 & 23 \\
Pegasus & 16-20 & 0.4-1.9 & 0.020 & 48 & 69 & 18 & 13 \\
Phoenix & 16-20 & 0.4-1.9 & 0.017 & 30 & 52 & 30 & 18 \\
 \enddata
 \end{deluxetable}

\begin{deluxetable}{l l c c }
\pagestyle{empty}
\tablewidth{0pc}
\tablenum{17}
\tablecolumns{4}
\tablecaption{\label{tab:reddenings}Reddenings}
\tablehead{
\colhead{Galaxy}
&\colhead{Total}
&\colhead{Foreground}
&\colhead{Internal} \\
&\colhead{$E(B-V)$}
&\colhead{$E(B-V)$\tablenotemark{a}}
&\colhead{$E(B-V)$ }
}
\startdata
M31 & $0.13\pm0.02$ & 0.06 & 0.07 \\
M33 & $0.12\pm0.02$ & 0.05 & 0.07 \\
LMC & 0.13\tablenotemark{b} & 0.08 & 0.05 \\
SMC & 0.09\tablenotemark{b} & 0.04 & 0.05 \\
IC10 & $0.81\pm0.02$ & 0.7-1.3 & \nodata \\
NGC 6822 & $0.25\pm0.02$ & 0.22 & 0.03 \\
WLM & $0.07\pm0.05$ & 0.03 & 0.04 \\
Sextans B & $0.09\pm0.05$ & 0.03 & 0.06 \\
Sextans A & $0.05\pm0.05$ & 0.04 & 0.01\\
Pegasus & $0.15\pm0.05$ & 0.06 & 0.09\\
Phoenix & $0.15\pm0.05$ & 0.02 & 0.13\\
\enddata
\tablenotetext{a}{Calculated from the Schlegel et al.\ 1998 100$\mu$m dust emission maps, using software kindly made publicly available by D. Schlegel and D. Finkbeiner,
via http://astro.berkeley.edu/davis/dust/local/local/html.  The estimates for the foreground reddening towards M31, the LMC, and SMC come directly from their web site.}
\tablenotetext{b}{Adopted from Massey et al.\ 1995b.}
\end{deluxetable}

\begin{deluxetable}{l c c r r r r r l l l l}
\pagestyle{empty}
\tabletypesize{\tiny}
\tablewidth{0pc}
\tablenum{18}
\tablecolumns{12}
\tablecaption{\label{tab:ic10mem}IC10 Members Confirmed by Spectroscopy}
\tablehead{
\colhead{LGGS}
&\colhead{$\alpha_{\rm 2000}$}
&\colhead{$\delta_{\rm 2000}$}
&\colhead{$V$}
&\colhead{$B-V$}
&\colhead{$U-B$}
&\colhead{$V-R$}
&\colhead{$R-I$}
&\colhead{Sp. Type}
&\colhead{Notes\tablenotemark{a}}
&\colhead{Cross-ID\tablenotemark{b}}
&\colhead{Ref.}
}
\startdata
J001956.99+591707.5 &00 19 56.99&+59 17 07.5&21.47& 0.64& -0.37& 0.42& 0.34&WC4-5+abs& &[MAC92] 1 &2,4\\
J001959.63+591654.7 &00 19 59.63&+59 16 54.7&21.02& 0.81& -0.29& 0.53& 0.53&WC4      &M&[MAC92] 2 &1,4\\
J001959.69+591654.9 &00 19 59.69&+59 16 54.9&21.58& 1.59& -2.10& 1.04& 0.56&WC4      &M&[MAC92] 2 &1,4\\
J002003.01+591827.0 &00 20 03.01&+59 18 27.0&22.20& 1.08& 99.99& 0.82& 0.79&WC4      & &RSMV6     &4  \\
J002004.25+591806.2 &00 20 04.25&+59 18 06.2&21.66& 0.86& -0.27& 0.56& 0.57&WC4-5+abs& &RSMV5     &4  \\
J002011.55+591857.9 &00 20 11.55&+59 18 57.9&19.62& 0.69& -0.39& 0.50& 0.51&WC4-5    & &[MAC92] 4 &2,4\\
J002012.84+592008.1 &00 20 12.84&+59 20 08.1&21.59& 0.95& -0.30& 0.44&-0.11&WNE/C4   & &[MAC92] 5 &2,4\\
J002020.31+591839.5 &00 20 20.31&+59 18 39.5&21.54& 0.43& -0.11& 0.70& 0.69&WNE+abs  &M&RSMV9     &4  \\
J002020.34+591840.1 &00 20 20.34&+59 18 40.1&22.19&-0.34&  0.04& 1.31& 0.72&WNE+abs  &M&RSMV9     &4  \\
J002020.56+591837.3 &00 20 20.56&+59 18 37.3&20.24& 0.68& -0.54& 0.61& 0.45&WN10     & &RSMV8     &4  \\
J002021.87+591741.1 &00 20 21.87&+59 17 41.1&19.09& 0.67& -0.63& 0.45& 0.45&WC4-5+abs&M&[MAC92] 7 &2,4\\
J002021.97+591741.2 &00 20 21.97&+59 17 41.2&20.79& 0.20&  0.46& 0.49& 0.14&WC4-5+abs&M&[MAC92] 7 &2,4\\
J002022.68+591846.8 &00 20 22.68&+59 18 46.8&23.22& 0.23& 99.99& 0.56&99.99&WN3      & &[MAC92] 9 &2,4\\
J002022.76+591753.4 &00 20 22.76&+59 17 53.4&22.82& 1.14& 99.99& 0.67& 1.23&WC4      & &RSMV11    &4  \\
J002023.35+591742.2 &00 20 23.35&+59 17 42.2&20.52& 0.41&  0.32& 0.48& 0.10&WC7      & &[MAC92] 10&2,4\\
J002025.70+591648.3 &00 20 25.70&+59 16 48.3&22.53& 1.00& 99.99& 0.60& 0.62&WNE      & &RSMV12    &4  \\
J002026.20+591726.3 &00 20 26.18&+59 17 26.3&21.15& 1.08& -0.46& 0.96& 1.34&WC4      & &[MAC92] 12&2,4\\
J002026.54+591705.0 &00 20 26.54&+59 17 05.0&22.63& 0.58& 99.99& 0.89&-0.48&WC4      & &RSMV10    &4  \\
J002026.69+591732.8 &00 20 26.69&+59 17 32.8&20.89& 0.54& -0.48& 0.55& 0.62&WC5-6    & &[MAC92] 13&2,4\\
J002026.94+591719.9 &00 20 26.94&+59 17 19.9&20.61& 0.58& -0.01& 0.64& 0.90&WC5-6    & &[MAC92] 14&2,4\\
J002027.02+591818.0 &00 20 27.02&+59 18 18.0&22.27& 0.68& 99.99& 0.58& 0.55&WC6-7    & &[MAC92] 15&2  \\
J002027.73+591737.3 &00 20 27.73&+59 17 37.3&18.56& 0.56& -0.11& 0.37& 0.34&WN       &M&[MAC92] 24&3  \\
J002027.75+591738.1 &00 20 27.75&+59 17 38.1&18.62& 0.58& -0.22& 0.47& 0.38&WN       &M&[MAC92] 24&3  \\
J002028.07+591714.3 &00 20 28.07&+59 17 14.3&21.54& 0.78& -0.26& 0.70& 0.71&WN7-8    & &RSMV2     &4  \\
J002029.08+591651.7 &00 20 29.08&+59 16 51.7&21.72& 0.91& -1.09& 0.14& 0.75&WNE+OB   &M&[MAC92] 17&2,4\\
J002029.12+591651.8 &00 20 29.12&+59 16 51.8&22.48& 0.02& -0.92& 0.85& 0.81&WNE+OB   &M&[MAC92] 17&2,4\\
J002031.05+591904.2 &00 20 31.05&+59 19 04.2&22.60& 0.34& 99.99& 0.56& 0.34&WN4      & &[MAC92] 19&1,4\\
J002032.80+591716.4 &00 20 32.80&+59 17 16.4&21.38& 1.44& 99.99& 1.29& 1.42&WN7-8    & &[MAC92] 23&3  \\
J002034.52+591714.6 &00 20 34.52&+59 17 14.6&22.17& 0.42&  0.29& 0.59& 0.49&WC5      & &[MAC92] 20&1,4\\
J002041.61+591624.8 &00 20 41.61&+59 16 24.8&22.75& 0.36& 99.99& 0.56& 0.46&WN4      & &[MAC92] 21&2,4\\
\enddata
\tablecomments{Notes---Units of right ascension are hours, minutes,
and seconds, and units of declination are degrees, arcminutes, and
arcseconds. Note that an entry of ``99.999" denotes no measurement.}
\tablenotetext{a}{``M" denotes multiple cross-identifications due to crowding.}
\tablerefs{For spectral types:
(1) Massey et al.\ 1992;
(2) Massey \& Armandroff 1995;
(3) Massey \& Holmes 2002;
(4) Crowther et al.\ 2003.
}
\end{deluxetable}

\begin{deluxetable}{l c c r r r r r l l l l}
\pagestyle{empty}
\tabletypesize{\tiny}
\tablewidth{0pc}
\tablenum{19}
\tablecolumns{12}
\tablecaption{\label{tab:n6822mem}NGC 6822 Members Confirmed by Spectroscopy}
\tablehead{
\colhead{LGGS}
&\colhead{$\alpha_{\rm 2000}$}
&\colhead{$\delta_{\rm 2000}$}
&\colhead{$V$}
&\colhead{$B-V$}
&\colhead{$U-B$}
&\colhead{$V-R$}
&\colhead{$R-I$}
&\colhead{Sp. Type}
&\colhead{Notes\tablenotemark{a}}
&\colhead{Cross-ID}
&\colhead{Ref.}
}
\startdata
J194422.24-144342.4 &19 44 22.24&-14 43 42.4&17.33& 1.51& 1.51& 0.94&\nodata&RSG       && N6822a-96\tablenotemark{d} & 6 \\
J194431.99-144409.1&19 44 31.99&-14 44 09.1&19.69&-0.18&-0.76& 0.28& 0.16&WN      &M & N6822-WR3\tablenotemark{b} & 2 \\
J194434.21-144219.9&19 44 34.21&-14 42 19.9&17.75& 0.40&-1.09& 0.07& 0.02&O9.5~I  &M & N6822ob3-7=F9\tablenotemark{c,g}   &   1 \\
J194437.97-145106.2&19 44 37.97&-14 51 06.2&19.83& 0.01&-0.76& 0.12& 0.10&WN       &    &N6822-WR4\tablenotemark{b} & 2\\
J194501.60-145440.0 &19 45 01.60&-14 54 40.0&16.88& 0.28&-0.59& 0.19& 0.21&B8-A0: I  & &E141\tablenotemark{c} & 5 \\
J194448.11-144518.1 &19 44 48.11&-14 45 18.1&18.48& 2.03& 2.08& 1.07& 0.97&RSG       & & N6822b-356\tablenotemark{d} & 6 \\
J194448.65-145025.9 &19 44 48.65&-14 50 25.9&18.85& 0.86& 0.16& 0.55& 0.53&Emission  & & C72\tablenotemark{c} & 5 \\
J194449.03-144526.9&19 44 49.03&-14 45 26.9&17.54& 0.12&-0.74& 0.11& 0.08&EarlyB   &     &N6822ob7-15=B17\tablenotemark{c,e} & 1 \\
J194449.31-144404.1&19 44 49.31&-14 44 04.1&18.20& 0.09&-0.79& 0.13& 0.16&B0 Ia     &     & N6822ob6-16 & 3 \\
J194449.36-144539.8&19 44 49.36&-14 45 39.8&19.72& 0.28&-0.63& 0.29& 0.40&WN        &     &N6822-WR5\tablenotemark{b} & 2\\
J194449.96-144333.5 &19 44 49.96&-14 43 33.5&18.07& 2.21& 2.09& 1.29&\nodata&M2.5 I    & & N6822b-395\tablenotemark{d} & 6 \\
J194450.21-144253.6&19 44 50.21&-14 42 53.6&16.93& 0.23&-0.71& 0.15& 0.14&B1.5 III  &     &N6822ob8F-2=D14\tablenotemark{c,f}  & 1    \\
J194451.10-144355.4 &19 44 51.10&-14 43 55.4&18.81& 1.98& 1.49& 1.40& 1.64&M2.5 I/M1 I     & &N6822b-434=V12\tablenotemark{d} & 6, 5\\
J194451.18-144919.8 &19 44 51.18&-14 49 19.8&17.50& 0.23&-0.44& 0.20& 0.17&B0-1 Ia         & &A66\tablenotemark{c} & 5\\
J194451.53-144429.5 &19 44 51.53&-14 44 29.5&18.70& 1.87& 1.31& 1.03& 0.94&K0-3 I    & & N6822b-395\tablenotemark{d} &  6\\
J194451.67-144351.8 &19 44 51.67&-14 43 51.8&17.70& 2.03& 0.15& 1.13&\nodata&K5 I/M0-1 I     & & N6822b-1133=C26\tablenotemark{d} & 6, 5\\
J194452.28-145220.6 &19 44 52.28&-14 52 20.6&16.45& 0.36&-0.16& 0.20& 0.21&B1 I      & & C74\tablenotemark{c} & 5\\
J194453.25-144640.3&19 44 53.25&-14 46 40.3&17.39& 0.32&-0.35& 0.21& 0.26&A3 Ia     &     &A13=CW185\tablenotemark{c} & 4\\
J194453.96-144424.3 &19 44 53.96&-14 44 24.3&18.19& 2.01& 1.94& 1.04&\nodata&RSG       &&N6822b-1134\tablenotemark{d} & 6 \\
J194454.54-145127.1 &19 44 54.54&-14 51 27.1&17.05& 2.25& 2.15& 1.19&\nodata&M0-1 I    &&C79\tablenotemark{c} & 5\\
J194454.81-144347.8 &19 44 54.81&-14 43 47.8&18.05& 2.24&-0.30& 1.28& 1.30&cM/RSG  & & N6822b-554=V14\tablenotemark{d} & 5,6\\
J194455.08-145213.1 &19 44 55.08&-14 52 13.1&15.99& 0.70& 0.02& 0.27&\nodata&B5 Ia     & & C84\tablenotemark{c} & 5 \\
J194455.27-144631.6&19 44 55.27&-14 46 31.6&18.96& 0.19&-0.39& 0.30& 0.33&Early BI  & M & N66822ob7F-40 & 1 \\ 
J194455.47-144930.0 &19 44 55.47&-14 49 30.0&17.72& 0.27&-0.28& 0.20& 0.23&B1-2 I    & & A73\tablenotemark{c} & 5 \\
J194455.70-145155.4 &19 44 55.70&-14 51 55.4&16.91& 2.20& 1.99& 1.17&\nodata&M1-2 I    & & V18 & 5 \\
J194456.19-144503.0 &19 44 56.19&-14 45 03.0&16.68& 0.35&-0.73& 0.18& 0.16&OB        & & B1\tablenotemark{c} & 5 \\
J194456.32-144612.1&19 44 56.32&-14 46 12.1&17.22& 0.80&-1.04& 0.19& 0.22&A2 Ia      &    &A101=CW173\tablenotemark{c} & 4\\
J194457.31-144920.2 &19 44 57.31&-14 49 20.2&17.41& 2.28&-0.04& 1.21&\nodata&M1 I      & & V19 & 5\\
J194457.44-144345.0&19 44 57.44&-14 43 45.0&19.43&-0.05&-0.73& 0.00&-0.02& Of:        &    &N6822ob9-20A& 1 \\
J194458.31-144446.9 &19 44 58.31&-14 44 46.9&17.89& 2.11& 1.34& 1.18& 1.15&cM/RSG    &M&N6822b-684=V15\tablenotemark{d} &5,6\\
J194459.78-144857.6 &19 44 59.78&-14 48 57.6&17.94& 0.06&-0.36& 0.10& 0.07&B5 I      & &A165\tablenotemark{c}& 5\\
J194459.86-144515.4 &19 44 59.86&-14 45 15.4&16.93& 2.00& 1.81& 1.00&\nodata &M0 I/RSG      &M&N6822b-210=B110\tablenotemark{d} & 5,6\\
J194500.31-144434.9&19 45 00.31&-14 44 34.9&18.46& 0.06&-0.66& 0.09& 0.06& B2 Ia       &    &N6822ob11-8 & 3 \\
J194500.42-144823.1 &19 45 00.42&-14 48 23.1&17.35& 0.31&-0.42& 0.34& 0.47&B5 I      &M&A153\tablenotemark{c}&5\\
J194501.91-144732.2 &19 45 01.91&-14 47 32.2&17.55& 1.01& 0.57& 0.61& 0.57&RSG       & & N6822c-108\tablenotemark{d} & 6 \\
J194505.25-144312.4&19 45 05.25&-14 43 12.4&18.20& 0.30&-0.91&-0.23&-0.45& Early O      &M& N6822ob13-9 & 1 \\
J194513.26-144508.0&19 45 13.26&-14 45 08.0&18.02&-0.09&-0.69& 0.01&-0.07& Early B     & &N6822ob15-9 & 1 \\
J194513.50-144512.9&19 45 13.50&-14 45 12.9&18.96&-0.17&-0.54& 0.16& 0.04& WNE          &M&N6822-WR12  & 2 \\
\enddata
\tablecomments{Notes---Units of right ascension are hours, minutes,
and seconds, and units of declination are degrees, arcminutes, and
arcseconds. Note that an entry of ``99.999" denotes no measurement.}
\tablenotetext{a}{``M" denotes multiple.}
\tablenotetext{b}{For the WR stars, we have retained the nomenclature of Massey \& Johnson 1998; the CDS lists these stars as ``[AM85] N".}
\tablenotetext{c}{Designations  from Kayser 1966 
(``A" refers to her inner-most field), some of which are
cross referenced to Wilson 1992 
(``CW",  following Massey et al.\ 1995a).}
\tablenotetext{d}{For the RSGs, we have used the designations of Massey (1998b), cross referenced when possible
to Kayser 1966.}
\tablenotetext{e}{Classified as ``A0 Ia" by Humphreys 1980a.}
\tablenotetext{f}{Classified as ``B5-8 I" by Humphreys 1980a.}
\tablenotetext{g}{Classified as ``B1-2 I" by Humphreys 1980a.}
\tablerefs{For spectral types:
(1) Massey et al.\ 1995a;
(2) Massey \& Johnson 1998; 
(3) Muschielok et al.\ 1999;
(4) Venn  et al.\ 2001;
(5) Humphreys 1980a;
(6) Massey 1998b.}
\end{deluxetable}

\begin{deluxetable}{l c c r r r r r l l l l}
\pagestyle{empty}
\tabletypesize{\tiny}
\tablewidth{0pc}
\tablenum{20}
\tablecolumns{12}
\tablecaption{\label{tab:wlmmem}WLM Members Confirmed by Spectroscopy}
\tablehead{
\colhead{LGGS}
&\colhead{$\alpha_{\rm 2000}$}
&\colhead{$\delta_{\rm 2000}$}
&\colhead{$V$}
&\colhead{$B-V$}
&\colhead{$U-B$}
&\colhead{$V-R$}
&\colhead{$R-I$}
&\colhead{Sp. Type}
&\colhead{Notes\tablenotemark{a}}
&\colhead{Cross-ID\tablenotemark{b}}
&\colhead{Ref.}
}
\startdata
J000153.19-152729.1&00 01 53.19&-15 27 29.1& 19.01& -0.15& -0.89& -0.05& -0.10&B2II      & &B10=SC27    & 1 \\
J000153.22-152839.5&00 01 53.22&-15 28 39.5& 17.97&  0.00& -0.67&  0.03&  0.01&B9Ia      & &A12=SC22    & 1 \\
J000153.33-152851.9&00 01 53.33&-15 28 51.9& 19.88& -0.19& -0.75& -0.07& -0.10&B2.5Ib    & &A13=SC21    & 1 \\
J000153.63-152829.8&00 01 53.63&-15 28 29.8& 18.87& -0.15& -0.85& -0.05& -0.10&B1Ia      & &B13=SC23    & 1 \\
J000154.06-152745.4&00 01 54.06&-15 27 45.4& 19.32& -0.17& -0.87& -0.05& -0.09&B0Iab     & &A10=SC26    & 1 \\
J000155.03-152659.7&00 01 55.03&-15 26 59.7& 19.55& -0.10& -0.66& -0.02& -0.07&B3Ib      & &A8=SC55     & 1 \\
J000155.69-152449.0&00 01 55.69&-15 24 49.0& 20.48& -0.11& -0.59&  0.01& -0.06&B5II      & &B3=SC58     & 1 \\
J000156.16-152624.5&00 01 56.16&-15 26 24.5& 19.79&  0.22&  0.01&  0.18&  0.20&A7Ib      & &A6=SC51     & 1 \\
J000156.34-152758.3&00 01 56.34&-15 27 58.3& 19.77& -0.02& -0.49&  0.08&  0.08&A0Iab     & &B11         & 1 \\
J000156.36-152606.7&00 01 56.36&-15 26 06.7& 19.97&  0.09& -0.08&  0.09&  0.14&A2II      & &B6=SC52     & 1 \\
J000156.45-152901.6&00 01 56.45&-15 29 01.6& 19.90& -0.04& -0.39&  0.01&  0.01&A0Ib      & &B14         & 1 \\
J000156.62-152501.5&00 01 56.62&-15 25 01.5& 21.09&  1.09& -0.05&  0.62&  0.54&G0I       & &A3          & 1 \\
J000156.75-152636.6&00 01 56.75&-15 26 36.6& 20.27&  0.11& -0.35&  0.15&  0.10&A3II      & &B8          & 1 \\
J000157.17-152614.3&00 01 57.17&-15 26 14.3& 20.31&  0.14& -0.02&  0.15&  0.17&A3II      & &B7          & 1 \\
J000157.20-152718.0&00 01 57.20&-15 27 18.0& 18.39&  0.20& -1.25& -0.03& -0.06&B1.5Ia    & &A9=SC35     & 1 \\
J000157.89-153013.1&00 01 57.89&-15 30 13.1& 18.38&  0.20& -0.73&  0.06&  0.05&A2Ia      & &A16=SC16    & 1 \\
J000158.12-152648.5&00 01 58.12&-15 26 48.5& 19.66& -0.09& -0.97& -0.05& -0.11&B1.5Ia    & &A7=SC37     & 1 \\
J000158.46-152433.8&00 01 58.46&-15 24 33.8& 20.88&  1.56&100.00&  0.83&  0.75&G5I       & &A1          & 1 \\
J000158.73-153001.5&00 01 58.73&-15 30 01.5& 20.03& -0.02& -0.32&  0.05&  0.04&comp      &M &B16=SC12    & 1 \\
J000159.04-152442.8&00 01 59.04&-15 24 42.8& 20.14& -0.01& -0.23&  0.05&  0.03&A0II      & &A2=SC68     & 1 \\
J000159.41-153046.7&00 01 59.41&-15 30 46.7& 21.13& -0.05& -0.93&  0.10&  0.08&comp      & &B18         & 1 \\
J000159.56-152926.1&00 01 59.56&-15 29 26.1& 18.37&  0.09& -0.16&  0.07&  0.12&A2II      & &A14=SC15    & 1 \\
J000159.61-153059.9&00 01 59.61&-15 30 59.9& 18.98&  1.78&  1.55&  0.91&  0.82&G2I       & &A18=SC4     & 1 \\
J000159.88-152528.3&00 01 59.88&-15 25 28.3& 20.89&  0.26& -0.01&  0.22&  0.19&A5II      & &B4          & 1 \\
J000159.95-152819.0&00 01 59.95&-15 28 19.0& 18.38& -0.11& -1.04& -0.07& -0.12&O9.7Ia    & &A11=SC30    & 1 \\
J000200.02-152545.0&00 02 00.02&-15 25 45.0& 20.85& -0.03& -0.37&  0.02& -0.04&A0II      & &B5          & 1 \\
J000200.03-152930.9&00 02 00.03&-15 29 30.9& 20.27&  0.19&  0.01&  0.16&  0.19&A5II      & &B15         & 1 \\
J000200.19-153014.1&00 02 00.19&-15 30 14.1& 20.19& -0.03& -0.52&  0.06&  0.04&comp      & &B17         & 1 \\
J000200.48-153108.1&00 02 00.48&-15 31 08.1& 19.74& -0.16& -0.73& -0.09& -0.09&B2Ib      & &B19         & 1 \\
J000200.52-152951.8&00 02 00.52&-15 29 51.8& 20.25& -0.28& -0.98& -0.12& -0.15&O7V((f))  & &A15=SC14    & 1 \\
J000200.62-152829.8&00 02 00.62&-15 28 29.8& 18.76&  0.08& -0.12&  0.09&  0.61&A2II      & &B12=SC31    & 1 \\
J000200.81-153024.8&00 02 00.81&-15 30 24.8& 19.31& -0.11& -0.59& -0.03&  0.45&B5Ib      & &A17=SC9     & 1 \\
J000200.81-153115.7&00 02 00.81&-15 31 15.7& 18.69&  1.78&  0.77&  0.91&  0.34&G2I       & &A19=SC6     & 1 \\
J000201.57-152527.0&00 02 01.57&-15 25 27.0& 20.18&  0.02& -0.17&  0.04&  0.06&A2II      & &A4          & 1 \\
J000201.91-152725.2&00 02 01.91&-15 27 25.2& 19.77&  0.01& -0.26&  0.04&  0.07&A0II      & &B9=SC42     & 1 \\
J000203.31-152552.6&00 02 03.31&-15 25 52.6& 19.41& -0.12& -0.58& -0.02& -0.08&B8Iab     & &A5=SC45     & 1 \\
J000204.38-152446.5&00 02 04.38&-15 24 46.5& 20.51& -0.10& -0.47& -0.02& -0.04&A2Ia      & &B2          & 1 \\
J000205.15-152422.9&00 02 05.15&-15 24 22.9& 20.96&  0.14&  0.05&  0.17&  0.16&A2II      & &B1          & 1 \\
\enddata
\tablecomments{Notes---Units of right ascension are hours, minutes,
and seconds, and units of declination are degrees, arcminutes, and
arcseconds. Note that an entry of ``99.999" denotes no measurement.}
\tablenotetext{a}{``M" denotes multiple.}
\tablenotetext{b}{Designations  ``A" and ``B" from Bresolin et al 2006; ``SC" from
Sandage \& Carlson 1985b.
} 
\tablerefs{For spectral types:
(1) Bresolin et al.\ 2006.
}
\end{deluxetable}
\end{document}


\title{ERRATUM: A Survey of Local Group Galaxies Currently Forming Stars: \\II. 
{\it UBVRI} Photometry of Stars in 
Seven Dwarfs 
and a Comparison of the Entire Sample (2007, AJ 133, 2474)}

\author{Philip Massey,\altaffilmark{1, 2}
K. A. G. Olsen,\altaffilmark{3}
Paul  W. Hodge,\altaffilmark{4}
George H. Jacoby,\altaffilmark{5}\\
Reagin T. McNeill,\altaffilmark{1}
R. C. Smith,\altaffilmark{3}
Shay B. Strong\altaffilmark{1}}

\altaffiltext{1}{Lowell Observatory, 1400 W. Mars Hill Rd.,
Flagstaff, AZ 86001; Phil.Massey@lowell.edu.}

\altaffiltext{2} {Visiting Astronomer,
Kitt Peak National Observatory, National Optical
Astronomy Observatory (NOAO), which is operated by the Association of
Universities for Research in Astronomy (AURA), Inc., under cooperative agreement
with the National Science Foundation (NSF).}

\altaffiltext{3} {Cerro Tololo Inter-American Observatory, NOAO, which is operated by AURA, Inc., under cooperative agreement
with the NSF; kolsen@noao.edu, csmith@noao.edu.}

\altaffiltext{4}{Department of Astronomy, University of Washington,
Seattle WA 98195; hodge@astro.washington.edu.}

\altaffiltext{5}{GMTO Corporation, 813 Santa Barbara St., Pasadena, CA 91101; gjacoby@mgto.org.}
    
Massey et al.\ (2006) presented the Local Group Galaxy Survey (LGGS) {\it UBVRI} photometry of stars in M31 and M33.
These catalogs were reissued with a slight change in the naming convention in Massey et al.\ (2007, hereafter Paper II), Tables 12 and 13.
 Last year Nelson Caldwell
(2009, private communication) called our attention to the fact that there appeared to be a slight over abundance of stars in the M31 color
magnitude diagram with a $U-B$ value near 0.00.
We established that this came about due to a coding error by the first author
in how the photometry from adjacent fields were averaged.  The mistake affects {\it only} the $U-B$ and $R-I$ values for a small
fraction of the stars in the M31 and M33 catalogs.  These two colors were the only ones affected as stars were required to have
$V$, $B-V$, and $V-R$ values to be included in the catalog but not $U-B$ and not $R-I$, and the averaging failed for these under
certain conditions.  In addition, the {\it errors} on the $V-R$ colors were found to be underestimated in some cases.

These problems affected {\it only} the M31 and M33 photometry  (Tables 12 and 13) in Paper II,
 and not the photometry of the 7 dwarf galaxies in our sample, as the latter
were each observed in a single field rather than in overlapping fields.   The on-line version of this erratum contains the revised versions
of Tables 12 and 13 (also available from http://www.lowell.edu/users/massey/Table12Revised.txt and \\ http://www.lowell.edu/users/massey/Table13Revised.txt).

For M31 the $V-R$ errors were underestimated for 2,906 stars out 371,781 (0.8\%).  The $U-B$ values had to be revised for 13,532 stars
out of 192,602 with $U-B$ measurements (7.0\%).  Of these, only 1,025  
stars (0.5\%) had to be revised by more than 0.05 mag in $U-B$.  For the 329,931 M31 stars with $R-I$ measurements, 4,225 values (1.3\%) had
to be revised, but again only a handful (333, or 0.1\%) had significant ($>$0.05~mag) differences.

For M33, the statistics are similarly encouraging.  The $V-R$ errors were underestimated for 2,696 stars out of 146,622 (1.8\%).
The $U-B$ values had to be revised for 2,244 stars out of 113,058 with $U-B$ measurements (2.0\%).  Of these, only 783 stars (0.7\%)
had to be revised by more than 0.05 mag in $U-B$.  For the 133,258 M33 stars with $R-I$ measurements, 3,211 values (2.4\%) had to be
revised, but again only a handful (714, or 0.5\%) had significant ($>$0.05~mag) differences.

We are grateful to Nelson Caldwell for calling our attention to this problem, and we apologize for any inconvenience this error may have caused.
The work involved in this revision was supported by the National Science Foundation under grant AST-1008020.